\documentclass[twocolumn]{aastex62}
\usepackage{graphicx}
\usepackage{longtable}
\usepackage{rotating}
\usepackage{latexsym}
\usepackage{lineno}
\usepackage{amsmath}

\DeclareRobustCommand{\ion}[2]{%
\relax\ifmmode
\ifx\testbx\f@series
{\mathbf{#1\,\mathsc{#2}}}\else
{\mathrm{#1\,\mathsc{#2}}}\fi
\else\textup{#1\,{\mdseries\textsc{#2}}}%
\fi}

\received{July 1, 2020}
\revised{July 1, 2020}
\accepted{\today}
\submitjournal{ApJ}

\shorttitle{$\textit{WISE}$ study of G23}
\shortauthors{H. F. M. Yao et al.}
\begin{document}

\title{Galaxy and Mass Assembly (GAMA): A $\textit{WISE}$ study of the activity of emission-line systems in G23}

\correspondingauthor{Harris Yao Fortune Marc}
\email{yao.fortune@gmail.com}

\author[0000-0002-0786-7307]{H. F. M. Yao}
\affil{Department of Physics and Astronomy, University of the Western Cape, Robert Sobukwe Road, Bellville, 7535, Republic of South Africa}

\author{T. H. Jarrett}
\affil{Department of Astronomy, University of Cape Town, Private Bag X3, Rondebosch, 7701, South Africa}

\author{M. E. Cluver}
\affiliation{Centre for Astrophysics and Supercomputing, Swinburne University of Technology, John Street, Hawthorn, 3122, Australia}
\affiliation{Department of Physics and Astronomy, University of the Western Cape, Robert Sobukwe Road, Bellville, 7535, Republic of South Africa}

\author{L. Marchetti}
\affiliation{Department of Physics and Astronomy, University of the Western Cape, Robert Sobukwe Road, Bellville, 7535, Republic of South Africa}
\affiliation{Department of Astronomy, University of Cape Town, Private Bag X3, Rondebosch, 7701, South Africa}

\author{Edward N. Taylor}
\affiliation{Centre for Astrophysics and Supercomputing, Swinburne University of Technology, John Street, Hawthorn, 3122, Australia}

\author{M. G. Santos}
\affiliation{Department of Physics and Astronomy, University of the Western Cape, Robert Sobukwe Road, Bellville, 7535, Republic of South Africa}

\author{Matt S. Owers}
\affiliation{Department of Physics and Astronomy, Macquarie University, NSW 2109, Australia}
\affiliation{Astronomy, Astrophysics and Astrophotonics Research Centre,  Macquarie University, Sydney, NSW, 2109}

\author{Angel R. Lopez-Sanchez}
\affiliation{Australian Astronomical Optics, Macquarie University, 105 Delhi Rd, North Ryde, NSW 2113, Australia}
\affiliation{Department of Physics and Astronomy, Macquarie University, NSW 2109, Australia}
\affiliation{Astronomy, Astrophysics and Astrophotonics Research Centre,  Macquarie University, Sydney, NSW, 2109}
\affiliation{ARC Center of Excellence for All Sky Astrophysics in 3 Dimensions (ASTRO-3D)}

\author{Y. A. Gordon}
\affiliation{Department of Physics and Astronomy, University of Manitoba, Winnipeg, MB R3T 2N2, Canada}

\author{M. J. I. Brown}
\affiliation{School of physics and Astronomy, Monash University, Clayton, Victoria 3800, Australia}

\author{S. Brough}
\affiliation{School of Physics, University of New South Wales, NSW 2052, Australia}

\author{S. Phillipps}
\affiliation{Astrophysics Group, School of Physics, University of Bristol, Tyndall Avenue, Bristol BS8 1TL, UK}

\author{B. W. Holwerda}
\affiliation{Department of Physics and Astronomy, 102 Natural Science Building, University of Louisville, Louisville KY 40292, USA}

\author{A.M. Hopkins}
\affiliation{Australian Astronomical Optics, Macquarie University, 105 Delhi Rd, North Ryde, NSW 2113, Australia}

\author{L. Wang}
\affiliation{SRON Netherlands Institute for Space Research, Landleven 12, 9747 AD, Groningen, The Netherlands}
\affiliation{Kapteyn Astronomical Institute, University of Groningen, Postbus 800, 9700 AV Groningen, The Netherlands}

\begin{abstract}

We present a detailed study of emission-line systems in the GAMA G23 region, making use of $\textit{WISE}$ photometry that includes carefully measured resolved sources. After applying several cuts to the initial catalogue of $\sim$41,000 galaxies, we extract a sample of 9,809 galaxies. We then compare the spectral diagnostic (BPT) classification of 1154 emission-line galaxies (38$\%$ resolved in W1) to their location in the $\textit{WISE}$ colour-colour diagram, leading to the creation of a new zone for mid-infrared ``warm'' galaxies located 2$\sigma$ above the star-forming sequence, below the standard $\textit{WISE}$ AGN region. We find that the BPT and $\textit{WISE}$ diagrams agree on the classification for 85$\%$ and 8$\%$ of the galaxies as non-AGN (star forming = SF) and AGN, respectively, and disagree on $\sim$7$\%$ of the entire classified sample. 39$\%$ of the AGN (all types) are broad-line systems for which the [\ion{N}{ii}] and [H$\alpha$] fluxes can barely be disentangled, giving in most cases spurious [\ion{N}{ii}]/[H$\alpha$] flux ratios. However, several optical AGN appear to be completely consistent with SF in $\textit{WISE}$. We argue that these could be low power AGN, or systems whose hosts dominate the IR emission. Alternatively, given the sometimes high [\ion{O}{iii}] luminosity in these galaxies, the emission lines may be generated by shocks coming from super-winds associated with SF rather than the AGN activity. Based on our findings, we have created a new diagnostic: [W1-W2] vs [\ion{N}{ii}]/[H$\alpha$], which has the virtue of separating SF from AGN and high-excitation sources. It classifies 3$\sim$5 times more galaxies than the classic BPT. 

\end{abstract}


\section{Introduction} \label{sec:intro}

Mapping galaxies at different redshifts provides important insights into how galaxies develop and transform throughout cosmic history. One of the major discoveries in astronomy is the existence of black holes at the centres of massive galaxies which  when associated with an active accretion disc are called active galactic nuclei (AGN). Powerful AGN such as quasi-stellar objects (QSO) have a much larger luminosity than that of  normal star-forming galaxies, which appears to be due to the accretion of cold materials onto the supermassive black hole at the centre of the host galaxy (Kormendy $\&$ Richstone  \citeyear{Kormendy1995}; Magorrian et al. \citeyear{Magorrian1998}; Ferrarese  $\&$ Merritt  \citeyear{Ferrarese2000}; Graham \citeyear{Graham2016}). But generally AGN have a large variety of luminosities that can sometimes be dominated by the star light from the host galaxy (e.g. Graham $\&$ Soria \citeyear{Graham2019}).  Therein lies the problem for a galaxy's global properties such as star formation rate (SFR) or stellar mass are indispensable for understanding galaxy evolution (see review by Kennicutt \citeyear{Kennicutt1998}). The SFR is derived based on the assumption that the radiation measured originates from hot young stars, subsequently dust reprocessing, and then adopting an IMF to estimate the SFR across the aggregate mass spectrum (Driver et  al. \citeyear{Driver2008}). 
Significantly, any AGN activity within the host galaxy could not only contaminate the flux measurement, but also have a direct impact on the star formation process itself via AGN feedback (Kauffmann $\&$ Haehnelt \citeyear{Kauffmann2000}; Croton et al. \citeyear{Croton2006};  Graham $\&$ Scott \citeyear{Graham2013}; Sahu et  al. \citeyear{Sahu2019}). Therefore, a reliable AGN selection method is of great importance for both accurate measurement of the host's properties and a better understanding of the AGN activity.
Several methods have been devised to this end based on the characteristics of AGN in different wavebands spanning radio to $\gamma$ rays.

The most common excitation mechanisms in galaxies are:  photoionization by O and B stars,  photoionization by a power-law continuum source, and shock-wave heating. There is also excitation by planetary nebulae, which are photoionized by stars that are usually much hotter than normal galactic OB stars.
 So, any classification scheme based on the mechanisms described above could delineate classes of galaxies according to the process taking place.
 Emission line diagnostics:  from observational studies and simulations the following emission features, [\ion{Ne}{v}]\,$\lambda$3426, [\ion{O}{ii}]\,$\lambda$3727. [\ion{He}{ii}]\,$\lambda$4686, H$\beta$\,$\lambda$4861, [\ion{O}{iii}]\,$\lambda$5007, [\ion{O}{i}]\,$\lambda$6300, H$\alpha$\,$\lambda$6563, and [\ion{N}{ii}]\,$\lambda$6584 were proposed as potentially useful lines for discriminating between different excitation mechanisms.  
 One of the earliest was Baldwin, Philips $\&$ Terlevich (\citeyear{Baldwin1981}). Their method was  based on the optical line ratios  ([\ion{O}{iii}]/H$\beta$ vs [\ion{N}{ii}]/H$\alpha$; 
 [\ion{O}{iii}]/H$\beta$ vs [\ion{S}{ii}]/H$\alpha$; [\ion{O}{iii}]/H$\beta$ vs [\ion{O}{i}]/H$\alpha$) combined to form different planes referred to as BPT diagrams. These line ratios are sensitive to the hardness of the ionizing extreme ultraviolet radiation (EUV), so that they can provide important constraints on the shape of the EUV spectrum, and may also be used to estimate the mean ionization parameter and metallicity of the galaxies. AGN have a much harder ionizing spectrum  than hot stars, they can thus be differentiated from star-forming (SF) and starburst galaxies. 
 
 As galaxy surveys became much larger with better data, the techniques to understand the BPT diagrams have also improved, and updated cuts for distinguishing SF vs. AGN ionization have been published (see Veilleux et al. \citeyear{Veilleux2001}; Kewley et al. \citeyear{Kewley2001}; Kauffmann et al. \citeyear{Kauffmann2003}; Kewley et al. \citeyear{Kewley2006}). For this study we follow the recipes provided by Kewley et al. (\citeyear{Kewley2006}), which are the most commonly used. Following Kauffmann et al. (\citeyear{Kauffmann2003}) we can classify galaxies into 3 categories: pure SF, pure AGN, and intermediate objects which lie between the other two categories.

\begin{figure*}
\centering
        \includegraphics[width= 15cm]{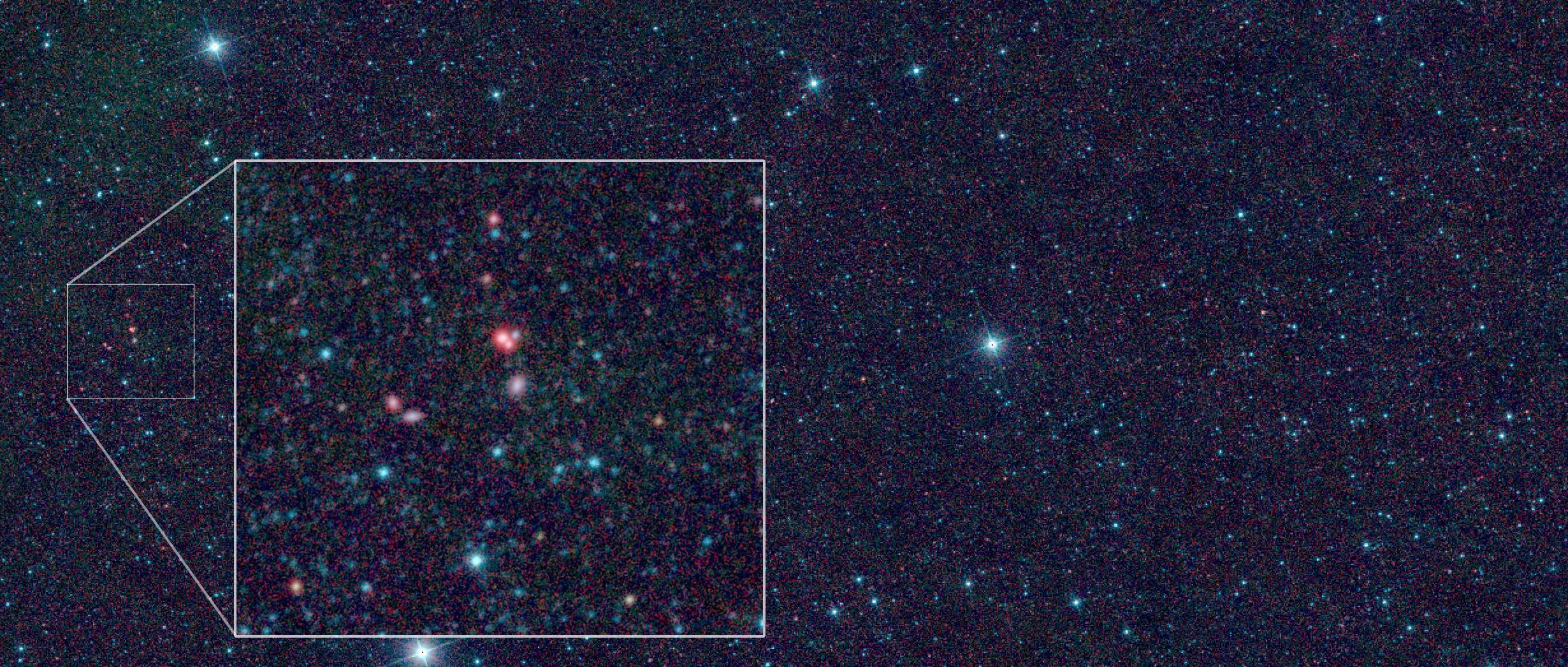}
    \caption{The $\textit{WISE}$ 50 deg$^2$ view of the G23 field. In the zoomed area (14$\times$11 arcmin) the reddish extended galaxies are easily distinguishable from foreground stars (blue in general). The image was created combining the 4 $\textit{WISE}$ colours: 3.4\,$\mu$m (blue), 4.6\,$\mu$m (green), 12\,$\mu$m (orange) and 22\,$\mu$m (red). The 50 deg$^2$ area contains about 700,000 $\textit{WISE}$ sources (galaxies + stars).}  \label{fig1}
\end{figure*}

The Wide-field Infrared Survey Explorer $\textit{(WISE)}$ is ideal to study the interplay between hot dust and AGN activity.
The AGN activity is most closely related to cold gas falling onto the central supermassive black hole  (SMBH) within the host galaxy. This flow starts from galactic scales beyond  100 pc to the sub-parsec environment. On the opposite side, the radiative pressure unbinds a large amount of the gas which is expelled via a radiative-pressure-driven wind (H{\"o}nig et al. \citeyear{Honig2019}; Leftley et al. \citeyear{Leftley2019}). Subsequently, this wind creates the 
conditions for dust formation (Sarangi et al.  \citeyear{Sarangi2019}).
 The radiation from the torus of dust around the SMBH and from evolved stars distributed throughout the disk and the bulge populations, can be detected in the mid-infrared.

Launched in December 2009, $\textit{WISE}$ surveyed the entire sky in four mid-infrared bands: 3.4\,$\mu$m,  4.6\,$\mu$m, 12\,$\mu$m, and 22\,$\mu$m  respectively, W1, W2, W3 and W4 (Wright et al. \citeyear{Wright2010}). 
While W1 and W2 are both sensitive to the continuum emission from evolved stars, the W2 band is additionally sensitive to hot dust; hence, this makes the 3.4\,$\mu$m - 4.6\,$\mu$m colour a good diagnostic to identify galaxies dominated globally by AGN emission (see e.g.,  Jarrett et al. \citeyear{Jarrett2011}; Stern et al. \citeyear{Stern2012}). 
The technique consists of projecting galaxies in the W1 - W2 vs W2 - W3  plane (eg.,  Wright et al. \citeyear{Wright2010}). Stars and early-type galaxies have colours near zero, while brown dwarfs are very red in W1-W2 and blue in W2-W3, and ultraluminous infrared galaxies (ULIRG) tend to be red in both colours. Different colour cuts have been proposed to separate AGN from star-forming galaxies, the most conservative being  W1 - W2 
= 0.8\,mag (Stern et al. \citeyear{Stern2012};  Assef et al. \citeyear{Assef2012}).

The different techniques show some disagreements from one waveband to the other. For example, a substantial number of galaxies classified as X-ray AGN appear to be star-forming
in the optical (Agostino $\&$ Salim \citeyear{Agostino2019}) and similar trends are also seen in the radio (Leahy et al. \citeyear{Leahy2019}), the mid-infrared (Ching et al. \citeyear{Ching2017}; Lam et al. \citeyear{Lam2018}) and others.

Given the relative scarcity of X-ray and optical spectroscopic data, $\textit{WISE}$ (an all-sky survey) is, therefore, a perfect tool for studying mid-infrared AGN in the local volume of space.
For example, in the study of Leahy et al. (\citeyear{Leahy2019}) $\textit{WISE}$ data in the G23 region is used in combination with the radio continuum fluxes and optical emission-line diagnostics to investigate AGN activity. 

However, the $\textit{WISE}$ survey only provided photometry optimised for point sources, with large and resolved systems often treated as agglomerations of point sources or missing significant flux. Therefore, any work based on nearby and extended galaxies requires reprocessing of all available $\textit{WISE}$ imaging and the careful extraction the resolved galaxies (see Jarrett et al. \citeyear{Jarrett2013}; Cluver et al. \citeyear{Cluver2014}; Jarrett et al. \citeyear{Jarrett2017}; Jarrett et al. \citeyear{Jarrett2019} for details). 

Our team earlier targeted the equatorial fields  of 
the Galaxy And Mass Assembly survey (GAMA, Driver et al. \citeyear{Driver2009};  Driver et al. \citeyear{Driver2011})  for which catalogues have already been generated and used for multi-wavelength studies (see Cluver et al. \citeyear{Cluver2014}; Jarrett et al. \citeyear{Jarrett2017}).
{\bf The current study represents the first $\textit{WISE}$ nearby galaxy catalogue in the southern sky (the GAMA G23 field), used in combination with the spectroscopic data available 
in GAMA to investigate the similarities and differences between the optical and mid-infrared activity.}

The structure of this paper is as follows: In Section~2 we describe the datasets and the generation of the $\textit{WISE}$ galaxy catalogue. In Section~3 we describe the properties of the cross-matched sample and the methods we use to divide SF galaxies and AGN. We also present some detailed case studies. The discussion of our results is presented in Section~4. Finally Section~5 summarises our conclusions.

The cosmology adopted throughout this paper is H$_0$ = 70\,km s$^{-1}$ Mpc$^{-1}$, $\Omega$$_{M}$ = 0.3 and $\Omega$$_{\Lambda}$= 0.7. The conversions between the luminosity distance and the redshift use the analytic formalism of Wickramasinghe $\&$ Ukwatta (\citeyear{Wickramasinghe2010}) for a flat, dark energy dominated Universe, assuming standard cosmological values noted above.
 All magnitudes are in the Vega system ($\textit{WISE}$ photometric calibration described in Jarrett et al. \citeyear{Jarrett2011}) unless indicated explicitly by an AB subscript. Photometric colours are indicated using band names; e.g., W1 - W2 is the [3.4\,$\mu$m]  - [4.6\,$\mu$m]  colour. Finally, for all four bands, the Vega magnitude to flux conversion factors are 309.68, 170.66, 29.05, 7.871\,Jy, respectively, for W1, W2, W3, and W4. Here we have adopted the new W4 calibration from Brown et al. (\citeyear{Brown2014b}), in which the central wavelength is 22.8\,$\mu$m and the magnitude-to-flux conversion factor is 7.871\,Jy. It follows that the conversions from Vega System to the monochromatic AB System  are 2.67, 3.32, 5.24 and 6.66\,mag.

\section{DATA AND METHOD}

The primary data for this study comes from the Galaxy and Mass Assembly (GAMA) Survey (see Driver et al.  \citeyear{Driver2011}). GAMA is a spectroscopic survey complete down to 19.8 mag in the r-band (for the 3 equatorial fields), covering a total area of $\sim$ 286 deg$^2$ . Its multi-wavelength coverage spans the optical (VST KiDS; Kilo- Degree Survey), near-infrared (VIKING; VISTA Kilo-degree Infrared Galaxy Survey), and far-infrared (Herschel). 

\subsection{The GAMA G23 region}

Located between Galactic longitude from 339$^\circ$ to 351$^\circ$ and Galactic latitude from -35$^\circ$ to -30$^\circ$,  G23 is the largest of the two GAMA southern fields (G23, G02). The main surface area after accounting for lost areas due to bright stars is about 50 deg$^2$.  
The optical spectra were obtained primarily with the 2dF instrument mounted on the 3.9-m Anglo-Australian Telescope  at Siding Spring Observatory (Australia), that feeds the AAOmega spectrograph. Additional 2dF data were extracted from the 2dFGRS (Colless et al. \citeyear{Colless2001}), and also from the 6dF Galaxy Survey (6dFGS, Jones et al. \citeyear{Jones2010}) that used the 1.2m UK Schmidt Telescope, also at Siding Spring Observatory. More details about the GAMA survey strategy and spectroscopic data products can be found in Baldry et al. (\citeyear{Baldry2010}); Baldry et al. (\citeyear{Baldry2018}); Robotham et al. (\citeyear{Robotham2010});  Hopkins et al. (\citeyear{Hopkins2013}); Gunawardhana et al. (\citeyear{Gunawardhana2013});  Liske et al. (\citeyear{Liske2015}). The spectroscopic redshifts used in this study are obtained from G23TilingCatv11.
Redshifts are associated with a quality number from 0 to 4. 0 represents a failure  of the data reduction,  1 means no redshift is found and 4 is the highest certainty with an associated probability $\geqslant$\,0.95 (Driver et al. \citeyear{Driver2011}).

The raw data  from the GAMA survey are processed using the software developed  at  AAO called 2DFDR (Sharp $\&$ Birchall \citeyear{Sharp2010}). The spectral output obtained from the 2DFDR is spectroscopically calibrated 
following the idlspec2d pipeline used for the SDSS DR6 (Adelman-McCarthy et al. \citeyear{Adelman2008}). Checking the calibrated data for the galaxies observed by both GAMA and SDSS shows good agreement  as presented in Hopkins et al. (\citeyear{Hopkins2013}). 

\subsubsection{Spectral line  measurements}  \label{sec1}

The optical line fluxes of the galaxies are extracted from the G23 GAMA II catalogue (``GaussFitSimplev05'' from within the SpecLineSFR DMU; Gordon et al. \citeyear{Gordon2017}) with redshifts $>$ 0.002  and a magnitude limit of i $<$ 19.2\,mag.  

The spectra are fitted  with the IDL code ``mpfitfun'' (Markwardt \citeyear{Markwardt2009}) which uses a Levenberg-Marquardt non-linear least squares minimisation to identify the best-fitting parameters for the model given the data and its associated uncertainties.

 Among the lines available in the catalogue, we use H$\beta$\,$\lambda$4861, [\ion{O}{iii}]\,$\lambda$4959, [\ion{O}{iii}]\,$\lambda$5007, [\ion{N}{ii}]\,$\lambda$6548, H$\alpha$\,$\lambda$6563, [\ion{N}{ii}]\,$\lambda$6583,  [\ion{S}{ii}]\,$\lambda$6716 and [\ion{S}{ii}]\,$\lambda$6731.

For lines that are always expected to be narrow such as [\ion{S}{ii}], the fitting can either be a straight line (in the case of non-detection) or a Gaussian. Fitting the H$\beta$ and H$\alpha$ in combination with their associated lines [\ion{O}{iii}]\,$\lambda$4959/$\lambda$5007 and [\ion{N}{ii}]\,$\lambda$6548/$\lambda$6583, respectively, can have more levels of complexity above the continuum. For example, fitting  H$\alpha$  +  [\ion{N}{ii}]\,$\lambda$6548/$\lambda$6583 can increase in complexity as follows:

\begin{itemize}

\item[-] no emission or absorption, just continuum\\

\item[-]  H$\alpha$ in absorption and no [\ion{N}{ii}]\,$\lambda$6548/$\lambda$6583 emission\\

\item[-]  [\ion{N}{ii}]\,$\lambda$6548 + H$\alpha$ + [\ion{N}{ii}]\,$\lambda$6583 all with narrow emission\\

\item[-]  [\ion{N}{ii}]\,$\lambda$6548 + [\ion{N}{ii}]\,$\lambda$6583 in emission + H$\alpha$  in absorption\\

\item[-]  [\ion{N}{ii}]\,$\lambda$6548 + [\ion{N}{ii}]\,$\lambda$6583 in emission + H$\alpha$  in emission and absorption\\

\item[-]  [\ion{N}{ii}]\,$\lambda$6548 +[NII]\,$\lambda$6583 in emission + H$\alpha$  in narrow plus broad emission\\

\end{itemize}

For each of the above fits, three different model selection methods are used to give a single global score based on which a more complicated model should be chosen versus a  simpler and vice-versa.
For models with the same number of fitted parameters, the model with the lowest $\chi$$^2$ value is chosen.

The following limits are applied  to the fitting model:\\
The line position is maintained within 200 km\,s$^{-1}$ of the expected position. The width of the Gaussian ($\sigma$) for the narrow emission lines is constrained by $0.75\sigma_{\text{inst}}  <\sigma<   \sqrt{500{^2} + \sigma{^2}_{\text{inst}}}$ where $\sigma_{\text{inst}}$ is the resolution of the spectrum in km\,s$^{-1}$.
The precedent constrains are made larger for broad emission-lines, $ \sqrt{500{^2} + \sigma{^2}_{\text{inst}}}  <\sigma<   \sqrt{5000{^2} + \sigma{^2}_{\text{inst}}}$ and the position of the line is maintained within 400 km\,s$^{-1}$. 

Only the AAOmega spectra are flux calibrated and can be used as individual line fluxes unlike in the line ratio used for both calibrated and uncalibrated spectra. For the doublets such as [\ion{O}{iii}]\,$\lambda$4959/$\lambda$5007, [\ion{N}{ii}]$\lambda$6548/$\lambda$6583 and [\ion{S}{ii}]\,$\lambda$6716/$\lambda$6731, the line with the longest wavelength is generally  the strongest (e.g., [\ion{O}{iii}]\,$\lambda$5007 can be up to 3 times stronger than [\ion{O}{iii}]\,$\lambda$4959) and is used to estimate the H$\alpha$/[\ion{O}{iii}] line ratio. H$\alpha$ and H$\beta$  must be corrected for stellar absorption as it was not accounted for in the fitting. 

\subsection{$\textit{WISE}$ image reconstruction}

The $\textit{WISE}$ image mosaics for G23 are reconstructed using the ``drizzle" resampling technique (Jarrett et al. \citeyear{Jarrett2012}; Masci \citeyear{Masci2013}). The final images have a native resolution of 5.9$^ {\prime \prime}$, 6.5$^ {\prime \prime}$, 7.0$^ {\prime \prime}$ and 12.4$^ {\prime \prime}$ in W1, W2, W3 and W4, respectively, as opposed to 8.1$^ {\prime \prime}$, 8.8$^ {\prime \prime}$, 11.0$^ {\prime \prime}$ and 17.5$^ {\prime \prime}$ given by the $\textit{WISE}$ all-sky `Atlas' imaging. For more information about the archival data, see Cutri et al. (\citeyear{Cutri2012}). The source measurement, characterisation and extraction processes
are amply described in Cluver et al. (\citeyear{Cluver2014}) and Jarrett et al. (\citeyear{Jarrett2013}, \citeyear{Jarrett2017}). 

Figure \ref{fig1} is the 4-band combination of all the mosaics in the G23 field ($\sim$ 50\,deg$^2$). Stars have not yet been removed and are mostly bluer than the extragalactic sources. Some red extended galaxies can be easily identified by eye in the zoomed area.

\subsection{Catalogue of $\textit{WISE}$  galaxies in G23} \label{sec23}

A positional cross-match between $\textit{WISE}$ sources from the $\textit{ALLWISE}$ catalogue and galaxies with known redshifts in the GAMA-G23 region was performed using a 3$ ^{\prime \prime}$ cone radius as suggested by Cluver et al. (\citeyear{Cluver2014}). This cross-match radius is motivated by the fact that the PSF-fitting of the $\textit{ALLWISE}$ pipeline can affect the resulting accuracy of the source's position particularly for resolved sources. This produces a catalogue of 40,843 $WISE$-GAMA galaxies.

We next briefly describe the extraction and cleaning method used to create our final catalogue of $\textit{WISE}$ galaxies (W1 resolved and point-like sources) in the GAMA G23 region, and the sample used for the current study.  This  is the fourth catalogue of its kind (after the GAMA G12, G09 and G15 fields) and the full procedure is described in  Cluver et al. (\citeyear{Cluver2014}) and Jarrett et al. (\citeyear{Jarrett2017}).

\subsubsection{Extraction and cleaning method}

The whole process of extracting the WISE resolved sources arises from the fact that the survey was optimised for the detection of point-source galaxies. Potentially resolved galaxies based on the W1 (3.4\,$\mu$m) band are selected from the $\textit{ALLWISE}$ catalogue using the PSF goodness of the fit metric, w1rchi2. We search for values of w1rchi2 $>$ 2, re-measure them with our special resolved source pipeline that characterises the global and surface brightness emission, revealing sources that are extended beyond the PSF.  These candidates are then characterized using the resolved-galaxy pipeline. 

Using the reprocessed $\textit{WISE}$ imaging described above, the photometry can be remeasured as appropriate for a resolved source. The different steps are done via a custom software  which adapted tools developed  for 2MASS extended sources  (Jarrett et al. \citeyear{Jarrett2000}) and the $\textit{WISE}$ pipeline (Jarrett et al. \citeyear{Jarrett2011}, \citeyear{Jarrett2013}; Cutri et al. \citeyear{Cutri2012}). The process is semi-automated such that each level can be visually inspected and corrected if necessary. The details are as follows:\\
(i) The point sources are removed by a combination of PSF-substraction and pixel masking. The bright stars in the vicinity are also masked out for a better estimation of the background. The shape of the galaxy is determined based on the 3$\sigma$ isophote (of the background RMS) which is considered to be constant at all isophotes. The first  set of photometry is derived at 1$\sigma$ isophote which encloses already more than 90$\%$ of the light (Jarrett et al. \citeyear{Jarrett2019}). A double S\'ersic fit (Graham $\&$ Driver \citeyear{Graham2005}) is used to estimate the light beyond the 1$\sigma$ isophote as well as computing the curve of growth asymptotic fluxes extending down to convergence.\\
Extracted sources are then removed from the images using their smoothed light distribution.\\
(ii) The process is repeated until all the sources are extracted.\\
(iii) The Rfuzzy parameter (see Cluver et al. \citeyear{Cluver2014}) is the 2nd order intensity weighted moment which allows a separation between faint, resolved and unresolved systems. Each type of object has a characteristic Rfuzzy value that is systematically smaller than that of the resolved galaxies. This step is important in order to get our final sample of resolved galaxies.\\
 
The final step is visual inspection to validate and identify blends and other complex scenarios. The whole area of study is divided into smaller areas centred on galaxies identified as resolved and processed by the pipeline. Each one 
of the galaxies is visualised (by the first author)  to make sure that it is definitely not a star or unresolved galaxy, that 
the de-blending is well performed (mostly for close proximity blend cases), and the apertures and position angle follow well the 2D light profile. The pipeline does well when the background is smooth and the signal to noise ratio (S/N) high (10 or higher). It needs human assistance in dense and low S/N environments, and with closely blended pairs. The parameters and source contamination can be interactively adjusted (see Jarrett et al. (\citeyear{Jarrett2017})

\subsubsection{Rest-frame correction}

To account for the redshifted emission across the optical and infrared bands, we applied a ``k-correction" to the magnitudes based on spectral energy distribution (SED) fitting before the derivation of any physical value. The SEDs are built combining flux densities spanning the optical, the near-IR and $\textit{WISE}$ mid-IR; for the optical and near-infrared data we have used preliminary photometry available in G23 (generated by ProFound; Robotham et al. \citeyear{Robotham2018}).  They are fitted to empirical templates of well-studied galaxies from Brown et al. (\citeyear{Brown2014b}, \citeyear{Brown2019})  and Spitzer-SWIRE/GRASIL (Polletta et al. \citeyear{Polletta2006}, \citeyear{Polletta2007}; Silva et al. \citeyear{Silva1998}). 
From the rest-frame-corrected catalogue, key parameters such as stellar mass were derived using the W1 in-band luminosity and the mass to light ratio (M/L; Cluver et al. \citeyear{Cluver2014};  Querejeta et al. \citeyear{Querejeta2015};  Kettlety et al. \citeyear{Kettlety2018};  Meidt et al. \citeyear{Meidt2012}).  The star formation rate (SFR$_{12 \textit{$\mu$} \text{m}}$)  and the specific star formation rate (sSFR) are estimated using $\nu$L$\nu$ luminosities and the mid-IR to total-IR calibration from Cluver et al. (\citeyear{Cluver2017}). The SEDs allow, to some extent, the differentiation of the AGN from normal star-forming galaxies. Examples of SEDs are presented in the case studies in the Appendix.

\subsection{KiDS}

The Kilo-degree Survey (KiDS) is an optical survey of $\approx$ 1500 square degrees (de Jong et al. \citeyear{deJong2013}; Bilicki et al. \citeyear{Bilicki2018}) using OmegaCAM inside the field imager on the VLT survey telescope. With its  268 Megapixel wide-field camera that provides a 1$^{\circ}\times 1^{\circ}$ field-of-view (with a pixel scale of  0.214 arcseconds/pixel), the exposure time is selected such that it can reach a median redshift of $z = 0.7$ which is well beyond the redshift limit used for the present study (z$<$0.3). The median seeing is about 0.7$^{\prime \prime}$ and the observations are done in 4 bands ($\it{u}$, $\it{g}$, $\it{r}$, $\it{i}$) with the r-band being the deepest reaching a depth of 25.4\,mag (AB system).
The resolution of the images is sub-arcsecond; it is therefore, a valuable data set to visualise fine details within the galaxies, usually not visible using $\textit{WISE}$ images. KiDS has several data releases starting from KiDS-ESO-DR1 and the release used in this study is KiDS-ESO-DR4 (see Kuijken et al. \citeyear{Kuijken2019} for more details).

\section{Results}

\subsection{Property of GAMA-WISE galaxies}

\begin{figure}
 \begin{center}
        \includegraphics[width= 8.5cm]{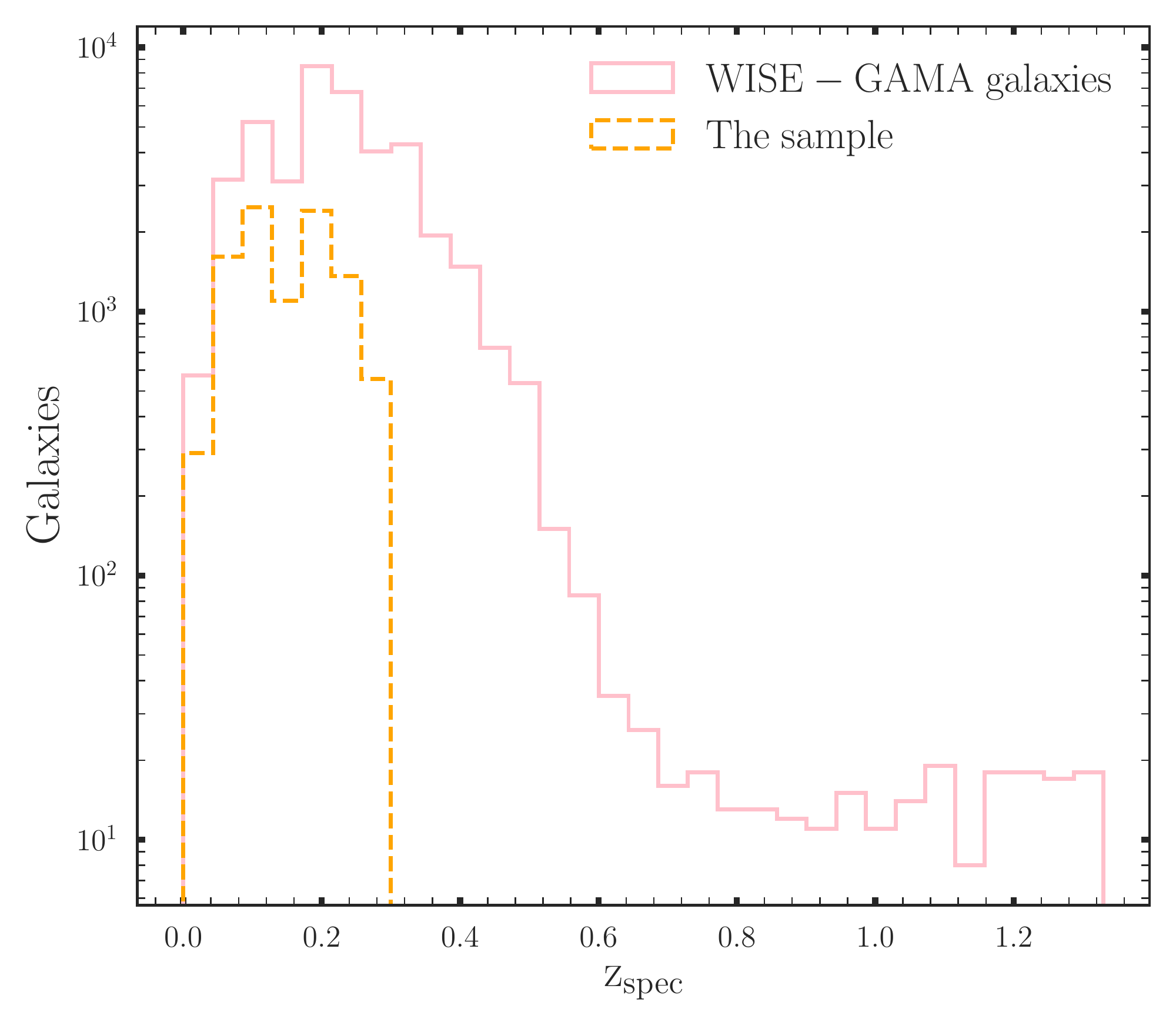} 
    \caption{The number distribution as a function of the redshift. All $\textit{WISE}$-GAMA G23 galaxies ($\sim$ 41,000 galaxies with GAMA redshifts) represented by the solid pink colour of which more than 75$\%$ have redshifts $<$0.3. The final sample (dashed-orange colour) has about 9,800 galaxies after applying a redshift and magnitude cuts of \textit{z} $<$ 0.3 and W1 $<$ 15.5\,mag, respectively. }   \label{fig2}
    \end{center}
\end{figure}

Figure \ref{fig2} presents the full $\textit{WISE}$-GAMA galaxy catalogue in G23 with redshifts (40,843 galaxies). The $\textit{WISE}$ resolved galaxies (total of 2,695 in G23) have been added to the point source photometry from $\textit{ALLWISE}$ to create a quality-controlled catalogue, which is well-suited for detailed study at low redshifts. The galaxy distribution  peaks around $z = 0.2$ with more than 75$\%$  located at z $<$ 0.3, and most of the resolved galaxies with redshifts less than 0.15.

\begin{figure*}
\centering
  \includegraphics[width= 13cm]{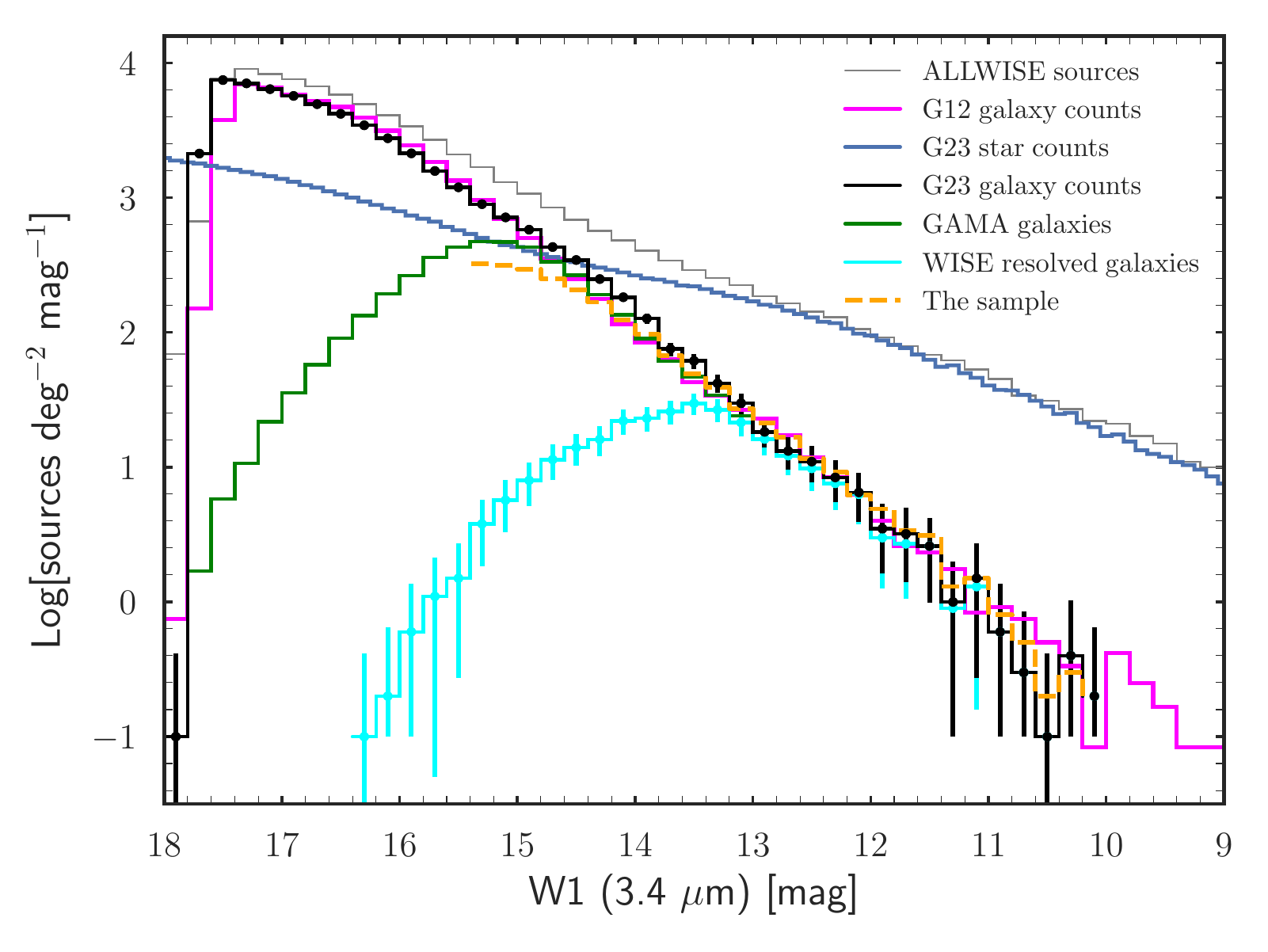}
    \caption{Differential W1 (3.4\,$\mu$m) source counts in the G23 region with magnitudes in the Vega system. The $\textit{ALLWISE}$  sources are shown in solid grey. The $\textit{WISE}$ star and galaxy counts are the dark-blue and black lines,
    respectively.  The $\textit{WISE}$ cross-matched GAMA galaxies and resolved galaxies are in solid green and light blue, respectively. The total number of $\textit{WISE}$ sources  is $\sim$ 600,000 (galaxies + stars) and $\sim$
    41,000 of the galaxies have GAMA redshifts. The orange dashed-dotted histogram represents our sample which has a magnitude limit in W1 $<$15.5\,mag (in Vega system) and redshifts $<$0.3.  The $\textit{WISE}$  cross-matched GAMA (G12, Jarrett et al.  \citeyear{Jarrett2017}) galaxies in magenta have been added for comparison. The vertical lines represent Poisson error bars.}    \label{fig4}
\end{figure*}

The differential source counts in W1 for the different $\textit{WISE}$ and/or GAMA sources are shown in Figure \ref{fig4}. The galaxy source counts for the Galactic North Pole and GAMA G12 from Jarrett et al. (\citeyear{Jarrett2017}) have been added for comparison.  The large majority of galaxies at the bright end are resolved up to a W1 magnitude of 13.5\,mag (1.2 mJy) where the distribution turns over.  
The galaxy counts in G23 and  G12  are similar all the way to the faint end ($\sim$ 17.5\,mag), which may be expected since both are focused near the Galactic poles. The $\textit{WISE}$-GAMA galaxy counts rises close to the $\textit{WISE}$ galaxy counts up to 15.5\,mag where it turns over, due to the G23 survey redshift limit. The proximity of the two curves shows that a very high fraction of 
$\textit{WISE}$ sources have redshifts in GAMA up to 15.5\,mag. The $\textit{ALLWISE}$ counts (galaxies + stars) shows the domination of Galactic stars at the bright end, although we only expect minimal contamination due to our filtering and classification.  The faint end, which is not relevant to the study presented here,  is expected to have unresolved galaxies with a higher probability of blending as the confusion limit of $\textit{WISE}$ is reached (Jarrett et al. \citeyear{Jarrett2017}).

  \begin{figure}
 \begin{center}
        \includegraphics[width= 8.6cm]{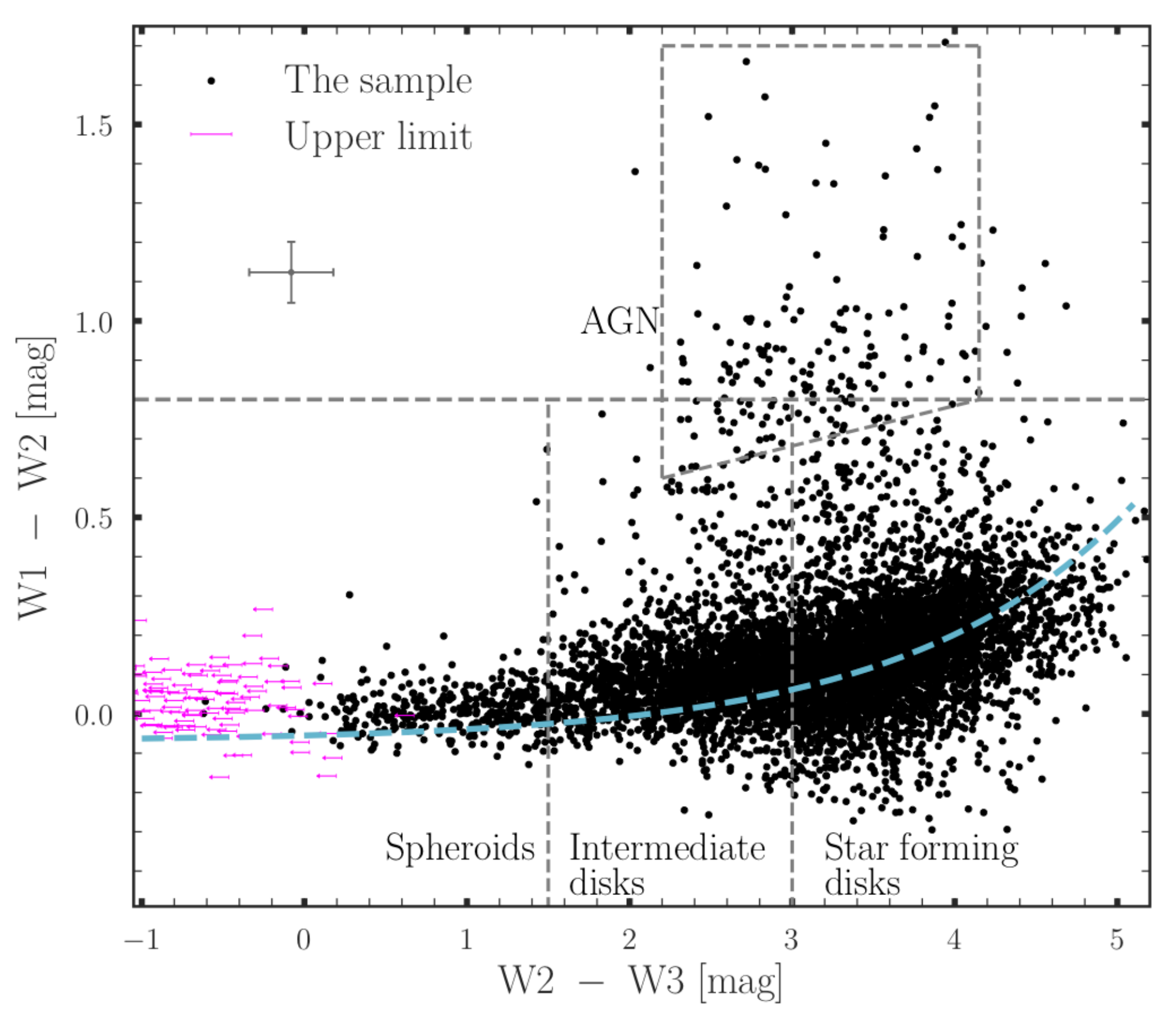} 
    \caption{The $\textit{WISE}$ k-corrected colour-colour plane W1-W2 vs W2-W3 (magnitudes are in the Vega system). The upper limits are represented by the magenta arrows and the dashed blue line is the SF sequence (Jarrett et al. \citeyear{Jarrett2019}) seen at low W1-W2 ($<$0.5\,mag). The AGN box is from Jarrett et al.  (\citeyear{Jarrett2011}). We also plot the mean W1 - W2 and W2 - W3 uncertainties with the error bars in the upper-left corner.}   \label{fig5}
    \end{center}
\end{figure}

 \subsubsection{The study sample}

\begin{table}[h!]
\small
\centering
\caption{Table showing the number of galaxies for different constraints applied. The letters in brackets represent a given sample (e.g: A) } \label{study_sample}
\begin{tabular}{lc}
\tablewidth{0pt}
\hline
Total WISE-GAMA G23 catalogue (A)& 41,000  \\ \hline
A with W1$<$ 15.5 mag, z $<$ 0.3 (B) &  9,800 \\ \hline
B with S/N\,(W1)$>$5,\,S/N\,(W2)$>$5,\,S/N\,(W3)$>$2 (C)& 6,493 \\ \hline
C with H$\alpha$,\,H$\beta$,\,[\ion{O}{iii}],\,[\ion{N}{ii}] all $>$0, all S/N$>$ 3  (D) & 1,173 \\ \hline
C with H$\alpha$,\,H$\beta$,\,[\ion{O}{iii}],\,[\ion{S}{ii}] all $>$0, all S/N$>$ 3  (E) & 1,037 \\ \hline
\end{tabular}
\end{table}

For the comparative study of $\textit{WISE}$ colours  and the optical BPT diagrams, we have applied a magnitude and a redshift cut of W1 $<$ 15.5\,mag (in Vega) and  \textit{z} $<$ 0.3, respectively, to the $\textit{WISE}$ catalogue of galaxies in G23 having redshifts, hereafter referred to as  ``the sample"  representing (B) in Table \ref{study_sample}.  For the resolved galaxies, this corresponds to an average S/N $\approx$ 30 in W1, and corresponding lower values for W2 and W3 bands. A S/N limit of 3 is imposed to all the optical spectral lines used in the BPT.
 Table \ref{study_sample} shows  how the number of galaxies  decreases from the initial WISE-GAMA G23 catalogue to  the final  number of galaxies used for 
the optical and infrared diagnostics. (A), (B), (C) and (D) represent each step.  As an example, ``A'' with W1$<$ 15.5 mag, z $<$ 0.3 means that  in the second row, a magnitude limit of  W1$<$ 15.5 mag and redshift  z $<$ 0.3 are applied to the total WISE-GAMA G23 catalogue.

In Figure \ref{fig2}, the sample (dashed grey) is presented along with the $\textit{WISE}$-GAMA galaxy catalogue, with more than 70$\%$  of galaxies found at redshifts less than 0.2.  The magnitude cut in W1 (the most sensitive band)  ensures detections with high signal to noise (S/N), but also selects well-deblended sources (based on $\textit{WISE}$ images), hence, good photometry quality in general.

 \begin{figure}
 \begin{center}
        \includegraphics[width= 8.6cm]{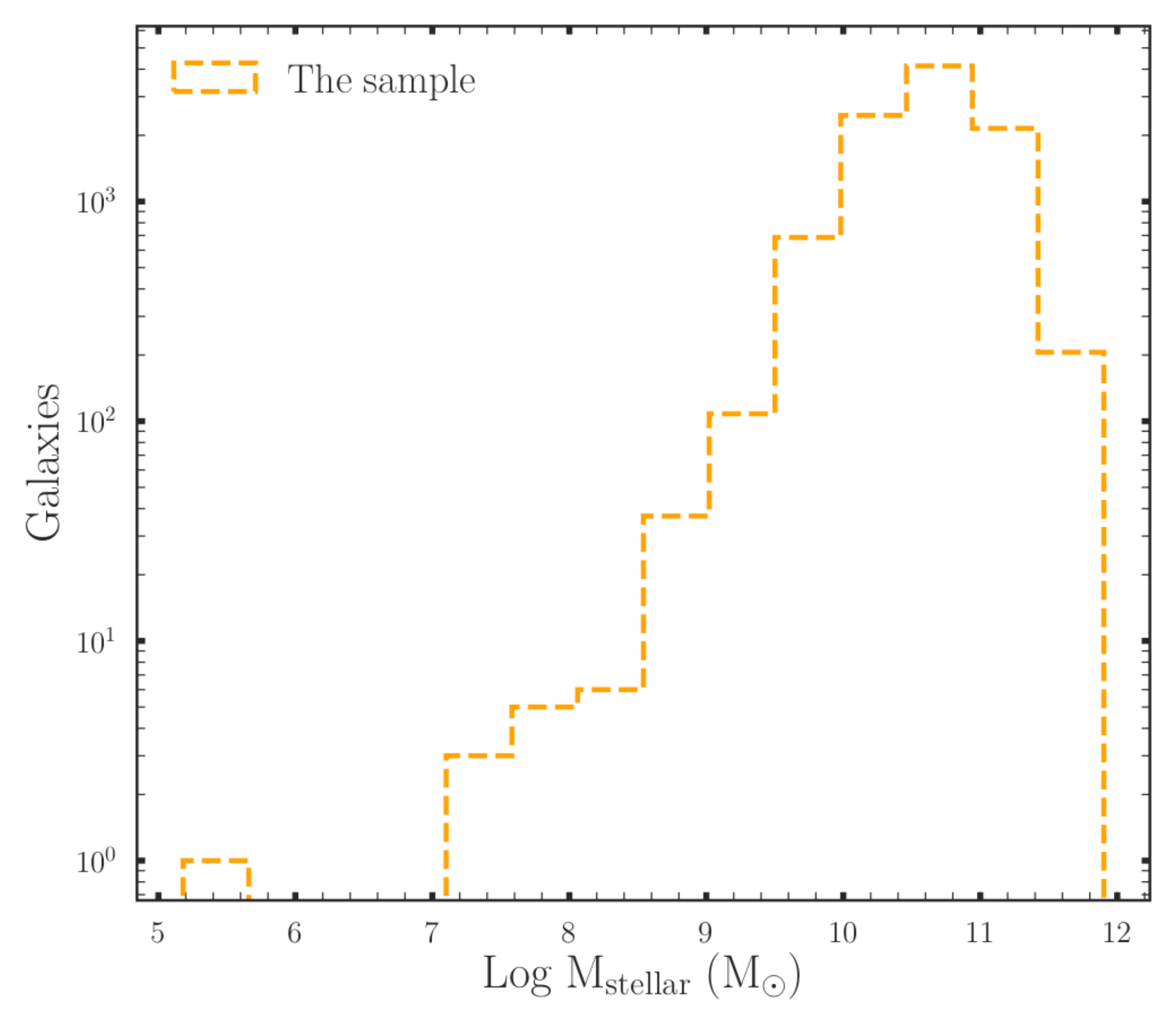} 
    \caption{The distribution of stellar masses for the galaxies in the study sample. The sample contains very few low mass galaxies with $<$ 2 $\%$  having  Log\,(M$_\text{stellar}$/M$_{\odot}$) $<$ 9.}   \label{fig5B}
    \end{center}
\end{figure}

 The magnitude limit  (in W1) of 15.5 mag eliminates most of the low-mass galaxies, Log(M$_\text{stellar}$/M$_{\odot}$) $<$ 9,  which are often misclassified using only the WISE colour-colour diagram (Hainline et al.  \citeyear{Hainline2016}). Using a sample  18,000 nearby dwarf galaxies
(M$_{*}$ $<$  3\,$\times$ \,10$^{9}$ M$_{\odot}$), selected from the Sloan Digital Sky Survey, they found that while the $\textit{WISE}$ colour-colour diagram is reliable in classifying moderate to high mass galaxies, the W1-W2 
colour alone shouldn't be used for dwarf galaxies. In Figure \ref{fig5B} less than $<$ 2 $\%$ of the galaxies in our sample have a stellar mass log (M$_\text{stellar}$/M$_{\odot}$)  $\leq$ 9. We therefore don't expect the result in this
study to be significantly affected by misclassifications related to dwarf galaxies.

As can be seen in Figure \ref{fig4}, a cut of 15.5\,mag offers the best of our carefully measured $\textit{WISE}$ resolved galaxies (2383, $\sim$90$\%$ of the $\textit{WISE}$ resolved sample) and also the brightest unresolved sources (7426 point-like or compact galaxies). A brighter magnitude cut will lead to fewer resolved galaxies and  a fainter cut will include more faint point source galaxies with  lower S/N. In this way, 15.5\,mag is the best balance between completeness and mid-infrared (W1, W2 and W3 bands) photometric quality giving a final sample of 9809 galaxies.

The W1 - W2 vs W2 - W3 colours  (colour-colour diagram)  of the sample with a S/N cut of 5, 5 and 2 in W1, W2 and W3, respectively, is presented in Figure \ref{fig5}. The lower S/N for W3 is to account for less sensitivity in this band relative to the W1 and W2 bands, as well as taking advantage of the large dynamic range (5 orders of magnitude) seen in the W2-W3 colour. The W1-W2 colour is the more critical of the two,  as its lower dynamic range requires higher accuracy.  The AGN box (Jarrett et al. \citeyear{Jarrett2011})  is the expected location of luminous AGN, while the W1 - W2 = 0.8\,mag is
a conservative limit for AGN; galaxies above are classified as $\textit{WISE}$ (mid-infrared) AGN and are frequently in the QSO class.
 The dashed blue line is the fit  to the sequence (equation \ref{e1})  seen in the colour-colour diagram for  all the catalogues of $\textit{WISE}$ resolved sources (Cluver et al. \citeyear{Cluver2017}; Jarrett et al. \citeyear{Jarrett2019}). The relation is described by Jarrett et al. (\citeyear{Jarrett2019}) as:

\begin{equation} \label{e1}
(\text{W1-W2}) = \bigg[0.015\times {\text{\Large{e}}}^{\frac{(\text{W2-W3})}{1.38}}\bigg] - 0.08
\end{equation}

\subsection{AGN classification}

\subsubsection{Optical Emission Line Diagnostics: BPT diagrams}

In this section we investigate the optical emission line properties of the sample derived from our $\textit{WISE}$-GAMA catalogue.
Special care has been taken to derive the photometry of the resolved sources, which  makes our sample a high-quality dataset for a  comparative study between the optical and mid-infrared diagnostics of activity. The constraints put on the magnitude and the redshift allow us to take advantage of the best photometry from $\textit{WISE}$ and avoid the required optical lines being shifted out of the wavelength range of the AAOmega spectrograph used in GAMA. All the optical spectra have been visually inspected to detect any systematic fitting anomalies. To this end, GAMA has a flag for the quality of each fit corresponding to each one of the optical lines of interest, but sometimes a human supervision is needed for the final decision, notably with broad-line systems. The selection criteria are that  we detect the emission lines of  H$\alpha$, H$\beta$,  [\ion{O}{iii}]  as well as  [\ion{N}{ii}]  or [\ion{S}{ii}]  depending on the diagnostic diagram used. H$\alpha$ and H$\beta$ are corrected for both reddening and stellar absorption using equation \eqref{e2}, and all the lines have a S/N $>$ 3 (note that this strict criteria eliminates a large fraction of sources from our sample).   
The Balmer line absorption correction is applied as follows:

\begin{equation} \label{e2}
\text{F}_{\text{corr}} =\bigg[ \frac{\text{\small{EW}}+2.5}{\text{\small{EW}}}\bigg] \times \text{F}_{\text{obs}}\large{\text{,}}
\end{equation}
where EW is the equivalent width of the line of interest, F$_{\text{obs}}$ is the observed flux and 2.5 is the approximate correction required for GAMA spectra (see Gunawardhana et al. \citeyear{Gunawardhana2011}; Hopkins et al. \citeyear{Hopkins2013}).

\begin{figure}
  \includegraphics[width= 8.8cm]{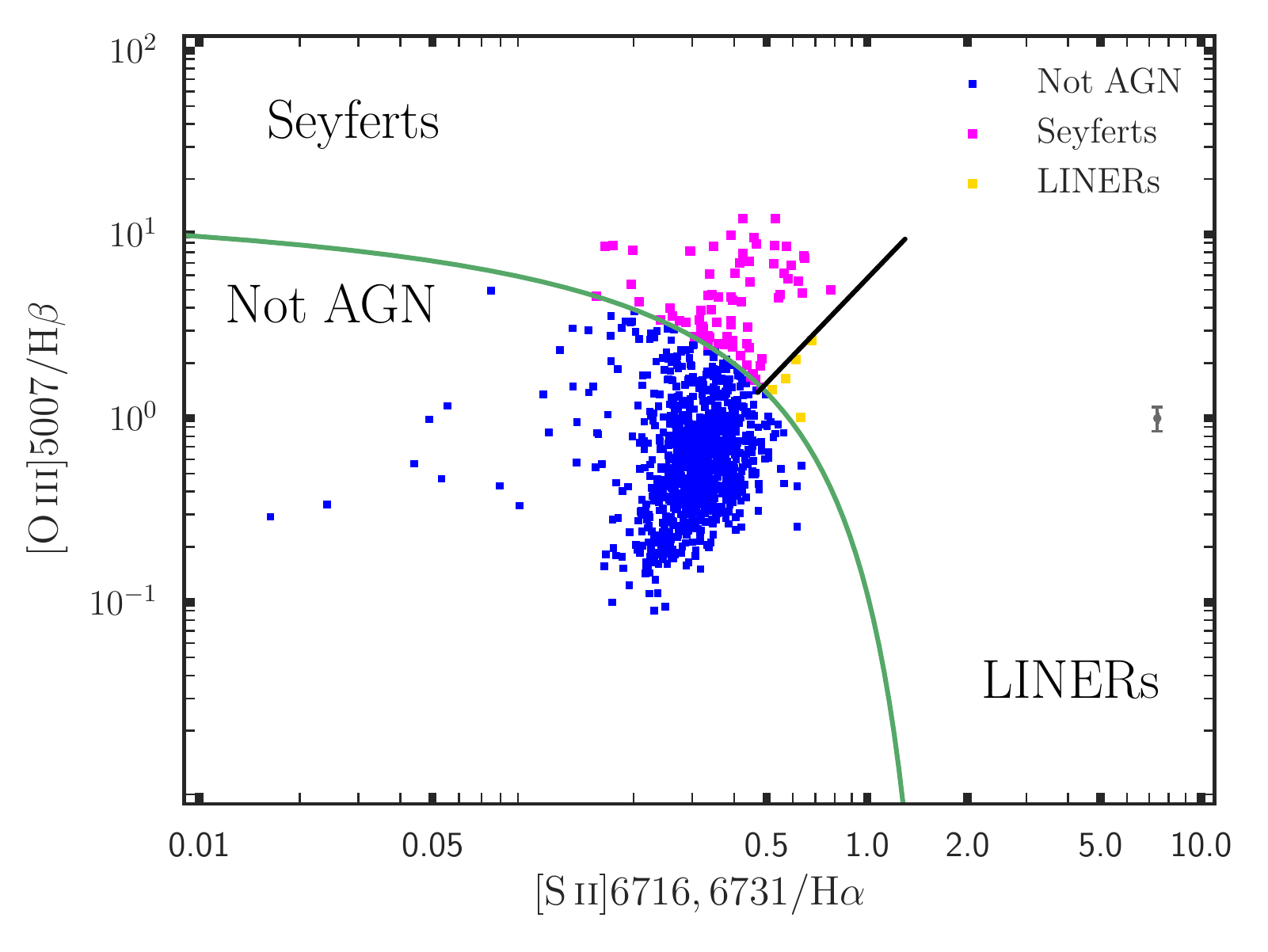}
    \caption{The BPT diagram  ([\ion{O}{iii}]5007/H$\beta$ vs.  [\ion{S}{ii}]6716,6731/H$\alpha$)  with all the required lines in emission. The extreme starburst classification line (green solid) and  the Seyfert-LINER  separation line (black solid)  by Kewley et al. (\citeyear{Kewley2006}).  Above the green line, lie the AGN (Seyferts or LINERs) and below the galaxies that are not AGN. There are very few LINERs in comparison to the Seyferts. We also plot the mean [\ion{S}{ii}]/H$\alpha$ and [\ion{O}{iii}]/H$\beta$ uncertainties with the error bars (middle-right).}   \label{fig6} 
\end{figure}

\begin{figure}
  \includegraphics[width= 8.8cm]{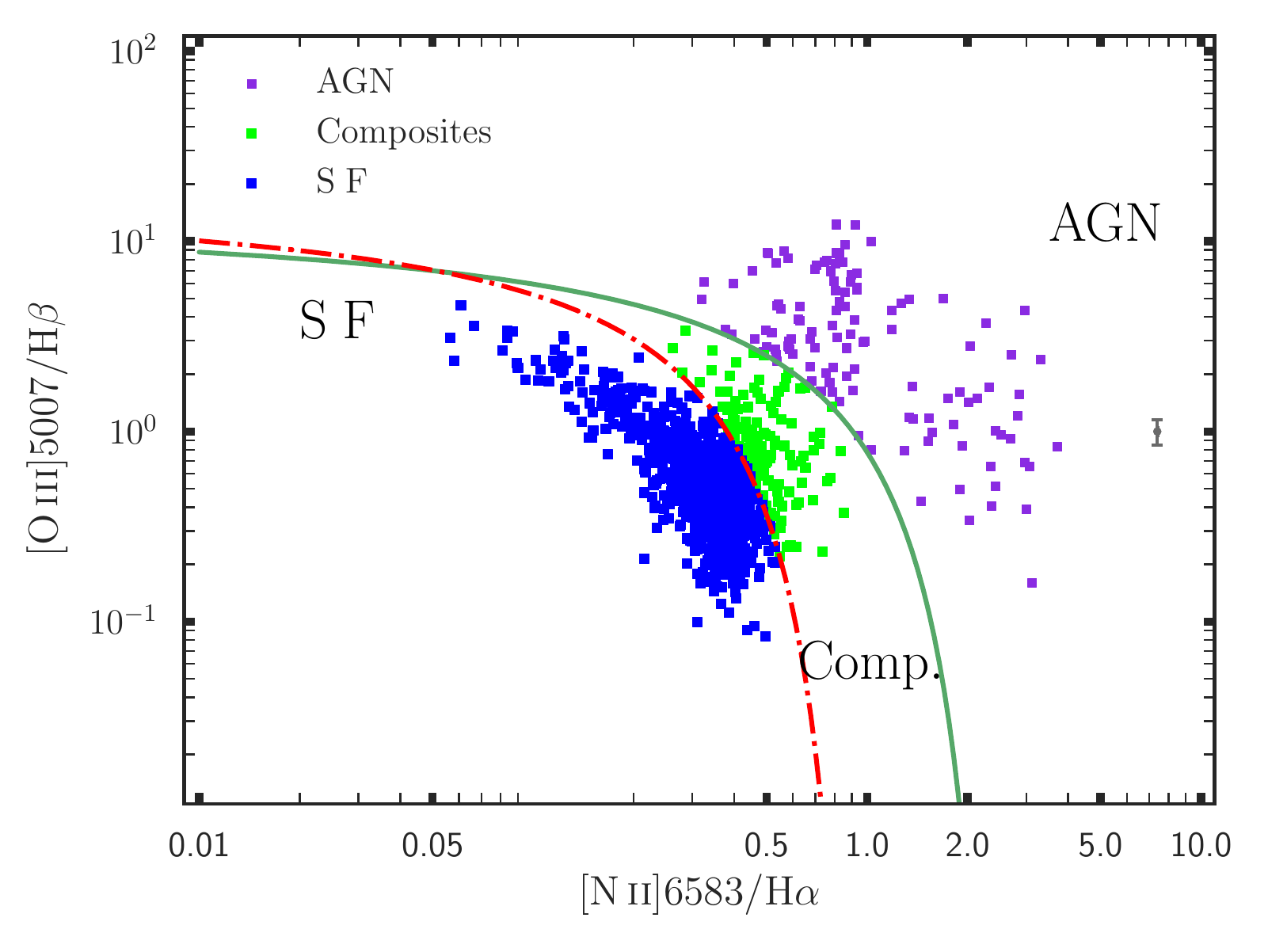}
    \caption{The BPT diagram ([\ion{O}{iii}]5007/H$\beta$ vs.  [\ion{N}{ii}]6583/H$\alpha$)  with all of the required lines in emission. The green line represents the limit between AGN and  Not AGN galaxies (Kewley et al. \citeyear{Kewley2006})  and the Kauffmann et al. (\citeyear{Kauffmann2003}) pure star-formation line (red dashed). 
 Above the green line, lie the AGN (violet), between the two lines are the Composite galaxies (green) and below the red dashed line lie the star-forming galaxies (blue). Note a clear separation of the AGN group into two distinct  subgroups. The second subgroup (right side) tends to be broad-line systems. We also plot the mean [\ion{N}{ii}]/H$\alpha$ and [\ion{O}{iii}]/H$\beta$ uncertainties with the error bars (middle-right).}   \label{fig7}
\end{figure}

 \begin{figure}
 \begin{center}
        \includegraphics[width= 8.6cm]{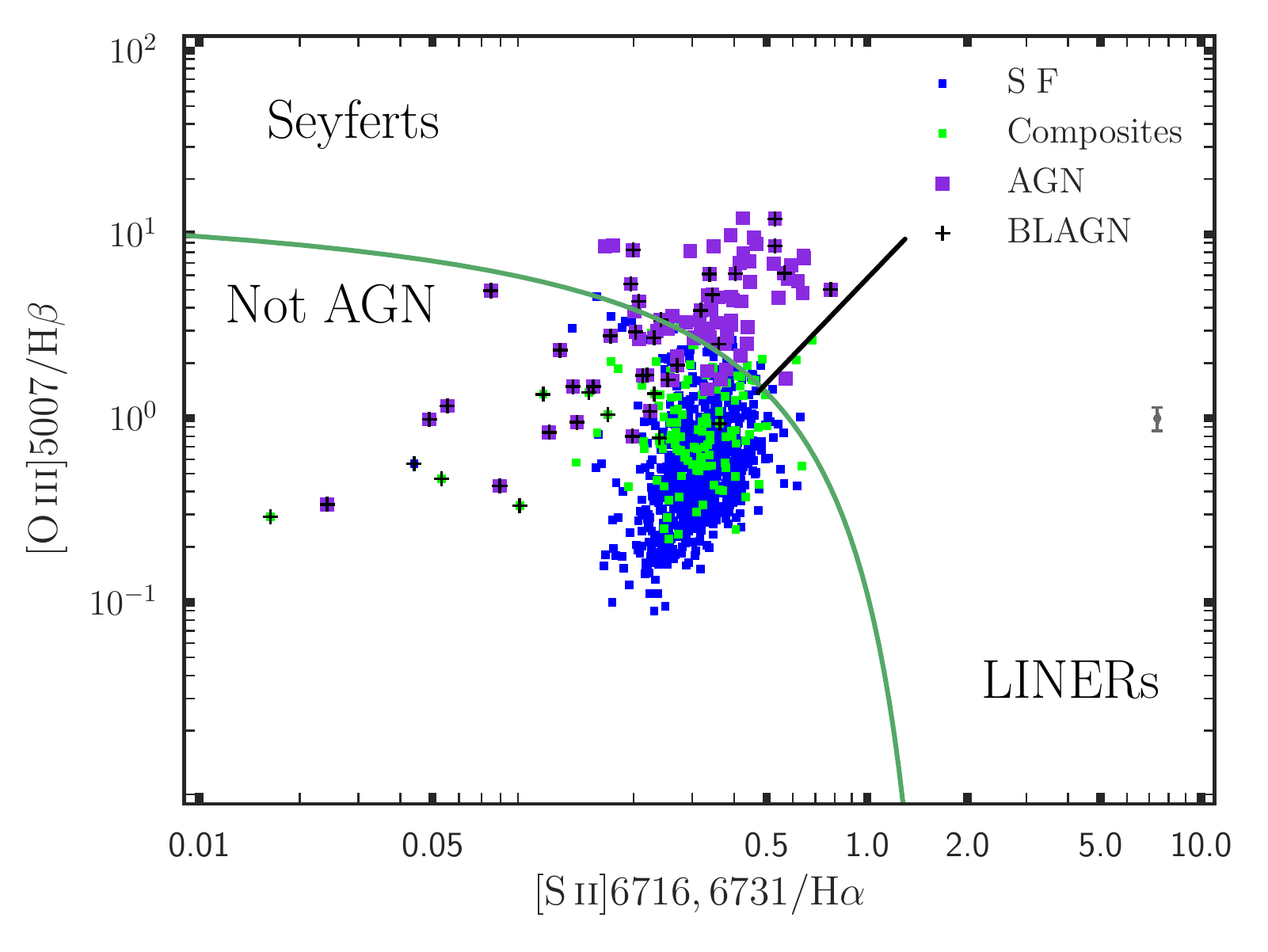} 
    \caption{Comparison beetween both optical classifications. About 85 $\%$ (998/1173) of the galaxies  represented in Fig. \ref{fig7} respect the conditions applied in Fig.  \ref{fig6}. In this figure one can see that the two BPT diagrams agree generally on the separation of AGN and SF galaxies. In most cases the AGN seen as outliers  are very broad-line system for which the measurement present challenges that will be addressed in the next sections. 
    We also plot the mean [\ion{S}{ii}]/H$\alpha$ and [\ion{O}{iii}]/H$\beta$ uncertainties with the error bars (middle-right).}   \label{fig7B}
    \end{center}
\end{figure}

The BPT classification on the plane [\ion{O}{iii}]5007/H$\beta$ vs.  [\ion{S}{ii}]6716,6731/H$\alpha$ is shown in Figure \ref{fig6}. A count of 1,037 out of the 9,809 initial sample satisfy the conditions mentioned-above. 75 galaxies are classified as Seyferts and 5 as LINERs.   

The redshift ($z$) cut of 0.3 is only for GAMA-derived spectra. But the catalogue also contains 6dF and 2dF spectra where \textit{z} = 0.13 and \textit{z}  = 0.2, respectively, should be used as upper limits  when using the [\ion{S}{ii}] line flux. 

The second BPT diagnostic diagram, [\ion{O}{iii}]5007/H$\beta$ vs.  [\ion{N}{ii}]6583/H$\alpha$, is shown in Figure \ref{fig7}. Similar constraints have been applied. In total, 1,173  galaxies are selected, 122  are classified as AGN, 170 composites, and 881 star-forming (SF).

 The [\ion{N}{ii}]6583/H$\alpha$ line varies as a function of metallicity which has a correlation with stellar mass (Wu et al. \citeyear{Wu2016}). Therefore, the 
low metallicity AGN could be found below the extreme starburst line presented in Kewley et al (\citeyear{Kewley2019}). The study by Kauffmann et al. (\citeyear{Kauffmann2003}) shifted the Kewley et al.  (\citeyear{Kewley2006})
 line to create a new 
limit for pure star-forming galaxies. The mixed region in between the two lines supposedly made up by galaxies exhibiting both AGN and star-formation activities is referred   to as composites (Kewley et al. \citeyear{Kewley2019}).
The current optical separation lines derived based on local galaxies are subject to variations with redshift and can be reliably used only at redshift  z $<$ 1 (Kewley et al. \citeyear{Kewley2013}). 

The redshift limit used in our study is 0.3 (local universe), where low metallicity AGN are extremely rare (Groves et al. \citeyear{Groves2006}).

This is further confirmed by comparing both optical methods in Figure \ref{fig7B}. The AGN seen as outliers in the figure have extremely broad H$\alpha$ line that are difficult to be accurately measured  as they are deeply entangled with the blended [\ion{N}{ii}]  lines. The broad-line (BL) AGN are specially addressed in the current work. One more noticeable feature is the quasi-presence of the composites in the SF region, Fig.  \ref{fig7B}, which appears to confirm indeed the rarity of low metallicity AGN in our sample, susceptible of being found in the composite region. Except for particular cases of BLAGN above-mentioned, the two diagrams largely agree on classifying AGN and star-forming 
galaxies. So we do not expect our result to be significantly affected by misclassification of AGN as SF caused by the variation in metallicity.

\subsubsection{Classifications derived from the combination of mid-IR and optical line properties}

 The [\ion{N}{ii}] emission lines are stronger than the [\ion{S}{ii}] lines and 
 using the [\ion{N}{ii}] rather than the [\ion{S}{ii}] line has the added benefit of including composite galaxies in the analysis. Based on these facts, we will be using the  [\ion{O}{iii}]5007/H$\beta$ vs.  [\ion{N}{ii}]6583/H$\alpha$ as the primary optical AGN diagnostic plane which will be compared to the W1-W2 vs W2-W3 colour plane throughout this work except where explicitly stated otherwise.

\begin{figure*}[!t]
\gridline{\leftfig{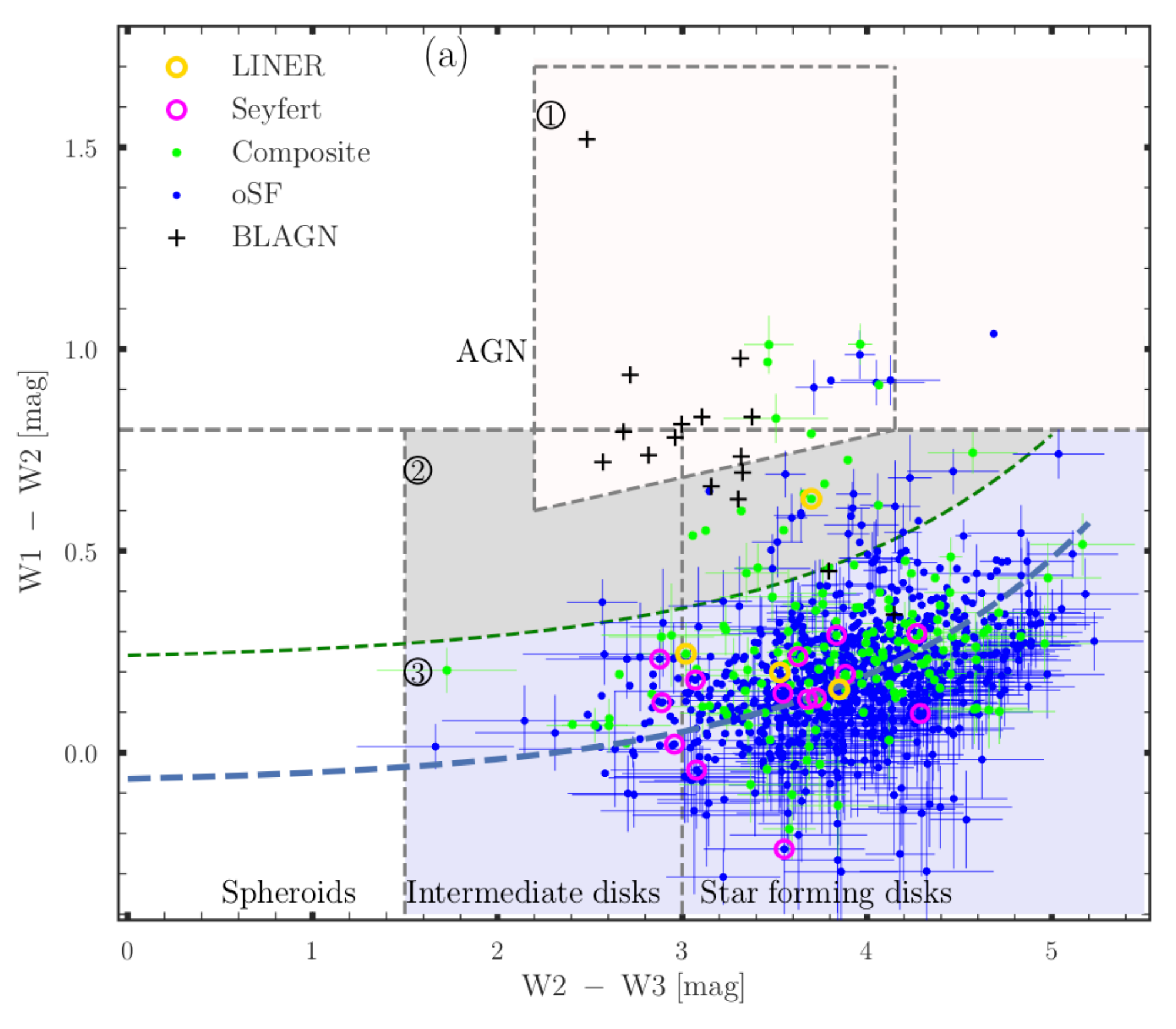}{0.5\textwidth}{(a) WISE colour-colour for optical SF and composites} 
              \rightfig{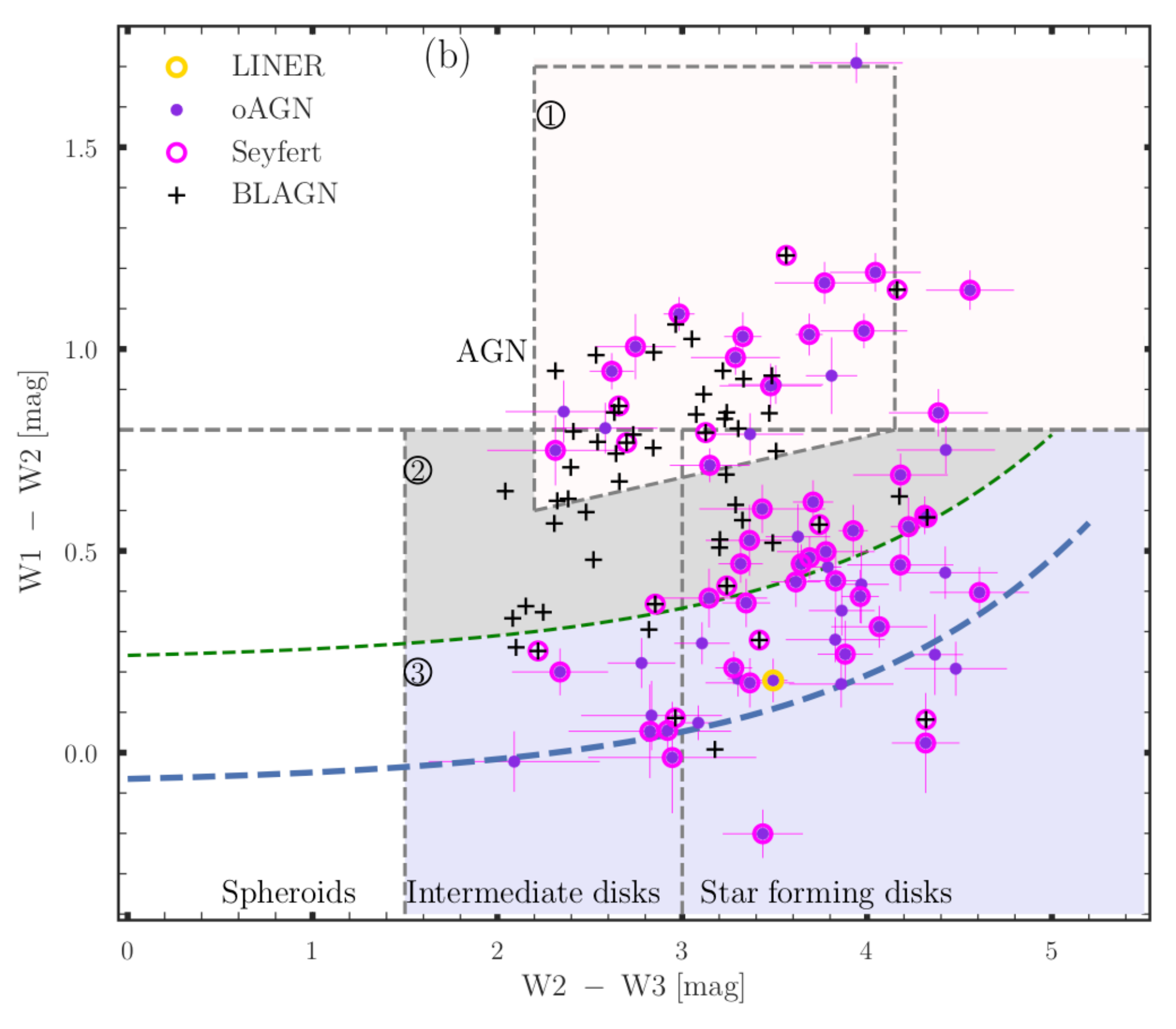}{0.5\textwidth}{(b)WISE colour-colour for optical AGN}}        
\caption{	The BPT [\ion{O}{iii}]5007/H$\beta$ vs. [\ion{N}{ii}]6583/H$\alpha$ classified galaxies from Figure \ref{fig7} plotted in the $\textit{WISE}$ colour-colour diagram. 
   Those also classified as Seyferts and LINERs from Figure \ref{fig6} are over-plotted in magenta and gold circles, respectively. The dashed blue and green lines are the fit to the sequence of galaxies (equation \ref{e1}) seen at low W1-W2 colour and the 2$\sigma$ RMS curve of the fit. The AGN box is from Jarrett et al. \citeyear{Jarrett2011}). The diagram is separated into 3 zones: The Zone 1, 2 and 3 are the respective locations of the $\textit{WISE}$ powerful AGN, the low power AGN and finally the non-AGN zone mostly populated by star-forming galaxies. The majority of the galaxies classified as LINERs using  [\ion{S}{ii}] are classified  as  composites or SF based on the  [\ion{N}{ii}] line while the Seyferts are generally classified as AGN using  [\ion{N}{ii}]. The composites (green) occupy the star-forming zone. The  majority of the optical non-AGNs in the $\textit{WISE}$ AGN zone are broad-lines for which using the line ratio is problematic. The optical star-forming and composites are presented in (a) and optical AGN in (b);   they are separated for clarity. \label{fig8}}               
\end{figure*}

For the comparison between $\textit{WISE}$ and the BPT, the detection S/N in W1, W2 and W3 are also required to be greater than 5, 5 and 2, respectively, giving a total of 1154 galaxies.
  We use the results from Figure  \ref{fig8}, that plots the optical classification from Figure \ref{fig7}  displayed in the $\textit{WISE}$ colour-colour diagram keeping the same colour-coding scheme. The violet dots represent the optically selected AGN (designated as oAGN hereafter), the blue points are the optical  star-forming galaxies (oSF) and the green points are the composites (referred to as ``composite'') which are in between the two. The crosses represent galaxies identified as broad-line AGN (BLAGN). The BLAGN selection is based on the method described in Gordon et al. (\citeyear{Gordon2017}). 
However, we only keep the galaxies clearly seen as BLAGN  without ambiguity after visual checking.  We overplot in magenta and gold circles the galaxies classified as Seyferts and LINERs (where available in the current diagram) from Figure $\ref{fig6}$.

We establish a  2$\sigma$ offset from the SF colour-colour sequence (Eq. \ref{e1}) delineated by the green dashed curve in Figure \ref{fig8}.  The colour-colour diagram is divided into 3 zones represented by the circled numbers:  1, 2 and 3.  Zone 1 (pink shade) is where the $\textit{WISE}$ obscured-AGN, QSOs, LINERs, and ULIRGs are located.  Zone 2 (grey shade) just below, is where the low power Seyferts and LINERs reside -- we call it the  ``mWarm'' zone signifying warmer W1-W2 colour due to greater nuclear activity and corresponding accretion disk emission.  Zone 3 (light blue shade) contains the sequence of $\textit{WISE}$ star-forming galaxies,  mostly a mixture of intermediates and star-forming disk galaxies having low (blue) W1-W2 colours. 

\begin{figure}
\begin{center}
  \includegraphics[width= 9cm]{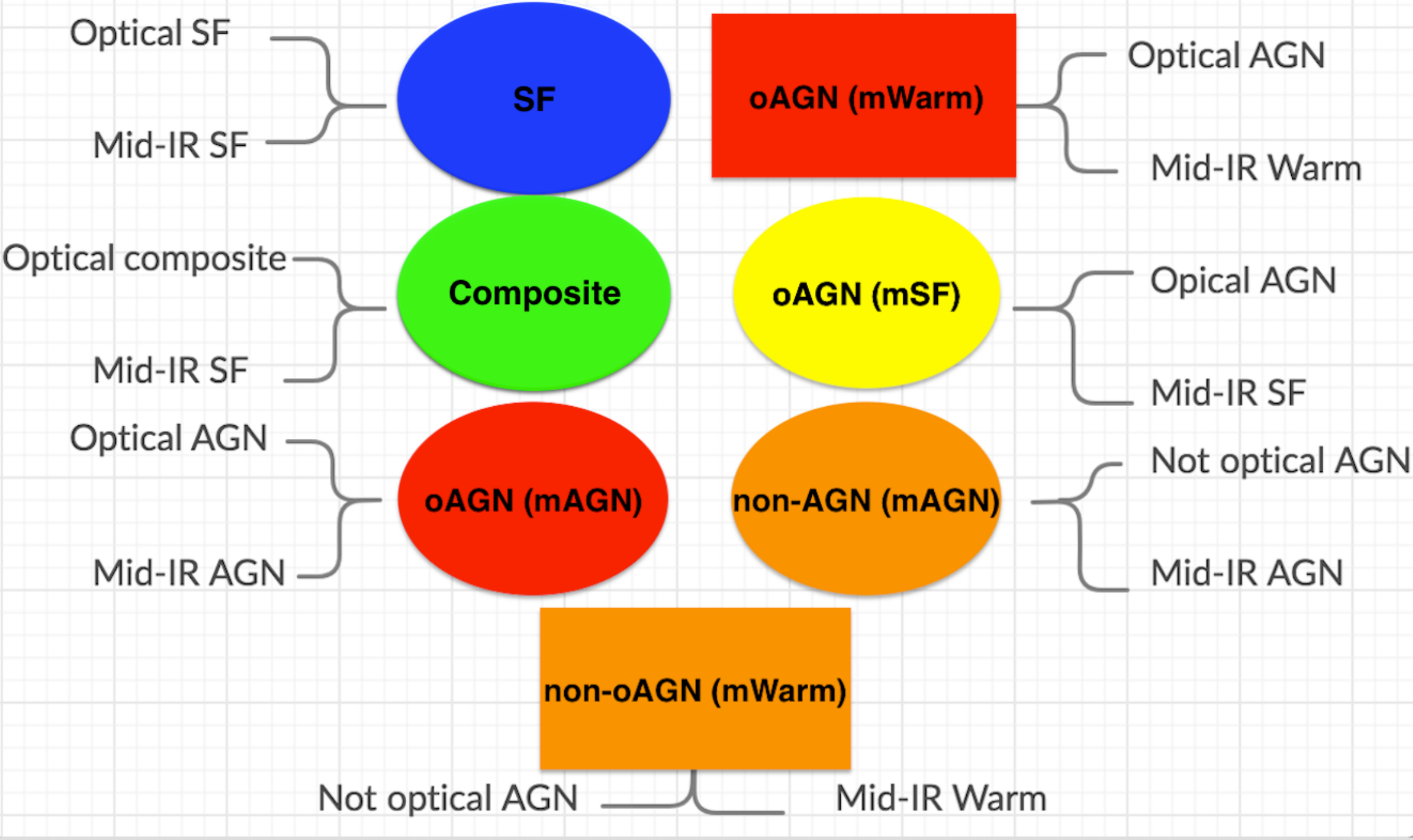}
    \caption{Summary of the classification combining both the optical BPT and the $\textit{WISE}$ mid-IR colour-colour diagrams. The ellipses, rectangles and colours are chosen to follow the legend as presented in Figure \ref{fig9}. Blue and red represent optical and infrared classifications that agree, while yellow and orange are in counter-agreement.} \label{fig8a}
\end{center}
\end{figure}

The diagram in Figure  \ref{fig8a} summarises the galaxy classification considering both their optical emission line properties and their infrared WISE colours.
The galaxies displayed in Figure \ref{fig8} are divided into different groups taking into account both their optical line properties and their mid-IR colours in $\textit{WISE}$. Galaxies classified as AGN in Figure $\ref{fig7}$ (BPT) which are in colour-colour Zone 1 (see Fig. \ref{fig8}b) are labeled as oAGN (mAGN). The optically classified star-forming or composites in Zone 1 are labeled non-oAGN (mAGN). The same applies to Zone 2 where the two groups are called respectively oAGN (mWarm)  and non-oAGN (mWarm). 
The mWarm are the galaxies warmer than the typical star-forming galaxies based on the W1-W2 colour.  
 Finally in Zone 3, all the galaxies classified as AGN in optical which reside in Zone 3 are called oAGN (mSF). The optical star-forming (SF) and composites in  Zone 3  (see Fig. \ref{fig8}a) are called ``SF" and ``Composite", respectively.

  There is a clear presence of oAGNs in the $\textit{WISE}$ star-forming zone. Some non-oAGN objects are located in both the mid-infrared warm (mWarm) and the obscured-AGN zone of which 25$\%$ in the mAGN zone are composites and 50$\%$ are broad-line AGN misclassified by the BPT (Fig. \ref{fig8}a). Having some optically classified composite galaxies in the mid-infrared warm or AGN zone is expected as they are a mixture of both AGN and star formation. However, the vast majority lie in the mid-infrared star-forming zone meaning that their host galaxies generally dominate over the AGN activity. The Seyferts are found in the mid-infrared  AGN zone as well as in the SF and are generally classified as oAGNs  while in most cases the LINERs (classification based on [\ion{S}{ii}] line) designate a non-oAGN (classification based on [\ion{N}{ii}] line) and are classified as  star-forming galaxies in the $\textit{WISE}$ colour-colour diagram. The nature of LINERs is not well understood, and while they may be associated with SF and shocks thereof, they may also harbour low luminosity AGN (Flohic et al.  \citeyear{Flohic2006}). Our data and analysis is not able to properly address this ongoing issue, and we provide no further analysis on LINERs. 
  
  Another trend is the presence of oAGNs in the $\textit{WISE}$ star-forming zone. There are even a few BLAGN with SF colours, which is intriguing since  $\textit{WISE}$ seems to be generally very sensitive to  broad H$\alpha$ lines. They constitute very good specimens to be followed up in radio and X-rays.

\begin{figure*}
\begin{centering}
\includegraphics[width= 9.15cm] {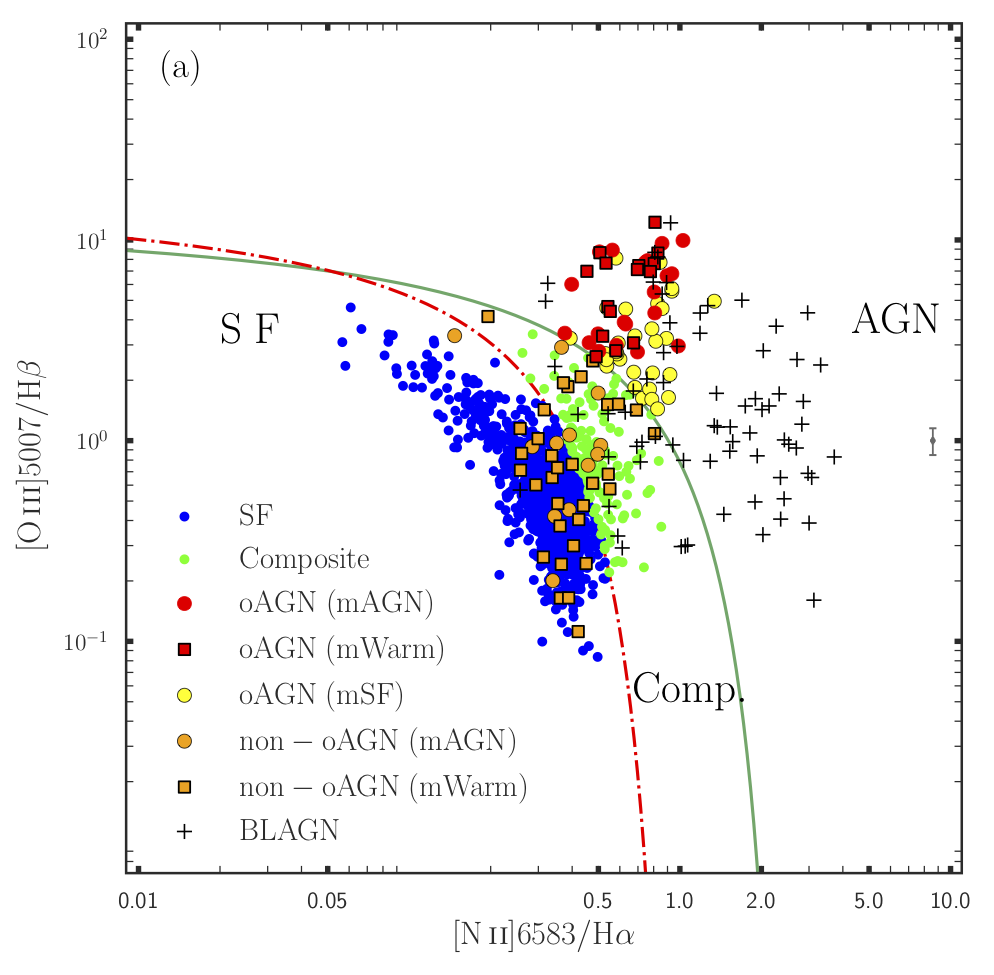}%
\includegraphics[width= 8.80cm] {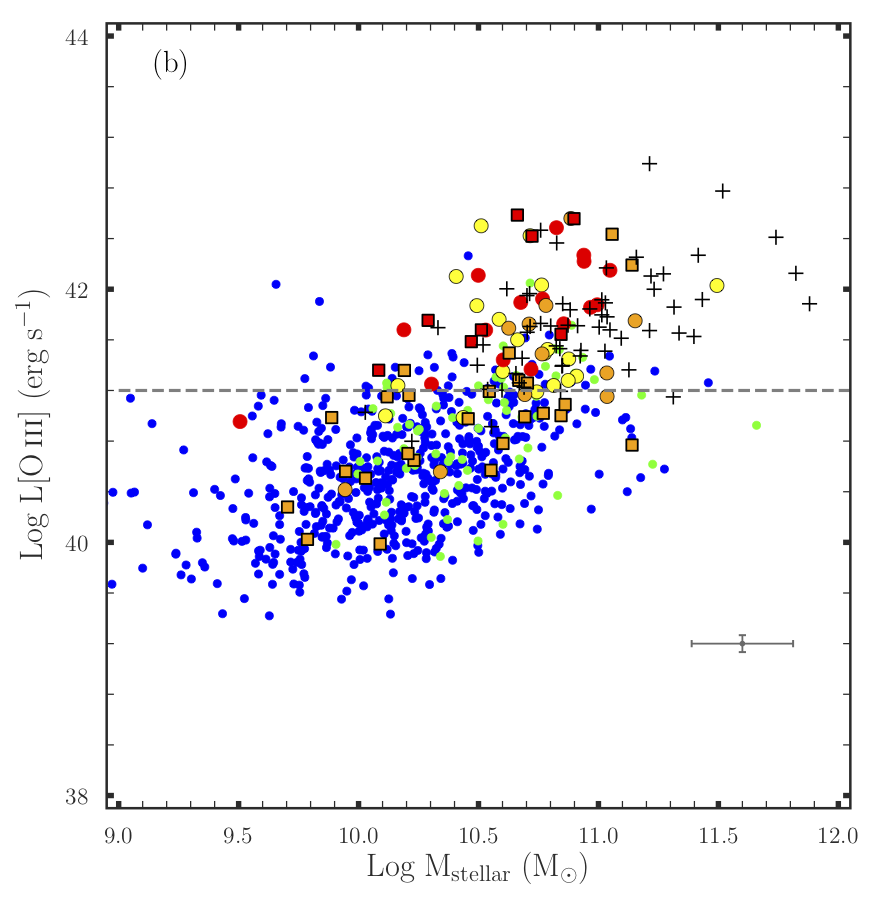}  
\caption{Combined optical-infrared BPT diagram, showing the different classes of galaxies. Panel (a): The galaxies classified in Figure $\ref{fig8}$ are displayed with different colours according to their categories. As can be seen, only 6 $\textit{WISE}$ AGN are classified as star-forming in the BPT while 6 are composites. Both the oAGN (mAGN) and oAGN (mSF)  are clearly divided into two parts. The AGN in the lower right corner are almost exclusively broad-line AGN whose [\ion{N}{ii}] line fluxes were overestimated giving, therefore, larger [\ion{N}{ii}]/H$\alpha$  ratio. The mean [\ion{N}{ii}]/H$\alpha$ and [\ion{O}{iii}]/H$\beta$ uncertainties with the error bars  are plotted (middle-right). Panel (b) shows the AGN strength represented by the [\ion{O}{iii}] line luminosity (not reddening-corrected) as a function of stellar mass.  In general, there is a clear separation between the AGN and SF galaxies except for the non-oAGN (mAGN) exhibiting low [\ion{O}{iii}] luminosities. All the data points including the broad-lines AGN are plotted using the simple Gaussian fit. Only the  [\ion{O}{iii}] luminosity derived from GAMA spectra (calibrated) are used  in (b). The dashed line (log L[\ion{O}{iii}] =  41.2\,erg/s) is the line above which the great majority of optical AGN is likely to be found. We also plot the mean $\mathrm{LogM_{stellar}\;[ M_{\odot}]}$ and Log (L[\ion{O}{iii}])  uncertainties with the error bars in the lower-right corner}  \label{fig9}
\end{centering}
\end{figure*}

We now reproject back into the BPT plane the new optical and mid-IR classification codes,
as shown in Figure $\ref{fig9}$a. The figure shows a clear separation of the oAGN (mAGN) into two subsamples. The lower right corner above the green curve, are broad-line AGN whose line ratios have been overestimated by the line fitting algorithm in GAMA. Here it is striking to see all the galaxies with [\ion{N}{ii}]/H$\alpha$ ratio $>$ 1 being broad-lines with generally overestimated [\ion{N}{ii}] fluxes. A larger sample would likely have shown a few non-BLAGNs for which [\ion{N}{ii}]/H$\alpha$ fluxes ratios are truly  $>$ 1, but these cases are rare. It is sometimes extremely challenging, if not impossible, to separate H$\alpha$ and [\ion{N}{ii}] when the H$\alpha$ line becomes broad.
Even a reassessment of the flux which will probably give a more reasonable ratio, lower than 1, will shift the galaxy (BLAGN) in the star-forming zone unless the [\ion{O}{iii}] line's flux is highly elevated. Those galaxies with an underestimated flux ratio constitute  50$\%$  of our non-oAGN (mAGN) sample.  The GAMA pipeline does not handle the broad lines well, notably the blending of H$\alpha$ and [\ion{N}{ii}]. Therefore, the BPT is not appropriate using the automated pipeline data from GAMA. We have displayed them in this diagram to show where they might rightly or wrongly be located in the BPT diagram.
 
 The oAGN (mSF) are generally located in the upper left side of the AGN zone along with the oAGN (mAGN). Although classified as AGN both in  $\textit{WISE}$ and optical, some of the latter galaxies are labeled as either peculiars or late-type spirals by their SED template fits. This could be explained by the fact that they are sharing similar properties with some of the oAGN (mSF) which are also classified as non-AGN based on the SED.
 Among the non-oAGN classified as mAGN only 6  are oSF.  The trend is that most of the  $\textit{WISE}$ AGN are also optical AGN or at least composite galaxies which need further attention using other diagnostic methods based on radio or X-ray data.  
 Zone 2  (mWarm) seems to be a transition zone  where we find the optical star-forming galaxies, the composites, and the BLAGNs in similar proportions. 
 
Figure $\ref{fig9}$b shows the host stellar mass as a function of [\ion{O}{iii}] line luminosity. In this figure, which requires the use of individual fluxes unlike the BPT based on line ratios, only the calibrated measurements are used (see Hopkins et al. \citeyear{Hopkins2013} for the data description). The number of galaxies, therefore, is reduced. The distribution can be separated into two regions, with optical AGN generally having luminosity greater than 10$^{41}\,$ erg s$^{-1}$ and the star-forming galaxies with lower values. The BLAGNs appear to have also strong [\ion{O}{iii}] lines. Most of the oAGN (mSF) present a strong [\ion{O}{iii}] luminosity confirming the presence of nuclear activity (if we  assume that [\ion{O}{iii}] lines are always related to the AGN or nuclear shocks). As expected,  the non-oAGN (mAGN) usually have  weak  [\ion{O}{iii}] luminosity, like the SF and composites, with a few exceptions. $\textit{WISE}$ classifying them as AGN probably has to do with other processes beyond that of a central AGN activity. Some interesting cases such as oAGN (mSF) with high [\ion{O}{iii}] luminosity are discussed in Section \ref{sec3}.

Table  \ref{table1.3bb} contains the detailed measurements used in this study, organised by their BPT-IR classification. W1, W2 and W3$_{\text{PAH}}$  are  rest-frame fluxes (using SED fitting, see Jarrett et al. \citeyear{Jarrett2017} for more details). The stellar emission has been subtracted in W3. W1-W2 and W2-W3 are the rest-frame colours used for the colour-colour diagram. The [\ion{O}{iii}]/H$\beta$ and [\ion{N}{ii}]/H$\alpha$ represent the optical line ratios used for the BPT diagnostic. Stellar masses and star formation rates (SFR$_{12 \textit{$\mu$} \text{m}}$) are derived using the calibration by Cluver et al.  (\citeyear{Cluver2014}) and Cluver et al.  (\citeyear{Cluver2017}), respectively  (see Table \ref{table1.3k}). The specific star formation rate is the ratio between the SFR and the stellar mass. The galaxies have been classified according to their different AGN groups.

 Table \ref{table1.4z} gives the number  of  galaxies and their average parameter values in the subsets for different conditions applied to the sample.  The number of  broad-line AGN included in each subset is also shown.   The ``condition A" presents only the infrared classification while in ``B"  the optical classification is presented. Both diagnostics are combined in ``C" giving a total number of 1154 classified galaxies. The ``condition D" is where we reclassify the broad-line AGN based on the fact that they are clearly optical AGN that were misclassified by the BPT diagram. Notably, the  2 BLAGN found in the composite group were reclassified as oAGN (mSF), the BLAGN (3 galaxies)  found in the  non-oAGN (mWarm) group become  oAGN (mWarm) and finally 11 non-oAGN (mAGN) become oAGN (mAGN).
We consider the galaxies in the mWarm region to have AGN activity, therefore in ``D"  both methods agree on the  final classification of 84.4$\%$ and  8.2$\%$  of the galaxies as non-AGN (SF + composite) and AGN, respectively, and disagree on the classification of  7.4$\%$. The redshift distribution of the different groups of galaxy is presented in the next section Figure \ref{fig12a}.

\newpage
 
\subsubsection{Galaxy properties in the different activity classifications}

\begin{figure*}
\begin{centering}
\includegraphics[width= 8.8cm] {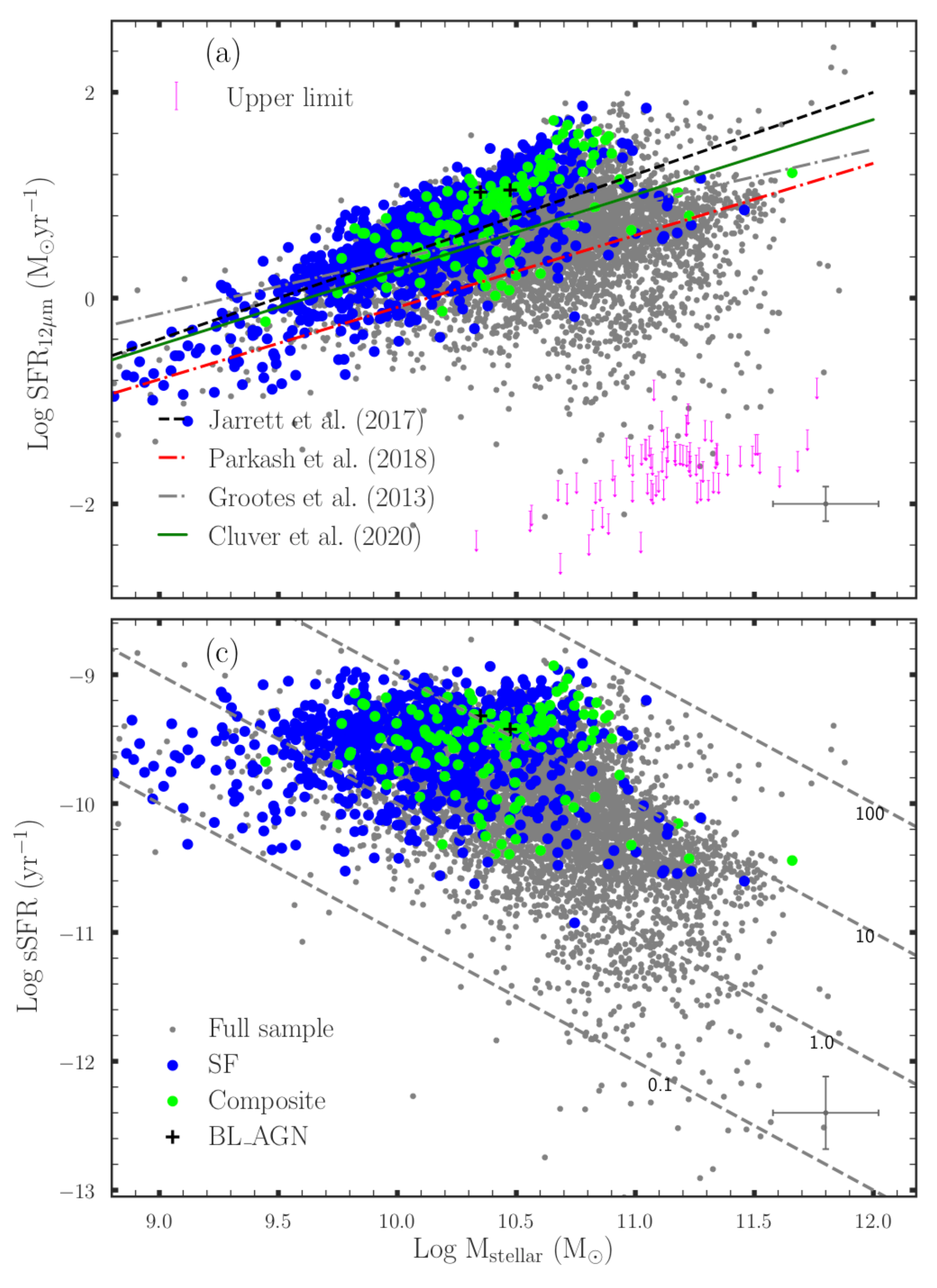} 
\includegraphics[width= 8.8cm] {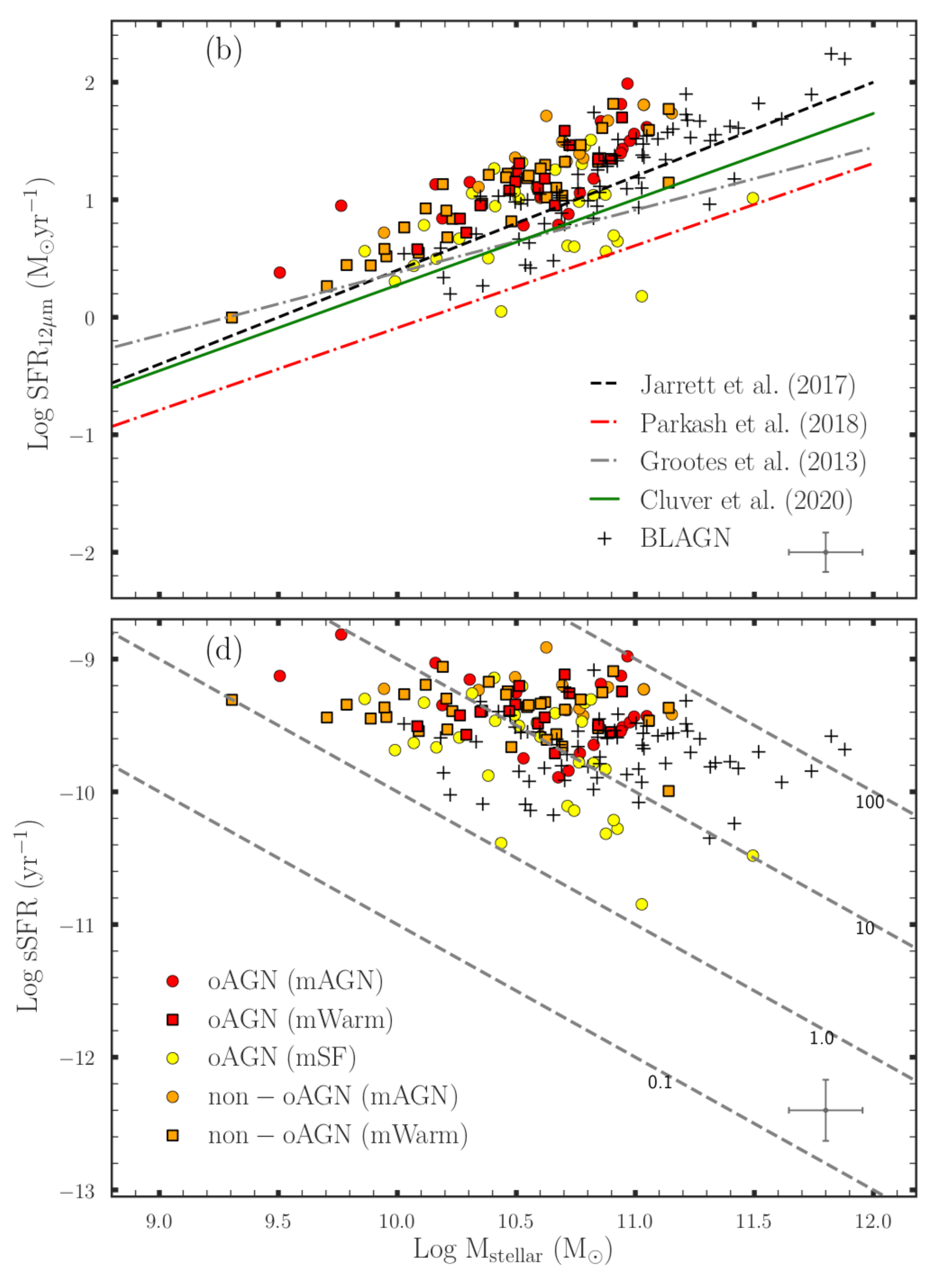}

\caption{Left panel:(a) Distribution of the SFR$_{12 \textit{$\mu$} \text{m}}$ as a function of stellar mass for the galaxies classified as non-AGN in both optical and mid-IR. (b) shows the sSFR as a function of stellar mass for the same non-AGN sample, with the dashed lines representing lines of constant  SFR (0.1, 1, 10, 100 M$_{\odot}$yr$^{-1}$).
The right panel is similar to the previous panel, but applied to our different groups of AGN as indicated in the legend.  We do expect the stellar mass and the SFR$_{12 \textit{$\mu$} \text{m}}$  to be overestimated due to the AGN activity within the galaxies (in the right panel). Notably for those in which AGN dominate (i.e., QSO and BLAGN). We also plot the mean uncertainties with the error bars for each parameter in the lower-right corner.}  \label{fig10}
\end{centering}

\end{figure*}

In Figure $\ref{fig10}$ we plot the SFR$_{12 \textit{$\mu$} \text{m}}$ as a function of stellar mass and include the main sequence relation of the local galaxies from Grootes et al. (\citeyear{Grootes2013}); Cluver et al. (\citeyear{Cluver2020}); Parkash et al. (\citeyear{Parkash2018}) and Jarrett et al. (\citeyear{Jarrett2017}).

The upper limits trace the galaxies with the lowest star formation rate corresponding to the passively evolving galaxies (also presented in Figure $\ref{fig5}$). All the groups share the same mass range from  10$^{9.8}$M$_{\odot}$ $<$ M$_\text{stellar}$ $<$10$^{11.5}$M$_{\odot}$. The low mass end (M$_\text{stellar}$ $<$10$^{9.8}$M$_{\odot}$) is exclusively populated by SF galaxies having low star formation along with a few non-oAGN (mWarm). The high mass end (M$_\text{stellar}$ $>$ 10$^{11}$M$_{\odot}$) is populated by galaxies with broad Balmer lines (panel b) which are classified as AGN, both in the optical and the mid-infrared. The linear fit to the selection of  Jarrett et al. (\citeyear{Jarrett2017}) which is similar to our sample (except that here the AGN have been separated from the star-forming)  follows our sequence especially the high mass end which is generally populated by AGN.  The new sequence fit by Cluver et al. (\citeyear{Cluver2020}), based on a sample of isolated SF galaxies, also traces our distribution with a slightly flatter slope.

Grootes et al. (\citeyear{Grootes2013}) UV selected a sample of nearby (\textit{z}  $<$ 0.13) spirals in GAMA that gives a flatter SFR-M$_{*}$ relation. 
 The relation given in Parkash et al. (\citeyear{Parkash2018}) traces best our low mass galaxies and also seems to follow the background distribution of galaxies (the grey points) which likely includes absorption-line features. 
 Our sample is limited to only bright galaxies (W1 $<$ 15.5\,mag), moreover it requires  at least a detection in W3 (dusty) which might preferably select more massive, star-bursting systems, compared to UV or optically-selected samples, as discussed  in Jarrett et al. (\citeyear{Jarrett2017}). Figure  \ref{fig10}c shows a flat distribution with the selected galaxies having generally log sSFR $>$ -11.4 yr$^{-1}$. The BPT based on emission-line galaxies might be preferentially selecting galaxies that are still building up  their disk and not probing enough intermediate (so-called ``green valley'') galaxies.
We note that we do expect a mass (and SFR$_{12 \textit{$\mu$} \text{m}}$) overestimation in galaxies having an AGN due to the additional mid-infrared flux coming from the active nucleus, rather than related to the star formation activity itself; this section is therefore intended to be illustrative.

The SF galaxies with the highest SFR$_{12 \textit{$\mu$} \text{m}}$ (logSFR$_{12 \textit{$\mu$} \text{m}}$ $>$ 1.6 M$_\odot$ $yr^{-1}$) are all located at W2 -W3 $>$ 4.6\,mag and  0.22\,mag $<$ W1-W2 $<$ 0.7\,mag. The majority of them (7/9, 77$\%$)  have  W1-W2 well above 0.43\,mag indicating extreme activity. Also, 6 in a total of 9  have a peculiar morphological type (by the SED template and confirmed through visual inspection). This could be a hint that the star formation has been triggered by external processes such as mergers or the tidal influence of the environment. Their masses range from log (M$_\text{stellar}$/M$_{\odot}$) = 10.6 to 11. But on the other hand, the star-forming galaxies with the highest stellar mass (logM$_\text{stellar}$ $>$ 11 M$_{\odot}$) are star-forming disk galaxies in the  $\textit{WISE}$ colour-colour diagram with  2.6$<$ W2 - W3 $<$3.9\,mag. They also have very low W1 - W2 colour ($<$ 0.13\,mag) except one galaxy (CATAID: 5256068, see Table \ref{table1.3bb}) which is warmer (W1-W2 $\sim$ 0.54 mag and W2-W3 $\sim$ 4.5 mag). They are among the lowest sSFR with a mean log  sSFR of -10.2\,$yr^{-1}$.

We show the specific SFR$_{12 \textit{$\mu$} \text{m}}$ (SFR/stellar mass) as a function of stellar mass in Figure $\ref{fig10}$c and $\ref{fig10}$d. The non-oAGN (mAGN) and non-oAGN (mWarm) have the highest log sSFR, or building rate, on average, $\sim$-9 \,$yr^{-1}$. In this case, the environmental influences such as tidal interactions or mergers could have triggered additional star formation. They are therefore warm enough to be classified as warm-AGN in $\textit{WISE}$.  

The non-oAGN (mAGN) group with average log  (sSFR) = -9.29\,$yr^{-1}$ can be divided into two sub-samples. The first is composed of galaxies having a very strong H$\alpha$ line compared to [\ion{N}{ii}] ($\approx$ 60$\%$ of galaxies in this group) where, the [\ion{O}{iii}] and H$\beta$  are almost non-existent. The galaxy 5200866 which is in the starburst zone, has one of the highest sSFR (see Table \ref{table1.3k}). The second sub-group concerns broad-line AGN that have been misclassified by the BPT diagnostic. Indeed in some cases of broad-line AGN, it is impossible to disentangle the H$\alpha$ and [\ion{N}{ii}], such that their ratio is either under or overestimated. The non-oAGN (mWarm) share similar properties as the first group. The third  highest mean sSFR  group is oAGN (mAGN) and the lowest sSFR are seen among the oAGN (mSF). It looks like the AGN classified by $\textit{WISE}$ with almost no/weak [\ion{O}{iii}] line exhibit higher sSFR compared to the galaxies classified as AGN in the optical and star-forming with very strong [\ion{O}{iii}] lines (Table \ref{table1.4z} gives a brief summary of the mean values). The fact that the oAGN (mSF) galaxies have the lowest sSFR might be a hint of a quenching activity (AGN feedback) occurring inside the galaxies. Once more the trend seen could be just related to the overestimation of the parameters due to the presence of the AGN itself. An alternative way of deriving the parameters (mass, SFR$_{12 \textit{$\mu$} \text{m}}$, sSFR), is with the AGN modelled and removed from the host emission
 (Assef et al. \citeyear{Assef2010}; Hainline et al.  \citeyear{Hainline2014}),  which should provide a more robust separation of AGN and ISM dust-emission.

\begin{longrotatetable}
\begin{deluxetable}{l@{\hspace{0.5cm}}cccccc}
\tablecaption{Subgroups and their different  constraints applied to the sample. The mean and median values of some properties have been added. The condition C  was applied to Figure \ref{fig8}. \label{table1.4z}} 
\tablewidth{750pt}
\tabletypesize{\scriptsize}
\tablehead{
\colhead{} &\colhead{Number} & \colhead{broad-line AGN} &\colhead{Redshift} & \colhead{ $\mathrm{Log\,M_{stellar}}$} &\colhead{SFR$_{12 \textit{$\mu$} \text{m}}$} & \colhead{ Log  (sSFR)} \\ 
\colhead{} &\colhead{} & \colhead{} & \colhead{} & \colhead{} & \colhead{}&\colhead{} \\
\colhead{} & \colhead{} & \colhead{} & \colhead{} & \colhead{($\mathrm{ M_{\odot}}$)}&\colhead{(M$_{\odot}$yr$^{-1}$)}&\colhead{(yr$^{-1}$)} \\
\colhead{Groups} &\colhead{No. ($\%$)} & \colhead{BLAGN} & \colhead{mean(med)} & \colhead{mean(med)} & \colhead{mean(med)}&\colhead{mean(med)} \\
} 
\startdata
$\textit{WISE}$/G23 &Condition A$^1$&&&&&\\ \hline
mid-IR SF (mSF) & 6133 (94.5 $\%$)&N/A & 0.14 (0.13) & 10.52 $\pm$ 0.003 (10.55) & 5.44 $\pm$ 0.041 (3.57) & -10.0 $\pm$ 0.004 (-9.95) \\ 
mid-IR``warm" (mWarm) & 181 (2.8 $\%$)&N/A & 0.19 (0.2) & 10.52 $\pm$ 0.014 (10.6) & 11.35 $\pm$ 0.425 (9.23) & -9.63 $\pm$ 0.019 (-9.61) \\
mid-IR AGN (mAGN) & 179 (2.8 $\%$)&N/A & 0.19 (0.2) & 10.6 $\pm$ 0.012 (10.71) & 23.98 $\pm$ 1.126 (13.84) & -9.49 $\pm$ 0.017 (-9.52) \\
Optical/G23 &Condition B$^2$&&&&&\\ \hline
Optical SF (oSF) & 881 (75.11$\%$)&1 & 0.11  (0.09) & N/A & N/A & N/A \\
Optical (composite) & 170 (14.49$\%$)&15 & 0.15  (0.14) & N/A & N/A & N/A \\
Optical AGN (oAGN) & 122 (10.4$\%$)&56 & 0.18  (0.19) & N/A & N/A & N/A \\ \hline
$\textit{WISE}$/Optical/G23 &Condition C$^3$&&&&& \\ \hline
SF & 838 (72.6 $\%$) & 0 & 0.10  (0.09) & 10.11 $\pm$ 0.01 (10.13) & 06.23 $\pm$ 0.13 (03.48) & -9.58 $\pm$ 0.01 (-9.52) \\
Composites & 138 (12.0$\%$) & 2 & 0.14  (0.12) & 10.39 $\pm$ 0.02 (10.41) & 10.43  $\pm$ 0.48(07.28) & -9.56 $\pm$ 0.02 (-9.48) \\
oAGN (mAGN) & 49 (4.2$\%$) & 30 & 0.20  (0.21) & 10.90 $\pm$ 0.02 (10.93) & 32.89 $\pm$ 2.67 (24.92) & -9.56 $\pm$ 0.03 (-9.58) \\
oAGN (mWarm) & 33 (2.9$\%$) & 18& 0.17  (0.18) & 10.61 $\pm$ 0.03  (10.62) & 14.84$\pm$ 1.25 (10.86) & -9.58 $\pm$ 0.04 (-9.55) \\
oAGN (mSF) & 37 (3.2$\%$) & 7 & 0.17  (0.18) & 10.61 $\pm$ 0.03  (10.60) & 09.33 $\pm$ 0.76 (08.82) & -9.78 $\pm$ 0.04 (-9.66) \\
non-oAGN (mAGN) & 23 (2.0$\%$) & 11& 0.21  (0.21) & 10.79 $\pm$ 0.03  (10.78) & 28.14 $\pm$ 2.53 (22.92) & -9.43 $\pm$ 0.04 (-9.43) \\
non-oAGN (mWarm) & 36 (3.1$\%$) & 3 & 0.18  (0.19) & 10.46 $\pm$ 0.03 (10.51) & 16.79 $\pm$ 1.43 (13.11) & -9.41 $\pm$ 0.04 (-9.37) \\ \hline
$\textit{WISE}$/Optical/G23 &Condition D$^4$&&&&& \\ \hline
SF & 838 (72.6$\%$) & 0 & 0.10 (0.09) & 10.11 $\pm$ 0.01  (10.13) & 06.23 $\pm$ 0.13 (03.48) & -9.58$\pm$ 0.01 (-9.52) \\
Composites & 136 (11.8$\%$) & 0 & 0.14  (0.12) & 10.39 $\pm$ 0.02 (10.41) & 10.42 $\pm$ 0.49 (07.02) & -9.56 0.02 (-9.49) \\
oAGN (mAGN) & 59 (5.1$\%$) & 41 & 0.20 (0.21) & 10.89 $\pm$ 0.02 (10.91) & 30.75 $\pm$ 2.23 (23.72) & -9.57 $\pm$ 0.03 (-9.58) \\
oAGN (mWarm) & 36 (3.1$\%$) & 21 & 0.18  (0.18) & 10.64 $\pm$ 0.03  (10.64) & 15.82 $\pm$ 1.26 (11.81) & -9.58 $\pm$ 0.04 (-9.55) \\
oAGN (mSF) & 39 (3.4$\%$) & 9 & 0.16 (0.18) & 10.60 $\pm$ 0.03 (10.59) & 09.41 $\pm$ 0.73 (09.17) & -9.76 $\pm$ 0.04 (-9.66) \\
non-oAGN (mAGN) & 13 (1.1$\%$) & 0 & 0.20 (0.20) & 10.74 $\pm$ 0.04 (10.77) & 34.18$\pm$ 4.23 (24.68) & -9.29$\pm$ 0.06 (-9.23) \\
non-oAGN (mWarm) & 33 (2.9$\%$) & 0 & 0.17 (0.19) & 10.41 $\pm$ 0.03 (10.46) & 15.90 $\pm$ 1.45 (12.64) & -9.39$\pm$ 0.04 (-9.36) \\ 
\enddata
\vspace{1ex}
     {\raggedright  $^1$Signal to noise (S/N):  W1 $>$ 5;  W2$>$5 and W3 $>$ 2;  redshift $<$ 0.3 and magnitude (W1) $<$ 15.5\,mag (in Vega) \par}
     {\raggedright  $^2$ S/N (optical lines) $>$ 3; redshift $<$ 0.3; H$\alpha$ $>$0; H$\beta$ $>$0; [\ion{O}{iii}]5007 $>$0; [\ion{N}{ii}]6583  $>$0  and magnitude (W1) $<$ 15.5\,mag (in Vega) \par}
     {\raggedright  $^3$ S/N (optical lines) $>$ 3; redshift $<$ 0.3; H$\alpha$ $>$0; H$\beta$ $>$0; [\ion{O}{iii}]5007 $>$0; [\ion{N}{ii}]6583  $>$0. S/N: W1 $>$ 5; W2$>$ 5 and W3 $>$ 2 and magnitude (W1) $<$   \par}
     {15.5\,mag (in Vega)  \par}
     {\raggedright  $^4$is when in ``condition c'' all the broad-line (BL) non-oAGNs are considered to be BL oAGNs and BL composites to be  BL  oAGN(mSF).  Recall that the position of the  BLAGNs are highly  uncertain on the BPT due errors related to the [\ion{N}{ii}]/H$\alpha$ flux ratio. \par}
\end{deluxetable}
\end{longrotatetable}

\begin{figure*}
\begin{centering}
\includegraphics[width= 8.8cm] {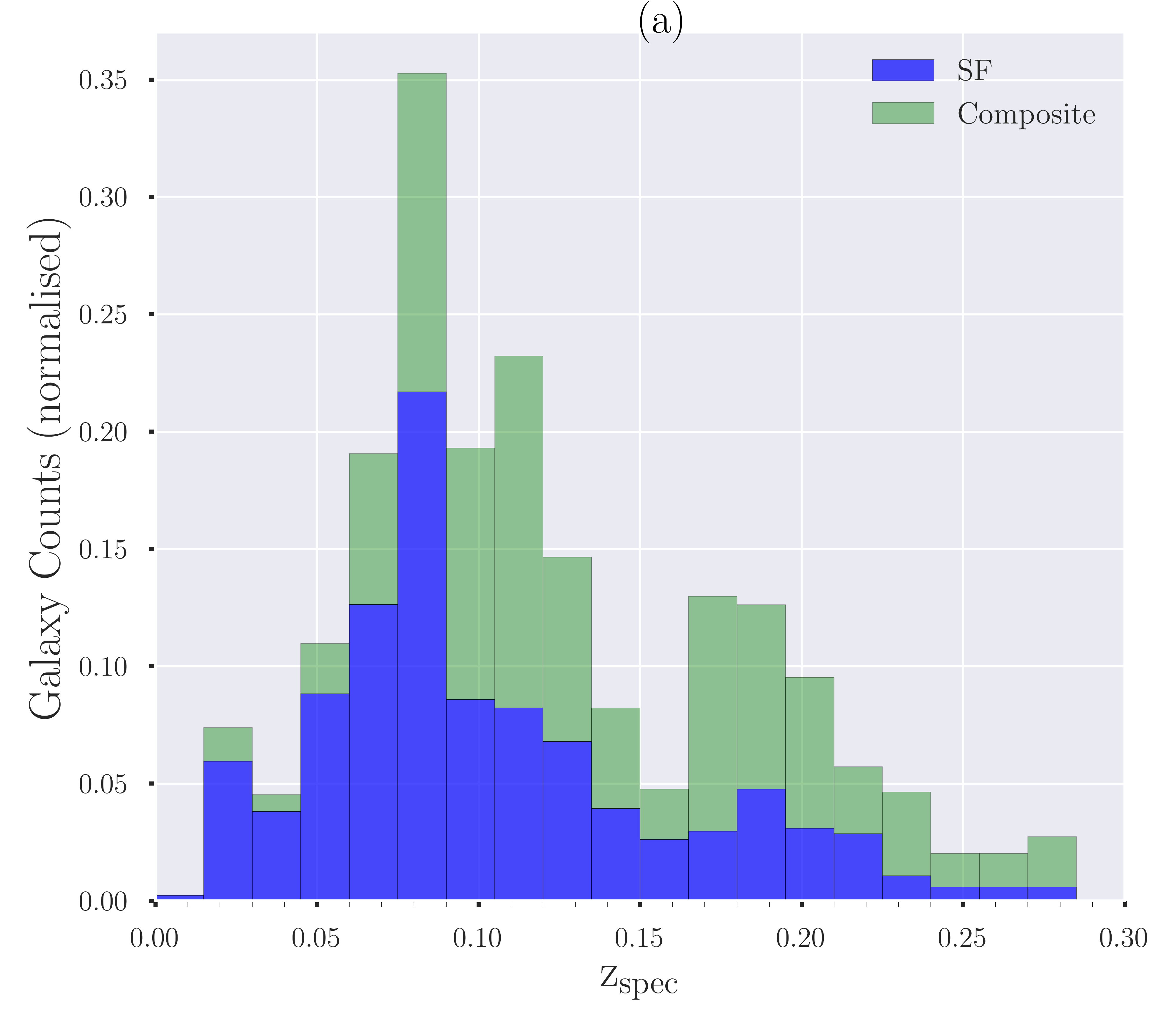} 
\includegraphics[width= 8.8cm] {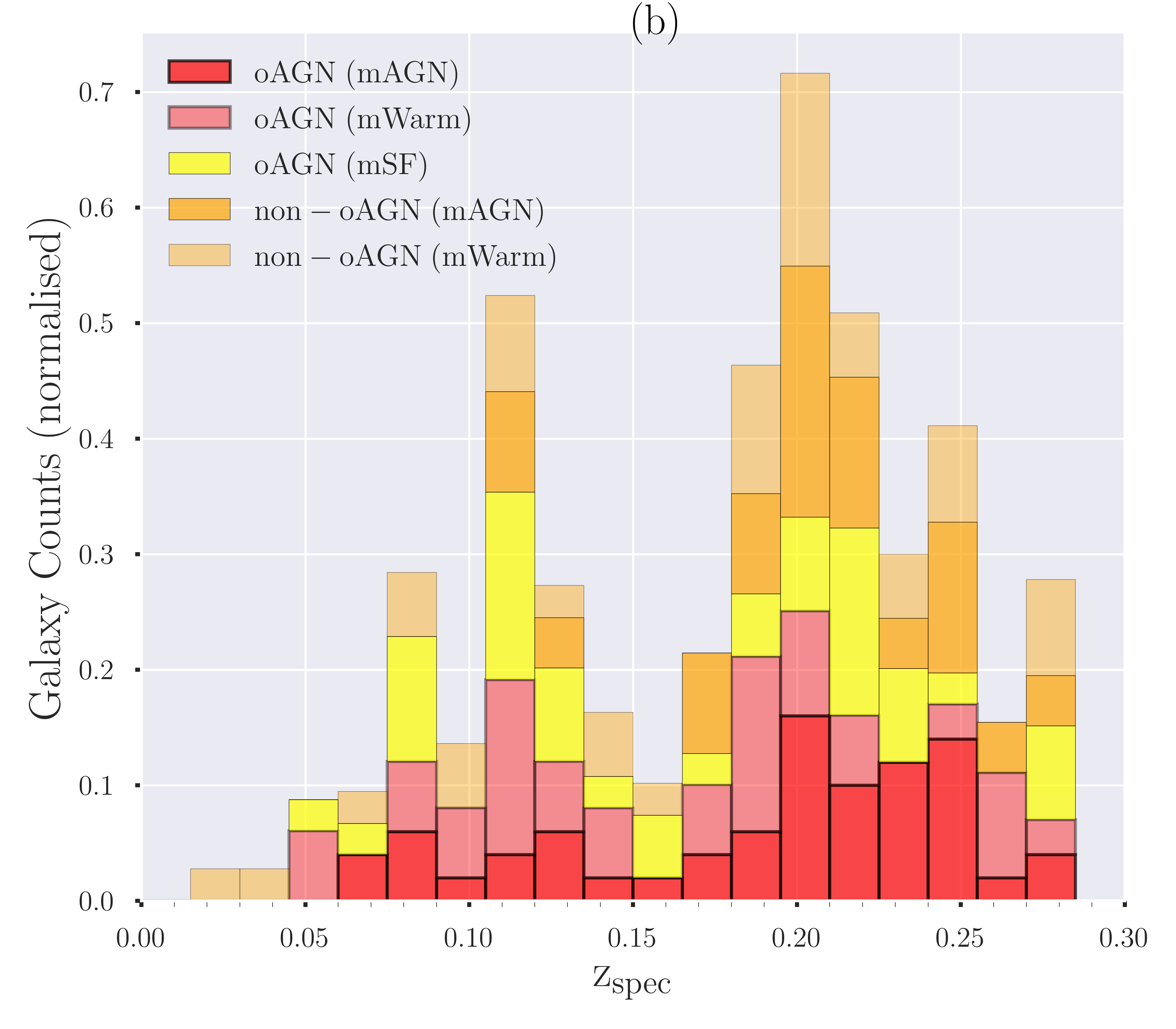}

\caption{The redshift distribution (fraction of galaxies per redshift bin, the BLAGNs are included in the different AGN category as presented in Table \ref{table1.4z} ``Condition d") for the SFs and composites (a) and the different groups of AGN (b). The colours and shades of the histograms have been kept consistent with Figures \ref{fig9}  and \ref{fig10}. Although, the redshift distribution could be only an observation bias due to the higher brightness of AGN compare to SFs, it is presented here to facilitate the description in this section. There are 2 main peaks located at redshifts of $\sim$ 0.08 and  $\sim$ 0.2.  The SFs and the Composites are concentrated at lower redshift (around \textit{z} = 0.08) while the oAGN (mAGN),  non-oAGN (mAGN) and non-oAGN (mWarm) seem to be generally located at redshift \textit{z} $>$ 0.2. The oAGN (mWarm) and  oAGN (mSF)  group look like a middle class with as many galaxies at low as at high redshifts. An inspection of the galaxy distribution using the entire G23 catalogue reveals the depletion around \textit{z} = 0.15 to be related to large scale structures. The highest peaks at a redshift \textit{z} $\geqslant$ 0.2 are seen among the  non-oAGN (mAGN) and non-oAGN (mWarm) which also contain the majority of  blended cases not visible in $\textit{WISE}$.}  \label{fig12a}
\end{centering}

\end{figure*}

Figure  \ref{fig12a} has been added to show the repartition of the different group of galaxies classified in the current study as a function of redshift. Our sample shows a high concentration of SF galaxies at low redshift that  decreases significantly toward higher redshift. This is probably due to an observation bias where fewer galaxies are detected at higher redshift as they get fainter. It is important to recognise that the oAGN (mSF) and the oAGN (mWarm)
are equally  distributed from low to high redshift as opposed to the non-oAGN (mAGN) generally found at higher redshift. Recall that the non-oAGN (mAGN) and the oAGN (mSF) groups represent galaxies for which the infrared and optical classification are contradictory.

\newpage

\subsection{Case studies} \label{sec3}

In this section, we showcase  galaxies selected from the different sub-groups created in Figure  \ref{fig8}. Each diagram from Figure \ref{fig12} to \ref{fig28}, presented in the Appendix, illustrates the investigation procedure adopted for the classification of the individual galaxies. Instead of making a comparison based only on the optical and the mid-IR  catalogues usually  available online, we adopted a more rigorous approach
in our study. For a given galaxy, a complete picture showing the BPT and mid-IR classification, the photometric SED with template fit, and the spectrum itself along with the $\textit{WISE}$ RGB image are presented. Finally, a deeper and more detailed picture is provided by the KiDS r-band imaging. The advantage here is that we see the photometric and physical properties, while also 
pinpointing any issues related to the data themselves. 
This combination  facilitates a consistency check of the data.
A list of the inferred parameter values and their uncertainties can be found in Table \ref{table1.3bb} (the selected galaxies as study cases are designated by the asterisk). Additionally, in order to better see the central regions, a zoomed KiDS r-band image of each one of the galaxies is presented in Figure \ref{fig29}. In this section, we will refer to the galaxies by their GAMA  catalogue ID ``CATAID"  or  ``Cat ID"  (in blue colour on the diagrams).  The $\textit{WISE}$ name, the fluxes in W1, W2, and W3 are also available on the diagrams.

\subsubsection{Optical and infrared Classified SF Galaxies} 

Here we present the optical star-forming galaxies and composites both classified as star-forming in the  mid-infrared (mid-IR). They are by far the most abundant type with a total (SF+composites) of $\sim$ 84$\%$ of the classification 
(see Table \ref{table1.4z}). 
They have the lowest averages of the stellar mass partly due to the fact that they are young galaxies which are still in the growing process, but on the other hand, unlike the galaxies hosting AGN whose fluxes are overestimated (as explained above), they have the most unaffected, reliable parameters (mass, SFR$_{12 \textit{$\mu$} \text{m}}$ and sSFR). 

The galaxies 5241095 and 5240983 (Figure \ref{fig12} and \ref{fig13}) are an interacting star-forming pair located at redshift \textit{z} = 0.027. Both galaxies are quite similar in their spectra and location in the BPT and $\textit{WISE}$ colour-colour diagram. This galaxy pair is atypical in the sense that each one of the galaxies shows the presence of a very high [\ion{O}{iii}] line which is stronger than the H$\alpha$ and a quasi absence of the [\ion{N}{ii}] line. The best-fit SED template classifies 5241095 as ``Irr" and  5240983 as a normal spiral ``Sc" galaxy. The KiDS image shows a clear interaction between the two galaxies. It appears as though  5240983 is pulling materials from 5241095 which is being ripped apart. Although the parameters here have much smaller values than the averages in their group (SF group, Table \ref{table1.4z}), the high values of stellar mass, SFR$_{12 \textit{$\mu$} \text{m}}$ and sSFR for 5240983 (9.14, 0.49 and -9.45) in comparison to 5241095  (9.05,0.2 and -9.75) seem to corroborate this scenario. It could also viewed as: galaxies get bigger with time through a series of mergers which in turn trigger occasional AGN activity (looking at the unusually high [\ion{O}{iii}] in this pair).

The galaxy 5306682  (Figure \ref{fig14}) is the perfect candidate (typical) to represent  the star-forming  group for having a very high H$\alpha$ line flux followed by  [\ion{N}{ii}],  [\ion{O}{iii}] and H$\beta$ with significantly smaller relative proportions. It is a good example that illustrates  how the deep KiDS image helps to decipher and reveal the host and it environment. Figure 
\ref{fig14}, shows one of our closest specimens in the total sample (redshift \textit{z} = 0.005). While in the $\textit{WISE}$ image, the magenta circle seems to enclose the entire galaxy, KiDS reveals a far larger galaxy that goes beyond the dusty part presented by $\textit{WISE}$ (recall that the two images have the same angular size).

The galaxy 5205726 in Figure \ref{fig16} is presented in the KiDS image as a  galaxy with clear spiral arms,  a bar, and a nuclear ring. It is a large galaxy ($\mathrm{log\;M_{stellar}}$ = 10.51 M$_{\odot}$) with a sSFR (sSFR= 10$^{-10.22}$ yr$^{-1}$). It is classified as a composite in optical with a [\ion{O}{iii}] luminosity of 10$^{40.29}\,$erg s$^{-1}$.
 
\subsubsection{Optical-infrared AGN} 

In this category, we have galaxies classified as AGN both in the optical (BPT) and the $\textit{WISE}$ colour-colour diagram (5.1$\%$ of the entire study sample).

The galaxy with GAMA ID 5339805 (Figure \ref{fig17})  is a typical case in the oAGN (mAGN) group. Its spectrum visually shows a broad H$\alpha$ line with a greenish colour in the $\textit{WISE}$ 3-band image characteristic of mAGNs. The KiDS image reveals a spiral galaxy with a bulge (or pseudo bulge) and a thin disk. It is also confirmed as an AGN by the SED fit. 

Although close to $\sim$ 70$\%$ of this sample is made up of broad-line AGN, they are mostly located at higher redshifts with a mean value of 0.2  (see Figure  \ref{fig12a}) as opposed to the current galaxy which is one of the nearest of its kind (redshift \textit{z} = 0.089). Generally the [\ion{N}{ii}] gets swallowed up by the broad H$\alpha$, which leads either to an underestimation/overestimation of  [\ion{N}{ii}]/H$\alpha$ line ratios. Indeed the BPT shows a [\ion{N}{ii}]/H$\alpha$ lines ratio $>$ 2 when in reality the [\ion{N}{ii}] line is barely visible next to the H$\alpha$.

In the same group, GAMA ID 5151978 presented in Figure  \ref{fig18} is located at higher redshift with the same typical $\textit{WISE}$ AGN colour. We can see a very strong [\ion{O}{iii}] line compared to H$\beta$ which places it well above the AGN dividing line in the BPT. It is located at the edge of the obscured-AGN box. The [\ion{O}{iii}] line luminosity $>$ 10$^{42}$\,erg s$^{-1}$ makes it a potential obscured quasar according to the criteria by Jarvis et al. (\citeyear{Jarvis2019}). Intriguingly, the SED template suggests a normal late-type galaxy, not an  AGN. The KiDS image provides a hint about the issue. Although the nearby source doesn't seem to interact with our main galaxy, it is close enough for $\textit{WISE}$ not to be able to disentangle both sources; they are instead treated as a single source due to the large beam of $\textit{WISE}$. The main galaxy is surely an AGN as shown by the spectrum and also the $\textit{WISE}$ colour, but the presence of the second might add extra features that somehow affect the SED fitting. Overall we classify this galaxy as an AGN as demonstrated by the diagnostics presented in the Figure.

\subsubsection{Optical AGN, infrared non-AGNs} 

This category is made up of galaxies that are classified as AGN based in the optical BPT, but seen as normal star-forming or as just getting warmer in $\textit{WISE}$ colour. We call warm galaxies in $\textit{WISE}$  for which W1-W2  is greater than a certain threshold, but below  the AGN zone (recall Figure \ref{fig8} for the different zones). One has to be careful during the classification for most of the misclassified cases are found in this category. Indeed taking into account the data quality flag in GAMA and also visually inspecting the spectra reduce the number of galaxies here by 25$\%$ not to mention that without a k-correction and a correction for stellar absorption, the number of false positives would have been much greater.  
The oAGN(mWarm) and oAGN(mSF) seem to belong to the same family which evolves with redshift. Both groups have average redshifts of \textit{z} = 0.16 and 0.18, respectively.

The oAGN(mWarm) group appears to be dominated by broad-lines (21/36) which unlike the general trend are not able to raise W1-W2 above the threshold required to be classified as infrared AGN or QSOs. Indeed $\textit{WISE}$ is very sensitive to the broad-lines systems and in most cases, such galaxies are classified as AGN. Maybe a parallel process is at work in the host making it difficult for $\textit{WISE}$ to have a clear view of the ongoing AGN activity. Galaxy 5347780 in Figure \ref{fig19} reveals broad H$\alpha$  and H$\beta$ lines, associated with a prominent [\ion{O}{iii}] line, all strong indications of AGN activity. However, W1-W2 is barely close to 0.6\,mag which falls in the ``warm" AGN zone.  The explanation of this behaviour might lie in the SED characteristic of a star-forming galaxy (i.e. the host dominates the observed emission).

 Galaxy 5294374 in Figure \ref{fig20} tells a slightly different story. Here in addition to the spectrum, the SED also presents the galaxy as an AGN while the $\textit{WISE}$ colour stills fall below the strong
AGN threshold. The change, in this case, clearly a late-stage merger, $\textit{WISE}$ colours could be dominated by a SF
galaxy (maybe the larger of the three in the merger) while AGN could come from anyone of these sources.

Galaxies 5249547 and 5155115 in Figure \ref{fig21} and \ref{fig22}, respectively are two examples of a clear disagreement between optical and mid-IR classifications. While the optical lines present them as AGN without ambiguity, the two galaxies sit exactly on the $\textit{WISE}$ star-forming main sequence, Eq. \ref{e1}. Furthermore, the SED confirms star formation as the dominant activity. Generally, the galaxies in this category are relatively nearby with the optical and the mid-IR giving evidence of AGN and SF activity, respectively. 
 We could argue that the galaxies here have AGN strong enough to be picked up by the optical, but with continuum (and mid-IR) emission dominated by the star formation of the host. The narrow optical fibres may see only the centre while $\textit{WISE}$ integrates the flux over the entire galaxy. It could also be that the AGN has recently turned off or shut down.  Generally, once the AGN turns off, the broad-line region (BLR), X-rays, and mid-IR shut down within a few decades, while the narrow-line region (NLR) could last thousands of years (Sartori et al. \citeyear{Sartori2018}).  This might be a plausible justification for some narrow-line optical AGN being classified as $\textit{WISE}$ star-forming galaxies.
Note that as it can be seen in Figure  \ref{fig21} (5249547, log L[\ion{O}{iii}] =  41.19\,erg/s), a substantial  number of the oAGN (mSF) present high luminosities (see Figure \ref{fig9}b) generally indicative of the presence of an AGN. The [\ion{O}{iii}] lines in these cases could originate from other processes different from the AGN activity. Possible explanations for this scenario are proposed in the discussion, below.

\subsubsection{Optical non-AGNs, infrared AGN} 

We address in this subsection galaxies with no AGN characteristic emission lines in the optical, but classified as either AGN or warm in the mid-infrared (the orange circles and squares in figure \ref{fig9}), named non-oAGN(mAGN) and  non-oAGN(mWarm), respectively.
 
 In our study, the tension between the optical vs infrared classification is mainly caused by two factors. The first is due to tightly blended galaxies for which the $\textit{WISE}$ image alone cannot separate them, but are clearly distinguishable in the deep KiDS r-band image. By comparing the KiDS vs $\textit{WISE}$  images in Figure  \ref{fig23},  \ref{fig24} and  \ref{fig27}, we can see that multiple sources are mistakenly being considered as single (in $\textit{WISE}$) owing to their angular proximity at the given redshift with mean redshift \textit{z} = 0.2 (among the most distant galaxies of the study sample),  while the optical (GAMA) which uses the KiDS galaxies location is only targeting the central galaxy.  The outcome of the classification in the mid-infrared will depend on the components of the merging system. A composition of several AGN will probably lead to a higher W1-W2 colour while the opposite could rather lead to a decrease. The system presented in Figure  \ref{fig27} (CATAID: 5204947) is an extreme case where apparently more than 3 galaxies seem to be merging. As we can see from these 3 examples (i.e., Figures \ref{fig23},  \ref{fig24}, and  \ref{fig27}), 5204947 which lies below the AGN region is yet the only one presented as AGN by the SED. 
 
 The second cause of misclassification is related to the broad-line measurement. It has been noted several times that for broad H$\alpha$ lines, it is sometimes challenging to distinguish between the 
 [\ion{N}{ii}] and H$\alpha$ to get the correct line ratios. In any case, a broad H$\alpha$ and H$\beta$ lines are already enough for the galaxy to be classified as AGN. Using the line ratio here leads to an underestimation of 
  [\ion{N}{ii}]/H$\alpha$, therefore, classifying the galaxy as non-oAGN. This picture is seen in Figure \ref{fig25}  and \ref{fig26} where the broad-line galaxies are correctly classified as AGN both using the $\textit{WISE}$ colour-colour diagram and the SED fitting, but lie in the composite area of the optical BPT.  
 
The last study case 5135645 (Figure \ref{fig28})  represents a blended system classified as optical SF and infrared SF (SF), but located at the extreme limit between the AGN and SF zone. We suspect the second galaxy to be an AGN justifying the high W1-W2 (0.74\,mag, very warm) or related to the fact that  the interaction is triggering AGN activity. The extreme W2-W3 $>$ 5.04\,mag is not unlike the class of HyperLirgs that $\textit{WISE}$ has recently uncovered (Tsai et al. \citeyear{Tsai2015}). The SED template fit using the famous starbust-AGN hybrid system  NGC3690 gives a glimpse of the probable internal processes taking place. These cases will be part of our next study which will bring in radio data from SKA pathfinders ASKAP/EMU and MeerKAT already in the reduction phase.

\subsection{The New  Diagnostic: \text{[W1-W2]} vs  \text{[\ion{O}{iii}]}/H$\beta$} 

Based on the optical emission lines and the mid-IR colours, we now attempt to combine them into one diagnostic, where we choose the best parameters for each:  namely, the strongest emission lines: [\ion{N}{ii}]/H$\alpha$,  and the highest S/N and most-AGN-sensitive colours: W1-W2. 
 The [\ion{N}{ii}]/H$\alpha$ line ratio is potentially sensitive to metallicity notably for lower stellar mass galaxies, but as noted in section 3.1.1 our sample will be minimally affected  by metallicity.

\begin{figure*} 
\begin{centering}
\includegraphics[width= 8.8cm] {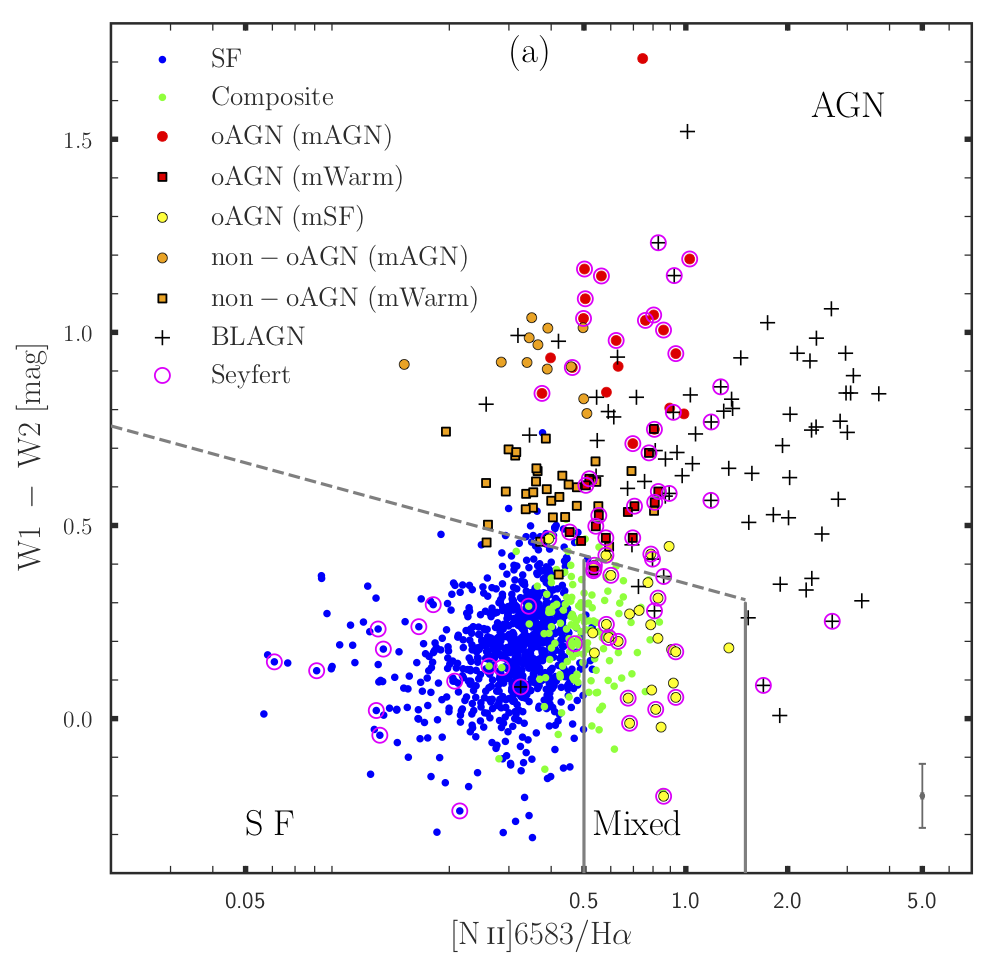} 
\includegraphics[width= 8.8cm] {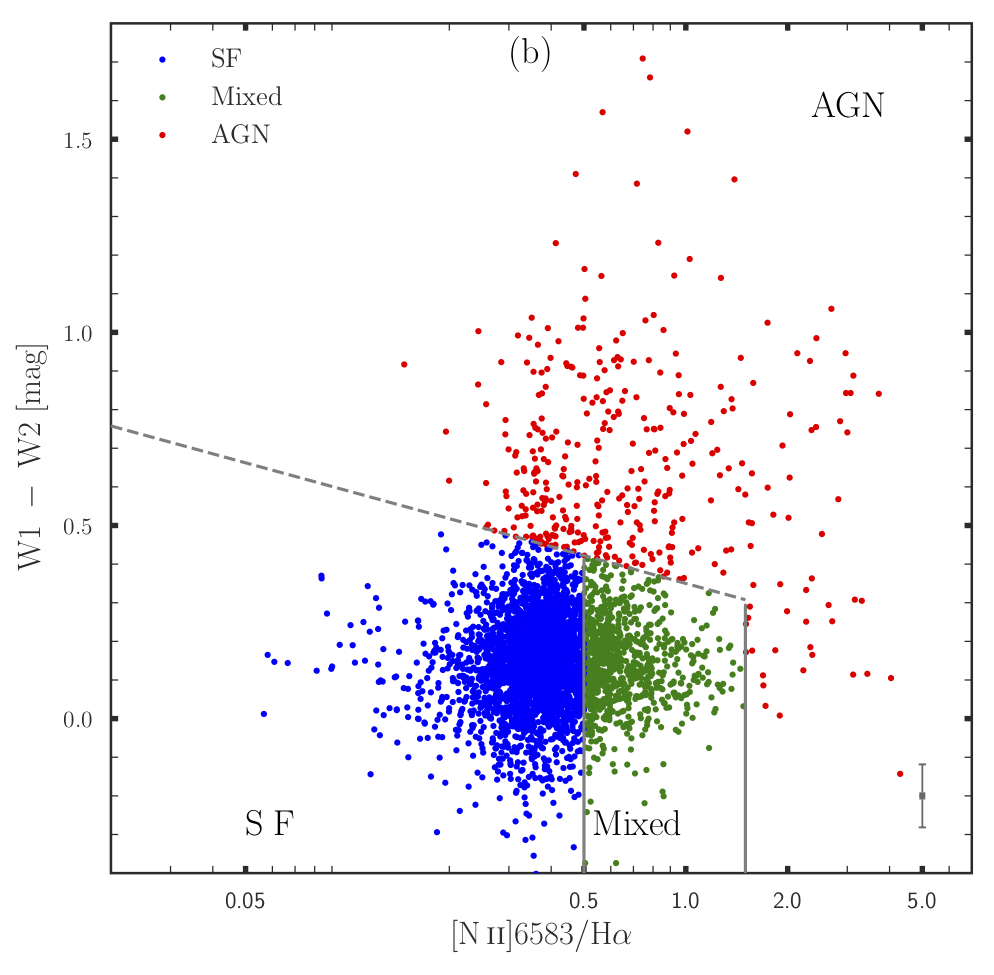}

\caption{Optical-IR SF-AGN diagnostic plot.  The new method used to classify galaxies based on both the optical BPT and the $\textit{(WISE)}$ colour-colour diagrams. In (a) we reproduce the classification of Figure \ref{fig9} keeping the same galaxy groups with their assigned colours, all the optical lines in emission ([\ion{O}{iii}], [\ion{N}{ii}], H$\beta$ and H$\alpha$ lines are all positive) and replacing the [\ion{O}{iii}]/H$\beta$ by the W1-W2 colour.  The Mixed  region (delimited by the solids and the dashed grey lines) is a mixture of the composites and the oAGN (mSF). The mean  [\ion{N}{ii}]/H$\alpha$ and W1-W2  uncertainties with the error bars are plotted in the lower-right corner.  (b) is similar to (a), but only  the fluxes of  [\ion{N}{ii}] and H$\alpha$ are required to be positive (emission lines). The H$\beta$ line could be either in emission or absorption. The statistic increases from 893 SF, 87 Mixed and  174 AGN in (a)  to 2818 SF, 853 Mixed and 364  AGN in (b).  The mean [\ion{N}{ii}]/H$\alpha$ and W2-W3  uncertainties with the error bars  are plotted in the lower-right corner.}   \label{fig12q}
\end{centering}
\end{figure*}

 In  Figure \ref{fig12q} we present our new classification scheme combining both the BPT and the WISE colour-colour diagrams. Figure \ref{fig12q}a is a modified version of Figure \ref{fig9}, where  the ratio [\ion{O}{iii}]/H$\beta$ is replaced by the W1-W2 colour, thus avoiding these emission lines that can be difficult to measure. Indeed the BPT is typically limited by the low S/N of the H$\beta$ line (and sometimes the oxygen line as well), which represents a strong and significant bias for a WISE-selected sample. 
The dashed grey line, crossing the [\ion{N}{ii}]/H$\alpha$ line ratio, empirically represents the best separation between AGN (upper region) and non-AGN,  with the latter divided into the star-forming (SF) and Mixed galaxies  region by the vertical grey line at [\ion{N}{ii}]/H$\alpha$ = 0.5.  The Mixed division is motivated by the large number of composites, SF and  oAGNs that inhabit relatively low (blue) W1-W1 colours. The Seyferts  found in the star-forming region  are the few star-forming galaxies that are found in the AGN side of the BPT as presented in Figure \ref{fig7B}. As both optical  diagnostics
agree generally, these  few outliers at the boundary are unavoidable and inherent to any empirical methods. But, the optical AGN found at low $\textit{WISE}$ colours (in the SF or Mixed zone)  are interesting  cases that need further investigation either in radio or X-rays. At [\ion{N}{ii}]/H$\alpha$ $>$ 1 lie generally broad line and Type-1 AGN. The study sample  does not include  low mass galaxies with AGN (a type of low luminosity AGN); and furthermore, is drawn from the local universe (z $<$ 0.3). Therefore, our work is relevant to z $<$ 0.3 and the limit boundaries here will require adjusting  at higher redshift especially when the impact of the metallicity on the BPT ([\ion{N}{ii}]/H$\alpha$) lines become more significant.

\begin{figure*} 
\begin{centering}
\includegraphics[width= 16cm] {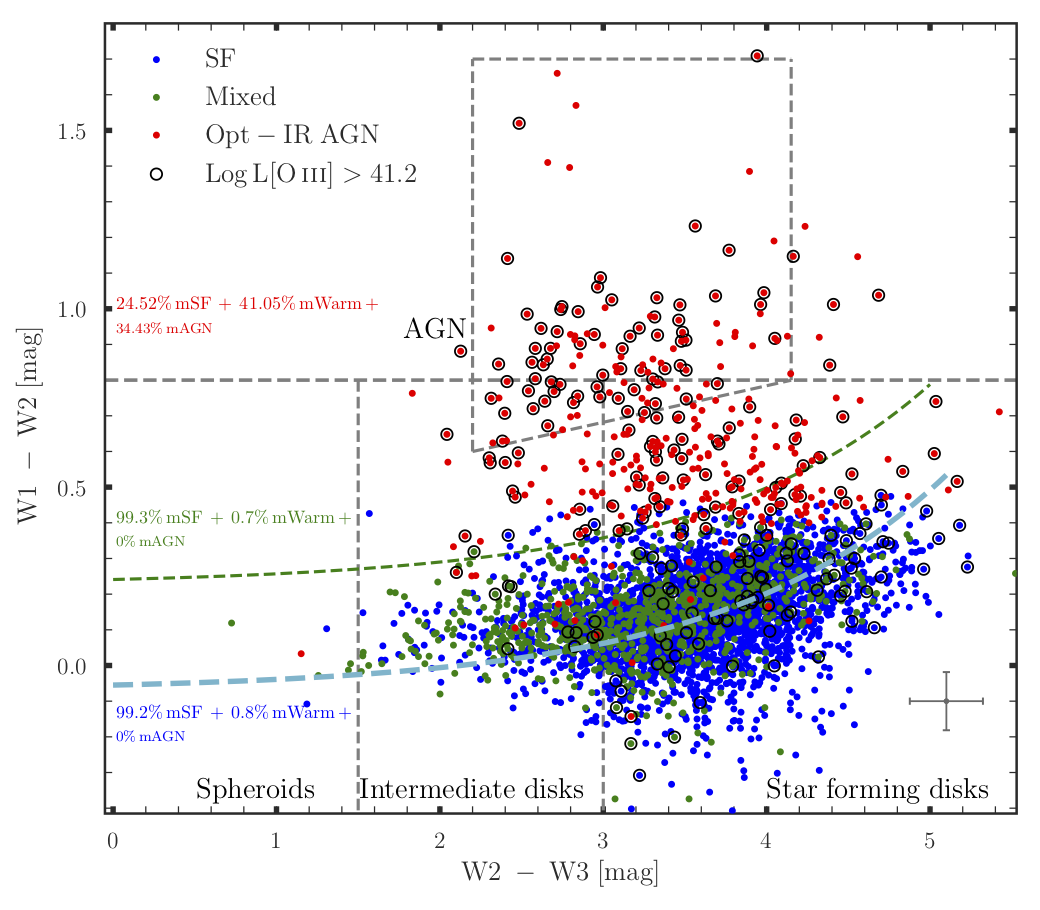} %
\caption{The $\textit{WISE}$ colour-colour representation of the galaxies classified in figure \ref{fig12q}b using the same key. The  AGN are well delimited by the  2$\sigma$  offset line up to  W2-W3 $\sim$3.8\,mag where their    W1-W2 colour tend to be slightly below.  The different galaxy groups (derived in Fig. \ref{fig12q}b) are presented with their relative proprtions, to the left. The strong [\ion{O}{iii}] emitters (see Fig. \ref{fig9}) are highlighted with black circles. We also plot the mean W1-W2 and W2-W3  uncertainties with the error bars in the lower-right corner.}  \label{fig12qb}
\end{centering}
\end{figure*}

W1-W2 $<$ 0.5 mag and the dashed grey line enclose all the SF and more than half of the composite galaxies. 
The Mixed region delimited by 0.5 $<$ [\ion{N}{ii}]/H$\alpha$ $<$ 1 and the dashed grey line described by equation \eqref{e3} is a mixture of oAGN(mSF) and composite galaxies and importantly, contains some
 strong [\ion{O}{iii}] emitters (see also Fig. \ref{fig9}b).

\begin{equation} \label{e3}
\text{Y} = -0.24 \times \text{log}(\text{X}) + 0.35
\end{equation}
where Y = W1-W2\,mag, X = [\ion{N}{ii}]/H$\alpha$.

 The remaining classes of galaxies are above the dashed grey line and represent the optical infrared AGN (Opt-IR AGN), which makes this new classification scheme an efficient way of disentangling pure
star-forming from galaxies having AGN activity. It allows us to get the best from both methods and the use of W1-W2 colour rejects far fewer galaxies than [\ion{O}{iii}]/H$\beta$. In Figure \ref{fig12q}b the constraints on the
 [\ion{O}{iii}] (flux $>$ 0) and H$\beta$ (flux $>$ 0) are removed leading to more classified galaxies. From the initial number of 1,154 classified galaxies in  (a),   4,035 galaxies are now classified in (b) where the W1-W2 is used instead of [\ion{O}{iii}]/H$\beta$,  hence nearly a 400\,$\%$ increase.

 The galaxies from Figure \ref{fig12q}b  are presented in the $\textit{WISE}$  colour-colour diagram in Figure \ref{fig12qb}.  It shows the position of the galaxies classified by the new diagram  (Fig. \ref{fig12q}b) on the  W1-W2 vs W2-W3 colour diagram with their associated fraction in blue (SF), green (Mixed) and red (AGN).  The new rendering of the colour-colour diagram may be directly compared with the relatively sparse results in Figure \ref{fig8}.
More than 99 $\%$ of the SF and Mixed from the new diagram are found in the star-forming region of the colour-colour diagram. About 75 $\%$ of the AGN are either mWarm or mAGN while 25 $\%$ are in the mSF zone. The galaxies in the latter case are, in general, close to the mWarm zone (at W1-W2 = 0.4\,mag) and have a high value of W2-W3 $>$ 3.3\,mag (i.e., red, dusty). In rare cases of extreme broad H$\alpha$  or where the [\ion{N}{ii}]  is stronger than the  H$\alpha$ line, the  AGN from the new diagram are classified as mSF. The galaxies with strong [\ion{O}{iii}] emission are highlighted by the black circles (see Figure \ref{fig9}). They seem to be distributed throughout the diagram but arguably represent higher proportion of the AGN (37\,$\%$) in comparison to the star-forming (2\,$\%$) and Mixed (5\,$\%$) galaxies.

\section{Discussion} 

We have conducted a careful analysis of the correspondence between optically-defined host and nuclear activity (using the BPT diagnostic) and mid-infrared colours in this study.  There is wide agreement between optical and mid-IR classification, but real differences that highlight the limitations and strengths of each, while also suggesting that a synergy of combination should provide new information. 

Agostino et al. (\citeyear{Agostino2019}) found a similar fraction of Seyferts and LINERs in their X-ray AGN sample. We find the opposite trend (BPT using [\ion{S}{ii}] line) where  very few galaxies are classified as LINERS in comparison to the Seyferts. Generally,  the LINERs from our sample are classified as mid-infrared star-forming galaxies. The large majority are also classified as optical star-forming using the BPT based on the [\ion{N}{ii}] line. This might be partly due to the very small number of LINERs (5 in total) found in our sample.  It is possible that the emission in these LINERs is not driven by AGN (Yan $\&$ Blanton \citeyear{Yan2012}, Belfiore et al. \citeyear{Belfiore2016}) or they could be simply LINERs with low-luminosity  AGN (Filho et al. \citeyear{Filho2006}; Flohic et al. \citeyear{Flohic2006}; Gonzalez-Martin et al. \citeyear{Gonz2009}). Additionally, strong shocks can significantly contribute to the AGN emission in these objects, creating the LINER-like emission lines (Dopita et al. \citeyear{Dopita1996}, \citeyear{Dopita1997}; Molina et al. \citeyear{Molina2018}).

Using the KiDS r-band images, we found that a large fraction of the non-oAGN (mAGN) have more than one galaxy in the aperture, some of which may be interacting or galaxy pairs close enough to be seen as a single galaxy in $\textit{WISE}$.
 Since in this case $\textit{WISE}$ can not distinguish the two galaxies then the companion of the targeted galaxy might have some AGN activity. It may also be that an interaction has triggered the AGN activity as suggested by Ellison et al.  (\citeyear{Ellison2019}). Using the Canada France Imaging Survey (CFIS) images they found that more than 60$\%$ of the mid-infrared AGN are interacting systems and concluded that the interaction might play an important role in the nuclear feeding process.

 The non-oAGN (mAGN) galaxies occasionally have their optical flux ratios underestimated due to the presence of broad H$\alpha$ lines, but there are still a few galaxies for which only the obscuration of the torus could explain their classification as non-AGNs in optical. The Compton-thick clouds such as those associated with galaxy mergers (Ricci et al. \citeyear{{Ricci2017}}; Satyapal et al. \citeyear{Satyapal2017}) can prevent the optical lines from being detected.

Our study shows about 21$\%$ of the optical AGN to be located in the $\textit{WISE}$ star-forming zone.  The simplest explanation is that they are low power AGN in which the starlight from the host dominates over the AGN. This could be verified through the [\ion{O}{iii}] line luminosity generally thought to be correlated to the AGN strength (Bassani et al. \citeyear{Bassani1999}; Heckman et al. \citeyear{Heckman2005}).
Unfortunately, not all the optical fluxes in the GAMA catalogue are calibrated to be used directly to determine luminosities (see section \ref{sec1}). Nevertheless, the available ones range from luminosity = 10$^{41}$  to  10$^{42.8}$\,erg s$^{-1}$  which appear to be high and AGN-like notably for those with  luminosity $>$ 10$^{42}$\,erg s$^{-1}$ (see Reyes et al. \citeyear{Reyes2008}; Yan et al. \citeyear{Yan2019}). So, the oAGN (mSF) with corresponding low [\ion{O}{iii}] luminosity  fall in the category described above where the starlight dominates the AGN.
On the other hand, an oAGN (mSF) presenting a high [\ion{O}{iii}] luminosity could be the consequence of a recent change in the nuclear emission. This implies that the accretion disk has shut down and stopped emitting in the infrared, but not enough time has passed to stop the narrow line region's (NLR) emission which corresponds to thousands of years (Sartori et al. \citeyear{Sartori2016}; Keel et al. \citeyear{Keel2017}; D. Stern and R. Assef, private communication). This seems to be the case for galaxy 5249547 (see the case study and Fig. A9) with a strong [\ion{O}{iii}] luminosity of 10$^{41.19}$\,erg s$^{-1}$, but whose AGN activity is invisible in $\textit{WISE}$ (W1-W2 $\sim$ 0\,mag).

An alternative explanation could be that the accretion disk is not powerful enough to generate winds that produce broad-line regions (BLR) or alternatively the [\ion{O}{iii}] lines are coming from non-nuclear shocks rather than from the central AGN. Berney et al. (\citeyear{Berney2015}) found a weak correlation of the hard X-ray fluxes with the fluxes of high-ionisation narrow lines such as [\ion{O}{iii}], [NeIII], [HeII], etc. not caused by factors like obscuration or slit size. This begs the question of whether the [\ion{O}{iii}] lines are always related to the central AGN. Indeed several studies have shown that while the optical line diagnostics can reasonably well separate AGN and SF, they are not able to differentiate between line emission arising from non-nuclear shocked gas and that of the AGN (Monreal-Ibero et al. \citeyear{Monreal-Ibero2010}; Rich et al. \citeyear{Rich2015}; D'Agostino et al. \citeyear{D'Agostino2019}; P. V\"ais\"anen, private communication)
  
These alternative scenarios may be investigated using X-ray data to check the presence of a torus and using long-slit spectroscopy  to map the [\ion{O}{iii}] distribution notably in the circumnuclear SF rings. It will also be useful to understand their radio properties using radio continuum data.

Few broad-line AGN are found in the $\textit{WISE}$ star-forming zone. An investigation revealed them to be part of the rare cases where the W1-W2 colour drops (becomes bluer) after the k-correction, examples of where the limited templates (notably for AGN) are poorly fit to the apparent fluxes.

The comparison between the  $\textit{WISE}$ and the BPT diagrams revealed   W1-W2 vs [\ion{N}{ii}]/H$\alpha$ to be a good diagnostic to separate star-forming galaxies from AGN. Our study shows the importance of the oAGN (mSF) class  which could provide crucial information about the AGN activity within galaxies. Fortunately, The oAGN (mSF) can be reliably found in the Mixed  region  (Figure \ref{fig12q}a) that
encloses about 90 $\%$ of their total number. They generally have low W1-W2 colour, high [\ion{N}{ii}]/H$\alpha$ ratios and are separated from the  rest of AGN by the dashed line (in Figure \ref{fig12q}) described by the equation \eqref{e3}. This new method does not require the H$\beta$ and [\ion{O}{iii}], thus allowing classification of galaxies with the [\ion{O}{iii}] and/or the H$\beta$ lines  either absent (low S/N) or in absorption which is not otherwise possible using the BPT. The Figure \ref{fig12q}b  shows 3 times more classified galaxies  than Fig. \ref{fig12q}a.  Using this new method, we have separated our larger line-emission sample into over 4000 SF and AGN galaxies (see Fig. \ref{fig12qb}, a 400$\%$ improvement over the traditional BPT).  As mentioned earlier, The accuracy of the $\textit{WISE}$ colour-colour and the BPT (using [\ion{N}{ii}]/H$\alpha$) classifications are potentially affected by the presence of dwarf 
galaxies and the metallicity, respectively.  This study has been conducted on a safe ground where those two caveats have been avoided by using intermediate to high mass galaxies LogM$_{\text{stellar}}$ $>$ 9\,M$\odot$ and a sample with a redshift limit of z $<$ 0.3. 
A greater sample combining several GAMA fields could be used in future with radio data to update our new diagnostic method. New and deeper spectroscopic surveys will be useful to extend the study to higher redshift
where the [\ion{N}{ii}]/H$\alpha$ line ratio is more susceptible to evolutionary effects.

\section{Conclusion} 

We investigated the optical emission lines properties of galaxies in GAMA G23 using mid-infrared $\textit{(WISE)}$ photometry. Unlike most preceding studies of the kind, special care was taken in extracting the nearby extended galaxies using our customised extended emission pipeline (see Section \ref{sec23}). The photometry was derived accordingly. A magnitude cut was applied to select the cleanest and the highest signal to noise galaxies leading to a high-quality dataset of  about 9,800 galaxies. They were  cross matched to the emission-line catalogue and the resulting sample (1154 galaxies) used for a comparison between the commonly 
used BPT  and $\textit{WISE}$ colour-colour diagrams. Additional visual inspection of all the spectra and the high-resolution KiDS r-band image of each one of the classified galaxies was carried out, with examples presented in Figure \ref{fig12} to \ref{fig28}.
The derived dataset and finding are presented as follows:\\

\begin{itemize}

\item[$\bullet$] The first $\textit{WISE}$ galaxy catalogue in the GAMA G23 field has been created including nearby galaxies extracted using the resolved pipeline.

\item[$\bullet$] The visual checking of the spectra revealed the BPT based on the [\ion{N}{ii}]/H$\alpha$ vs  [\ion{O}{iii}]/H$\beta$ to be more reliable than the one using [\ion{S}{ii}]/H$\alpha$ vs  [\ion{O}{iii}]/H$\beta$. Most of the strong  [\ion{S}{ii}] 
detections are artefacts. This mostly affects galaxies classified as LINERS.

\item[$\bullet$] Galaxies were classified based on their mid-IR and optical properties in Figure \ref{fig8} and Figure \ref{fig9}a. There is good agreement between the two methods in classifying the non-AGN galaxies. Only 4$\%$ of the mid-IR star-forming are classified as optical AGN and 5$\%$ of the optical star-forming are classified as mid-IR AGN. But 29$\%$ of the optical AGN are non-mid-IR AGN while 33$\%$ (if we consider the mWarms as midIR mAGNs) of the mid-IR AGN are not optical AGN. The optical composites share similar properties to the optical star-forming galaxies in the $\textit{WISE}$ colour-colour diagram, but more than 60$\%$ of the non-oAGN(mAGN) are optical composites.

\item[$\bullet$] Optical AGN are well-selected by high [\ion{O}{iii}]/H$\beta$ ratios while the  [\ion{N}{ii}]/H$\alpha$  ratios are generally $<$ 1. A visual inspection of the spectra revealed the AGN with [\ion{N}{ii}]/H$\alpha$  ratio $>$ 1 in Figure \ref{fig9}a to be overestimated due to the broad-line feature of the H$\alpha$. 

\item[$\bullet$] A detailed study of the different groups of galaxies shows a scenario in which SF galaxies have the lowest redshifts followed by the oAGNs(mSF) which are a combination of AGN and SF activity. $\textit{WISE}$ is often blind to their AGN leading sometimes to very low W1-W2 colours. For these galaxies relatively nearby, the spectroscopic fibre might be capturing only the central part while $\textit{WISE}$  is sampling the flux over the entire galaxy dominated by the SF activity.
At higher redshifts are the oAGN(mAGN) and the non-oAGN(mAGN). The latter group is made up of galaxies with underestimated [\ion{N}{ii}]/H$\alpha$ ratio (broad H$\alpha$  line) and by blended systems that are proximal-close and small enough to be seen as a single source by $\textit{WISE}$. We were able to identify them using the high-resolution images provided by KiDS.

\item[$\bullet$] We created a new diagnostic diagram that combines the optical lines and the $\textit{WISE}$ colour  in the form of W1-W2 vs  [\ion{N}{ii}]/H$\alpha$. The W1-W2 colour which is sensitive to the AGN emission can reliably replace the [\ion{O}{iii}]/H$\beta$ ratio (in the BPT diagram) which is limited by the H$\beta$ and [\ion{O}{iii}] detection. It has the distinct advantage of increasing the number of classified galaxies by more than 3 to 4 times.

\end{itemize}

\section{ACKNOWLEDGEMENTS}

We thank D. Stern, R. Assef and P.  V\"ais\"anen for useful discussion about AGN and shocks.  HFMY would like to acknowledge the support given from the NRF (South Africa) through the Centre for Radio Cosmology (CRC). MC is a recipient of an Australian Research Council Future Fellowship (project number FT170100273) funded by the Australian Government. T.H.J. acknowledge support from the National Research Foundation (South Africa).  This publication makes use of data products from the Wide-field Infrared Survey Explorer, which is a joint project of the University of California, Los Angeles, and the Jet Propulsion Laboratory/California Institute of Technology, funded by the National Aeronautics and Space Administration.\\
GAMA is a joint European-Australasian project based around a spectroscopic campaign using the Anglo- Australian Telescope. The GAMA input catalog is based on data taken from the Sloan Digital Sky Survey and the UKIRT Infrared Deep Sky Survey. Complementary imaging of the GAMA regions is being obtained by a number of independent survey programs including GALEX MIS, VST KIDS, VISTA VIKING, WISE, Herschel- ATLAS, GMRT and ASKAP providing UV to radio coverage. GAMA is funded by the STFC (UK), the ARC (Australia), the AAO, and the participating institutions. The GAMA website is http://www.gama- survey.org/.

\appendix

\section{Some Examples for the cases study}
 
 In Section 3.3 we presented detailed case studies of the different sub-group classifications.  The graphics for the case studies are presented here in the Appendix, and to follow (A1 to A16).   The final figure, A17, presents the inner regions as viewed with the KIDS r-band imaging.
 
\begin{figure*}[hb!]
\centering
  \includegraphics[width= 17cm]{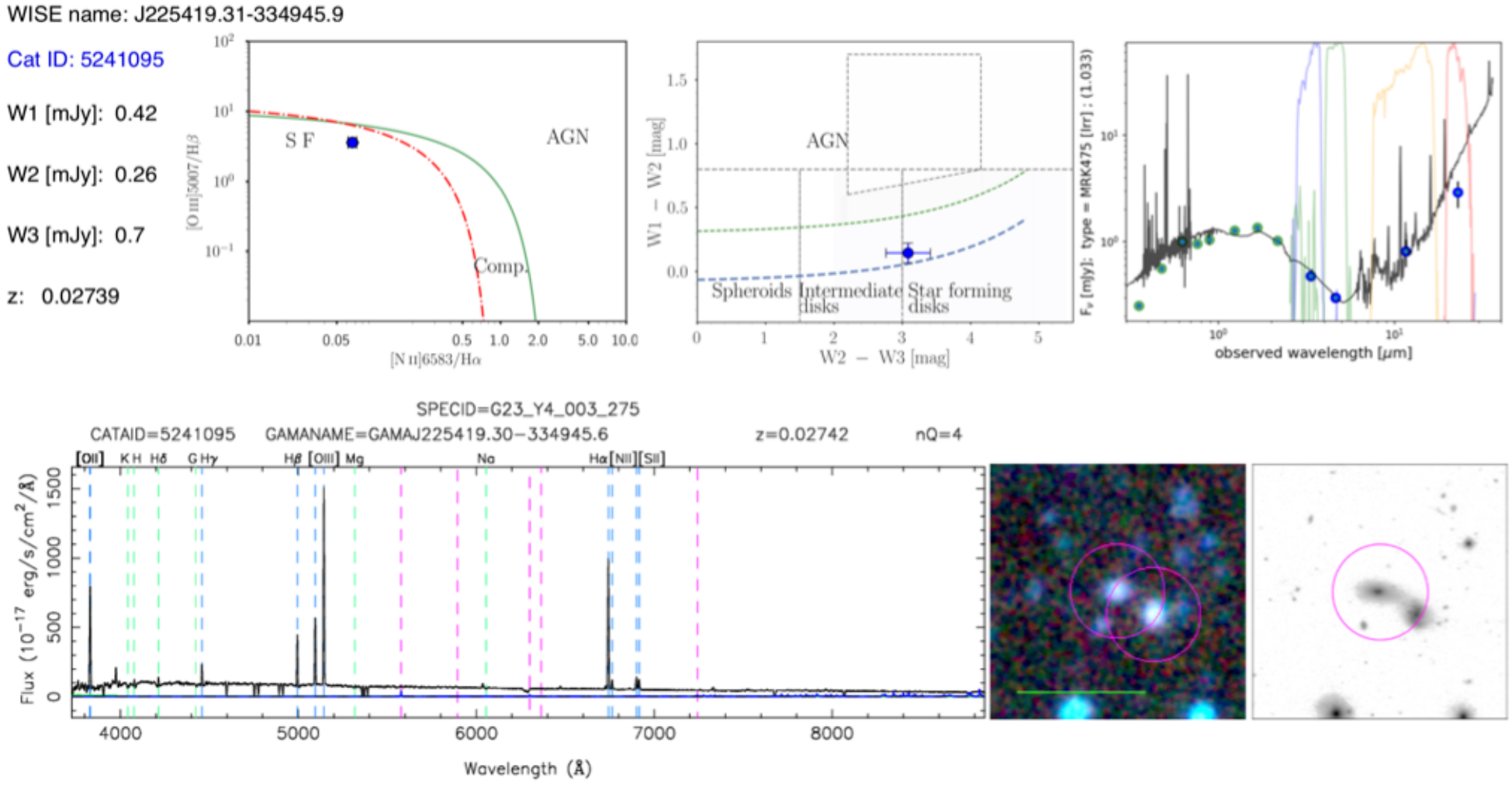}
    \caption{Classification of the galaxy (GAMA ID: 5241095) based on its optical (GAMA) and mid-IR (WISE)  properties. The first row, from left to right shows the BPT diagram, the $\textit{WISE}$  colour-colour 
    diagram and the photometric SED (with best-fit template). The second row shows the spectrum, the $\textit{WISE}$  RGB stamp (2 arcmin, the horizontal green line represents 1 arcmin) and the KiDS stamp (2 arcmin).  A zoomed in  KiDS stamp (30 arcsec)  is given in Figure \ref{fig29}  to  show a detailed picture of the galaxy and its nearby environment. This galaxy is an example of a tidal-interacting galaxy which is classified as star-forming by both the  optical (BPT) and the mid-infrared ($\textit{WISE}$ colours). The classification is also confirmed by the shape of the SED, but the galaxy exhibits an uncommonly high [\ion{O}{iii}]  luminosity (10$^{41.14}$ erg s$^{-1}$) that could be related to the interaction with its neighbours.}    \label{fig12}
\end{figure*}

\begin{figure*}
\centering
  \includegraphics[width= 17cm]{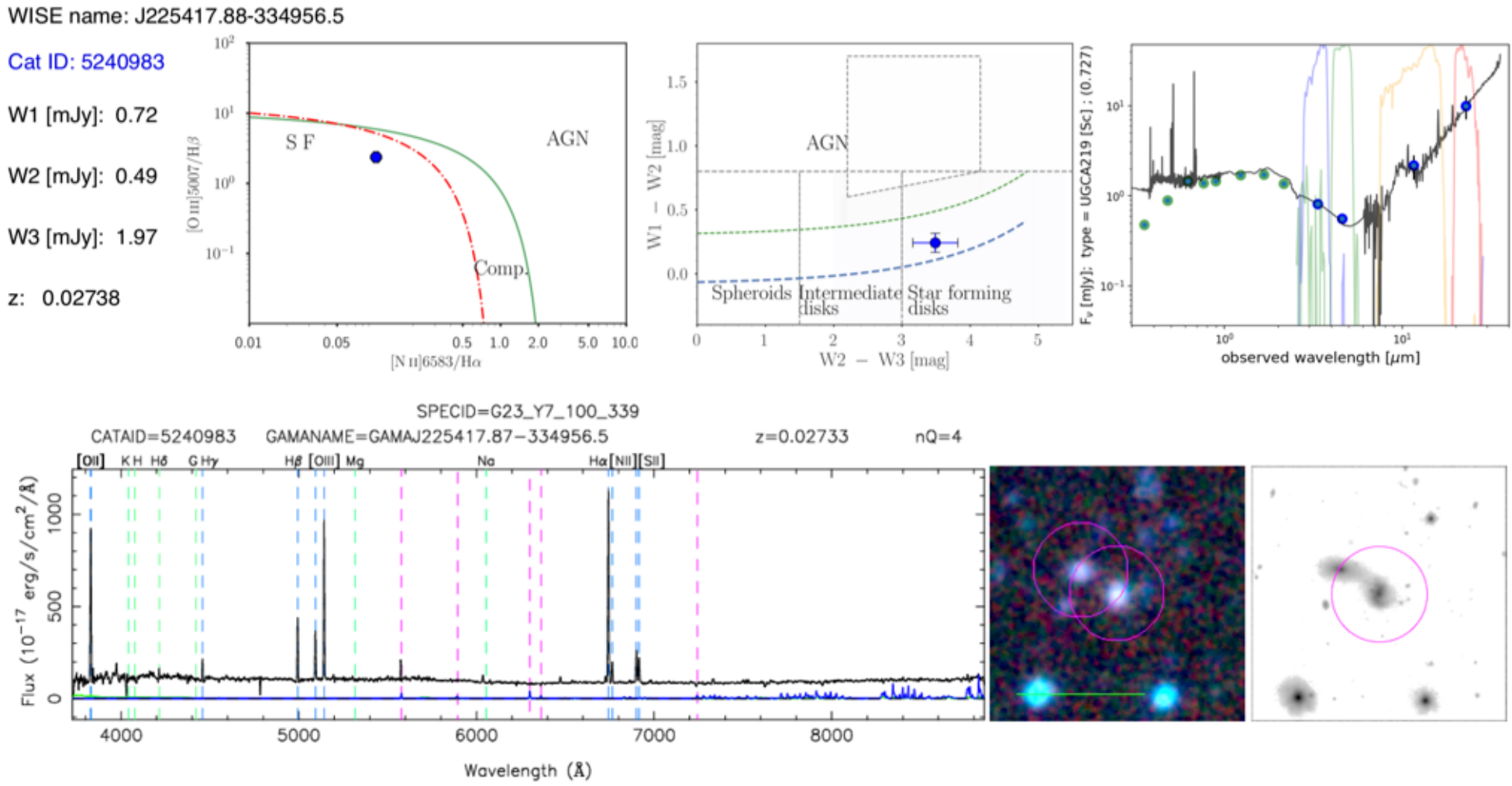}
    \caption{Galaxy 5240983. Classified as a star forming (blue). }    \label{fig13}
\end{figure*}

\begin{figure*}
\centering
  \includegraphics[width= 17cm]{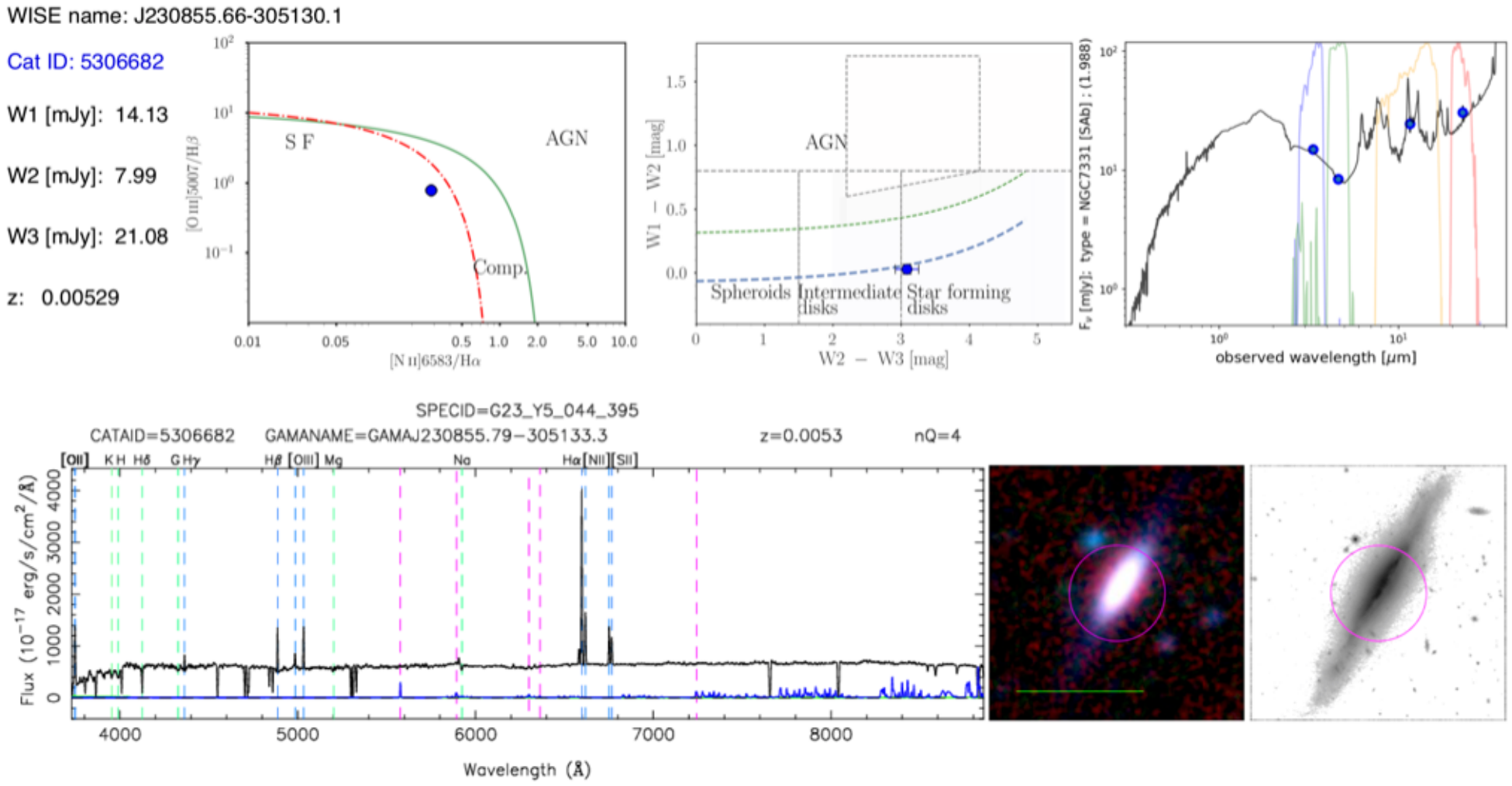}
    \caption{Galaxy 5306682. Classified as a star forming (blue). }    \label{fig14}
\end{figure*}

\begin{figure*}
\centering
  \includegraphics[width= 17cm]{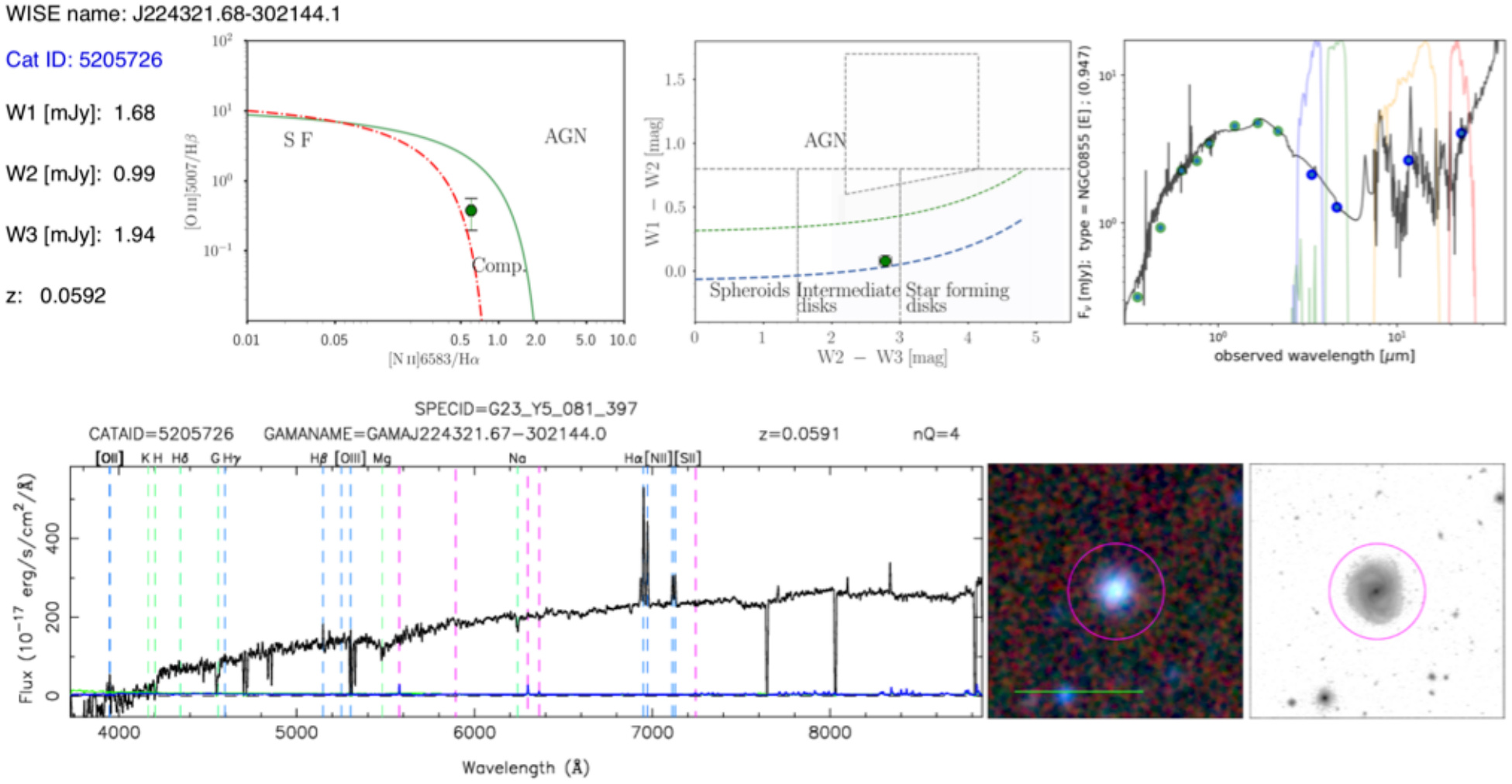}
    \caption{Galaxy 5205726. Classified as a composite  (green). }    \label{fig16}
\end{figure*}

\begin{figure*}
\centering
  \includegraphics[width= 17cm]{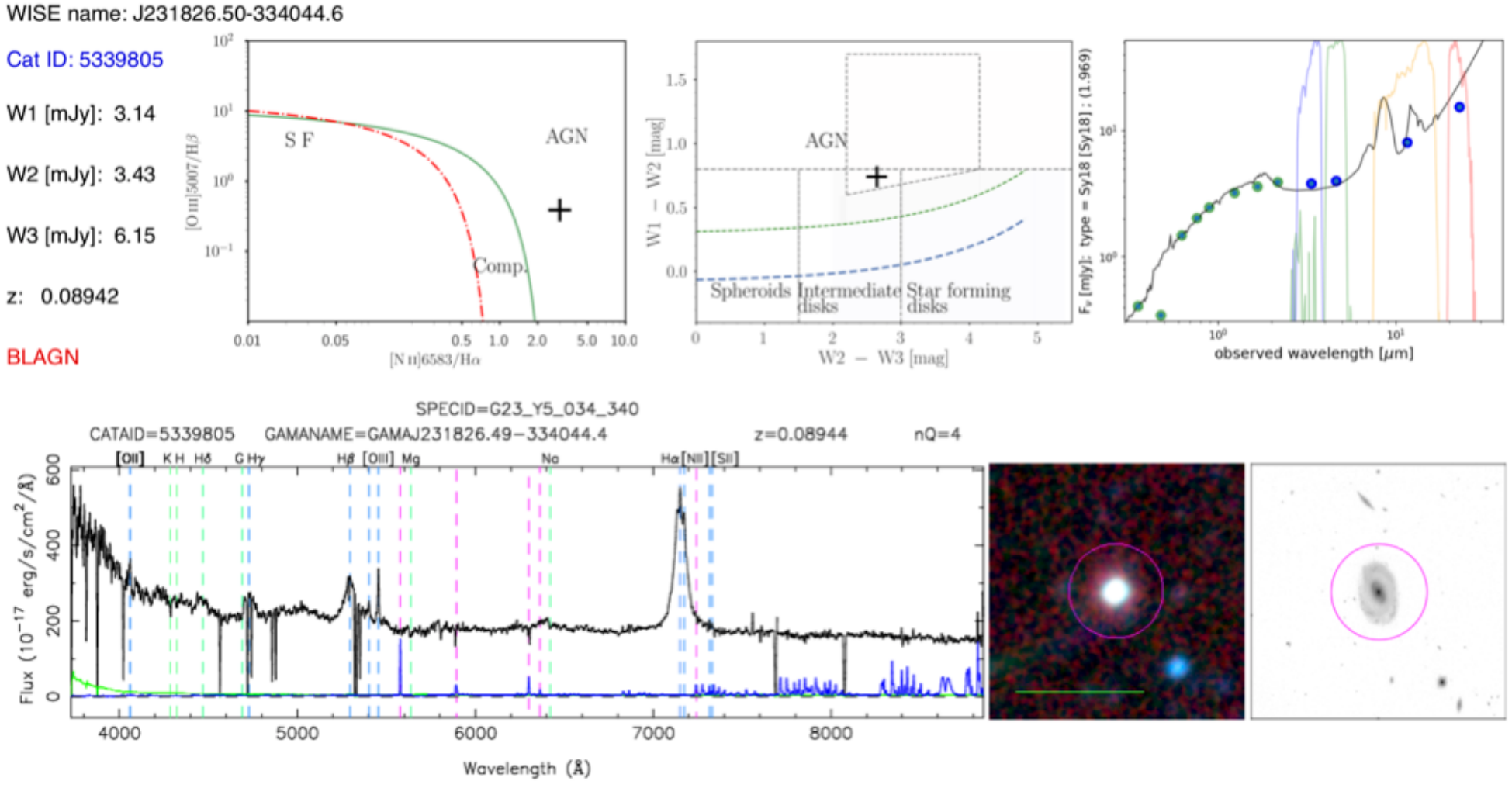}
    \caption{Galaxy 5339805. A broad-line AGN classified as a oAGN (mAGN).}    \label{fig17}
\end{figure*}

\begin{figure*}
\centering
  \includegraphics[width= 17cm]{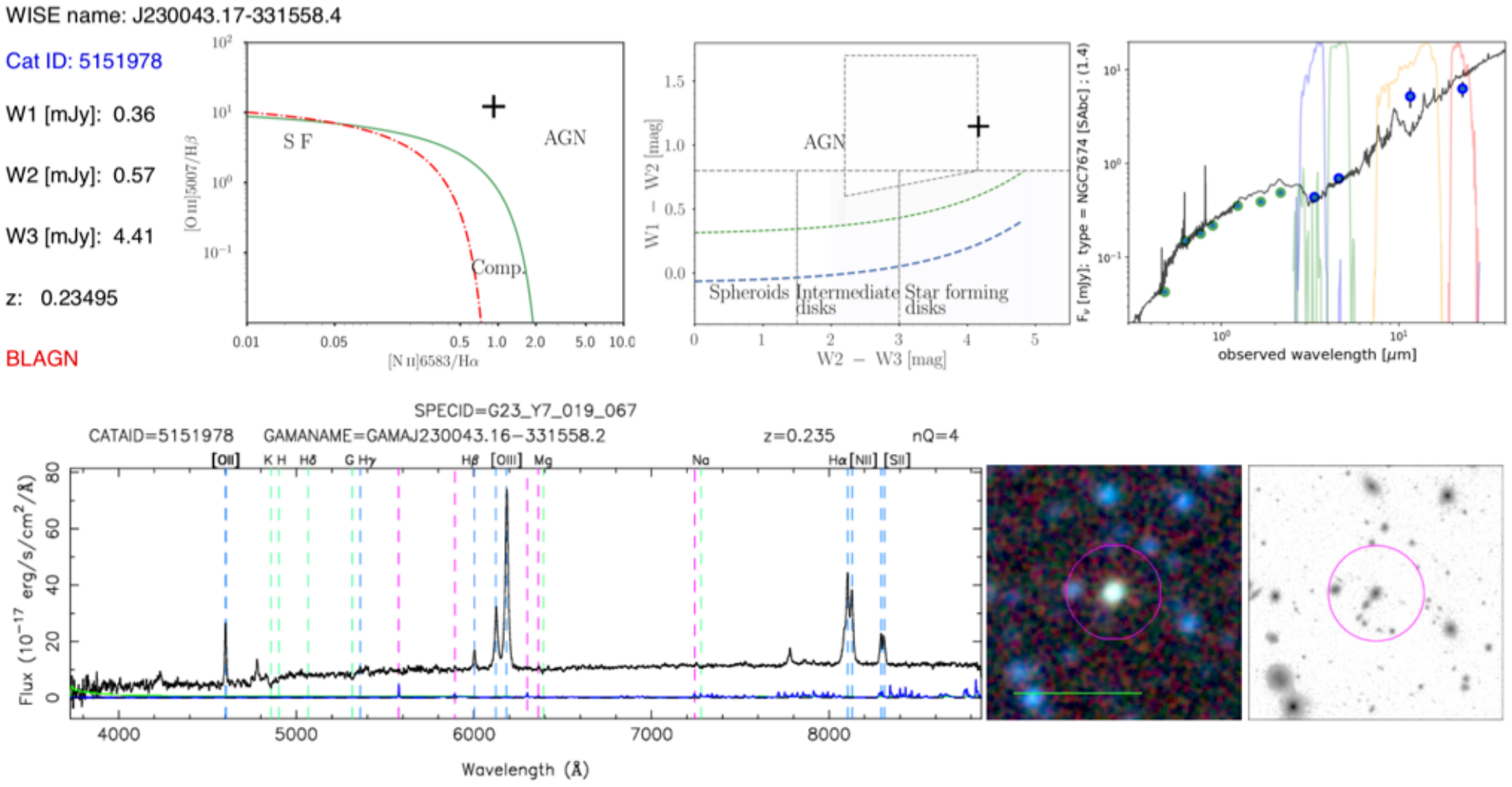}
    \caption{Galaxy 5151978. A broad-line AGN classified as  a oAGN (mAGN)}    \label{fig18}
\end{figure*}

\begin{figure*}
\centering
  \includegraphics[width= 17cm]{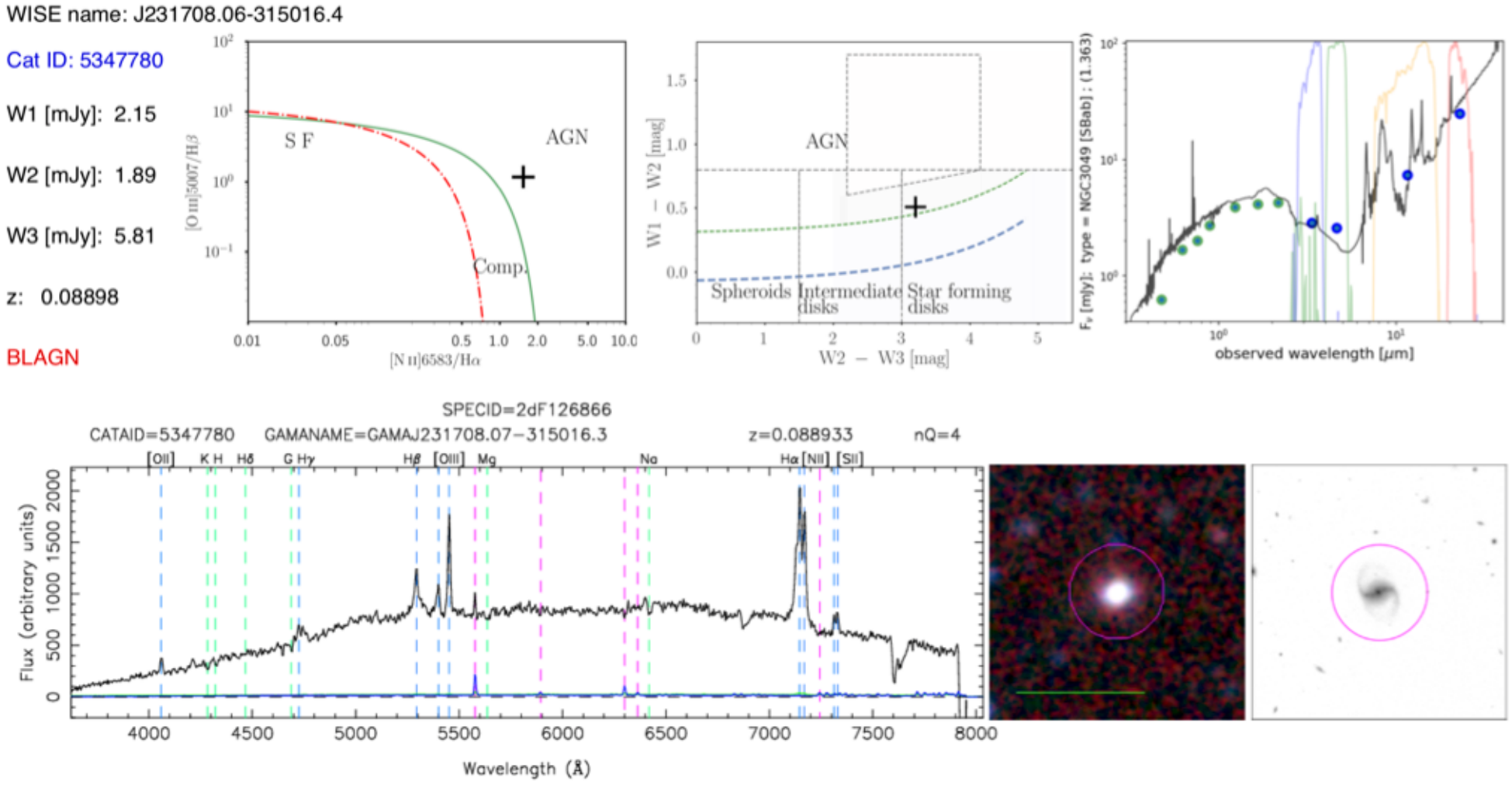}
    \caption{Galaxy 5347780. A broad-line AGN  classified as a  oAGN (mWarm).}    \label{fig19}
\end{figure*}

\begin{figure*}
\centering
  \includegraphics[width= 17cm]{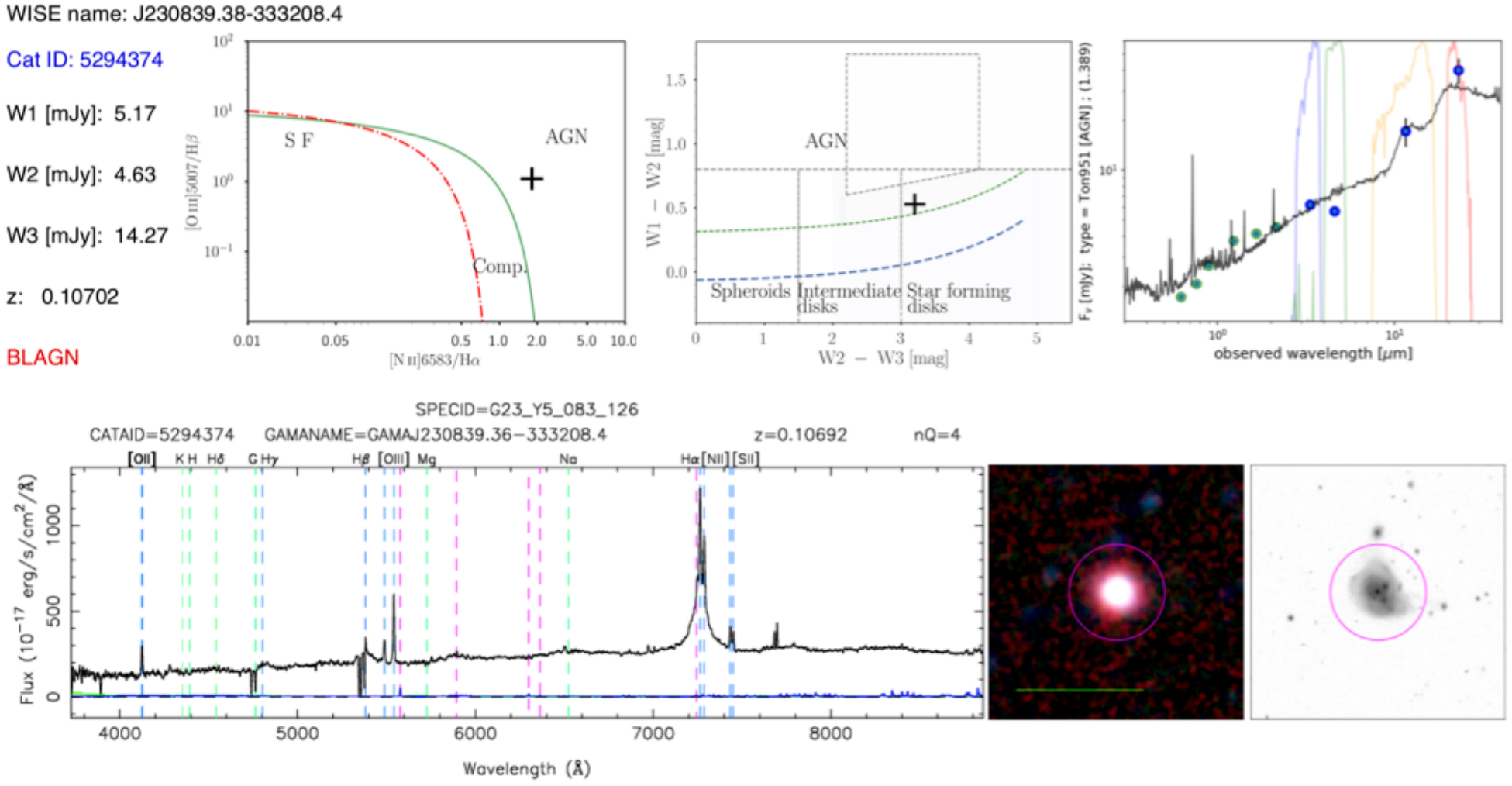}
    \caption{Galaxy 5294374. A broad-line AGN  classified as a  oAGN (mWarm).}    \label{fig20}
\end{figure*}

\begin{figure*}
\centering
  \includegraphics[width= 17cm]{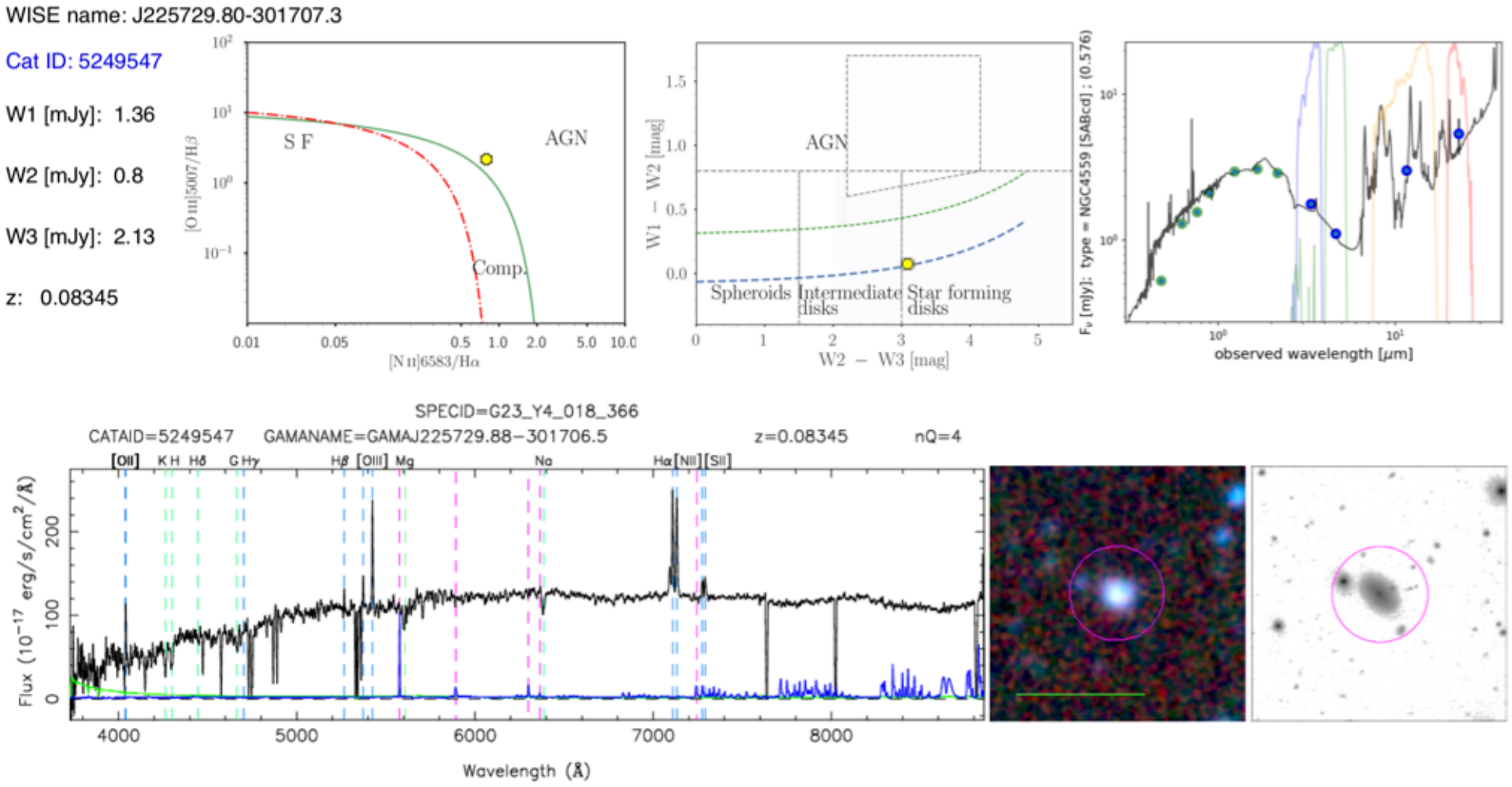}
    \caption{Galaxy 5249547. Classified as a oAGN (mSF). It is an all-important yellow case.}    \label{fig21}
\end{figure*}

\begin{figure*}
\centering
  \includegraphics[width= 17cm]{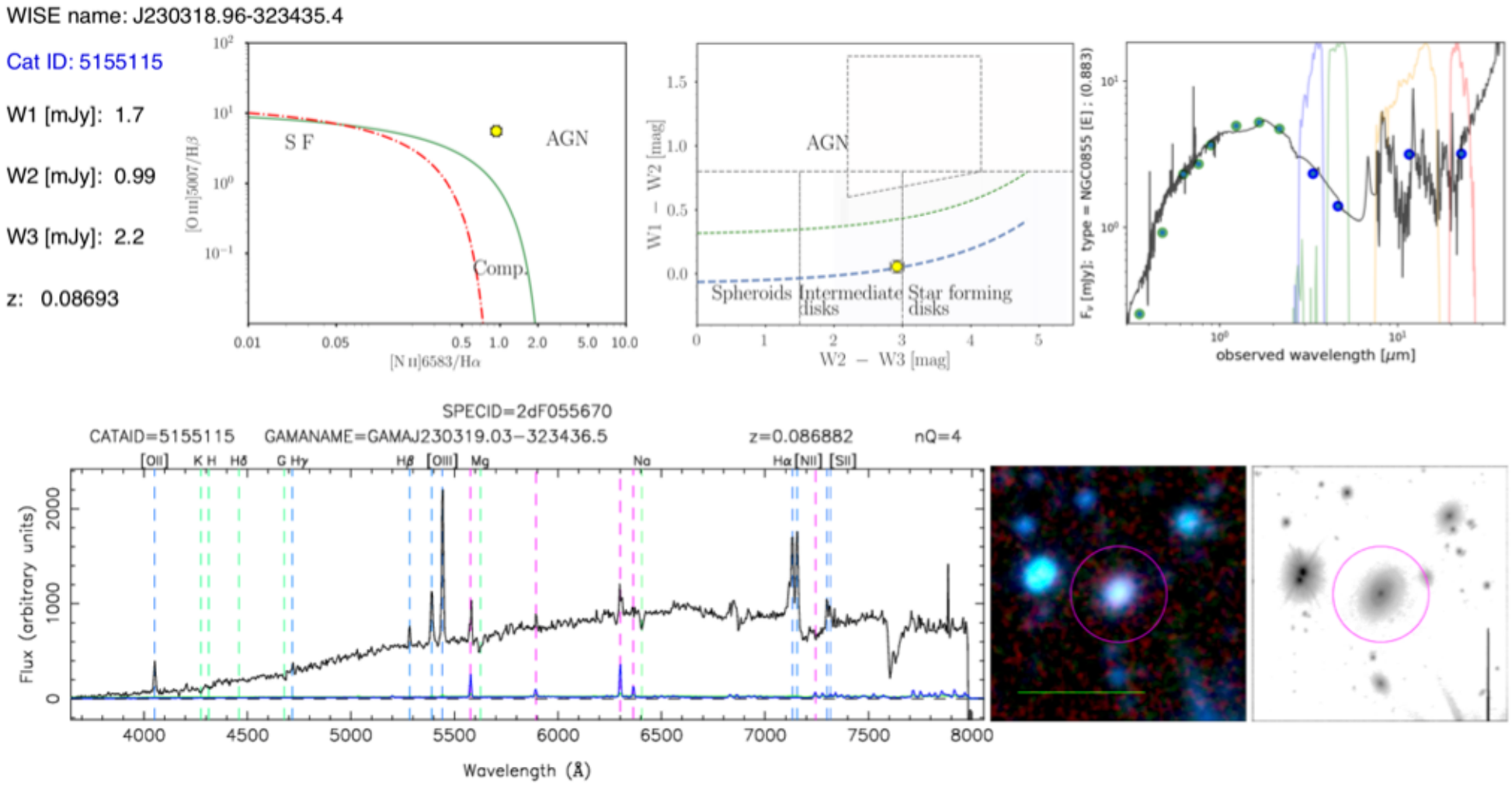}
    \caption{Galaxy 5155115. Classified as a oAGN (mSF) (yellow). }    \label{fig22}
\end{figure*}

\begin{figure*}
\centering
  \includegraphics[width= 17cm]{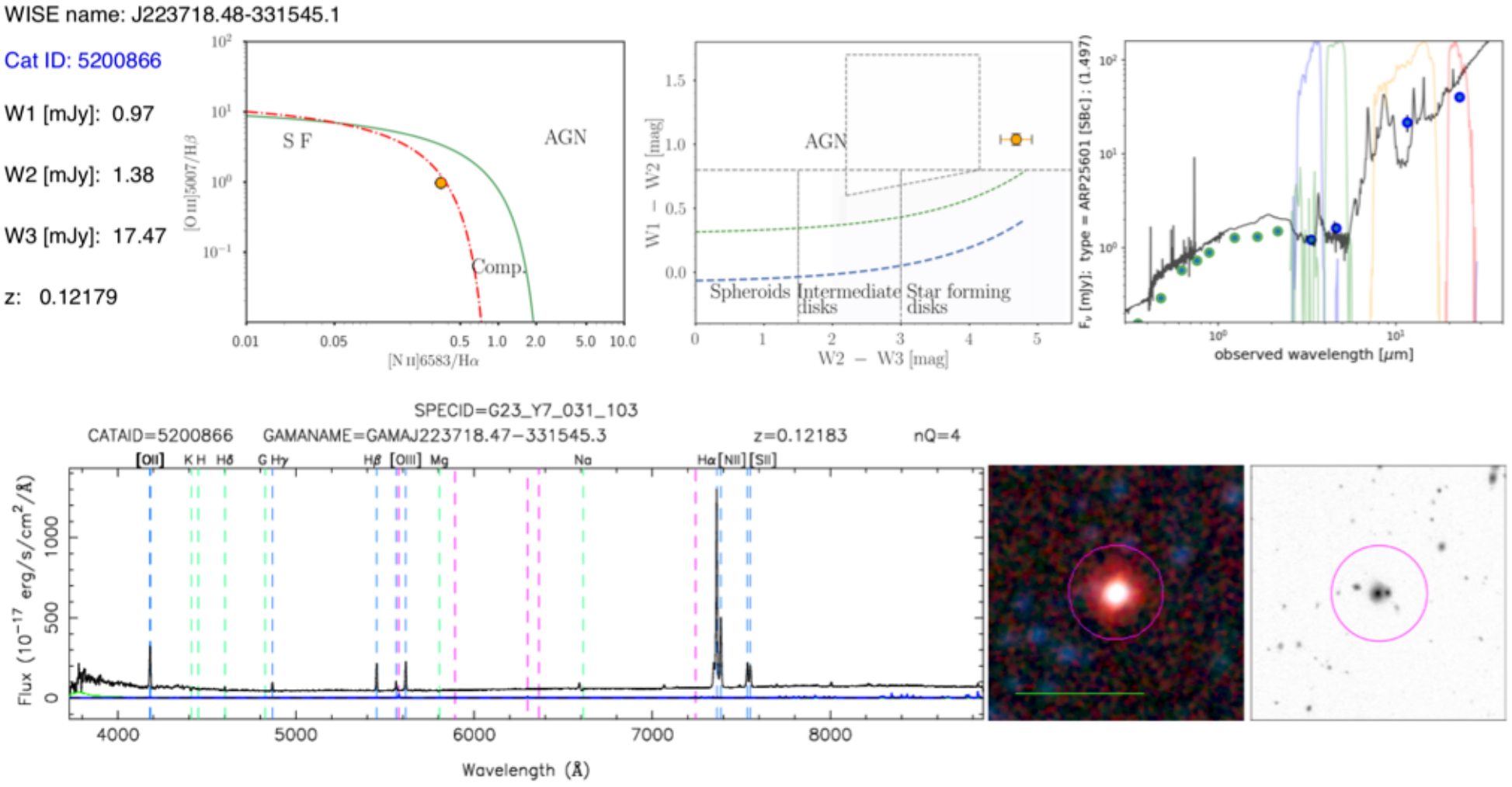}
    \caption{Galaxy 5200866. Classified as a non-oAGN (mAGN)  (orange).}    \label{fig23}
\end{figure*}

\begin{figure*}
\centering
  \includegraphics[width= 17cm]{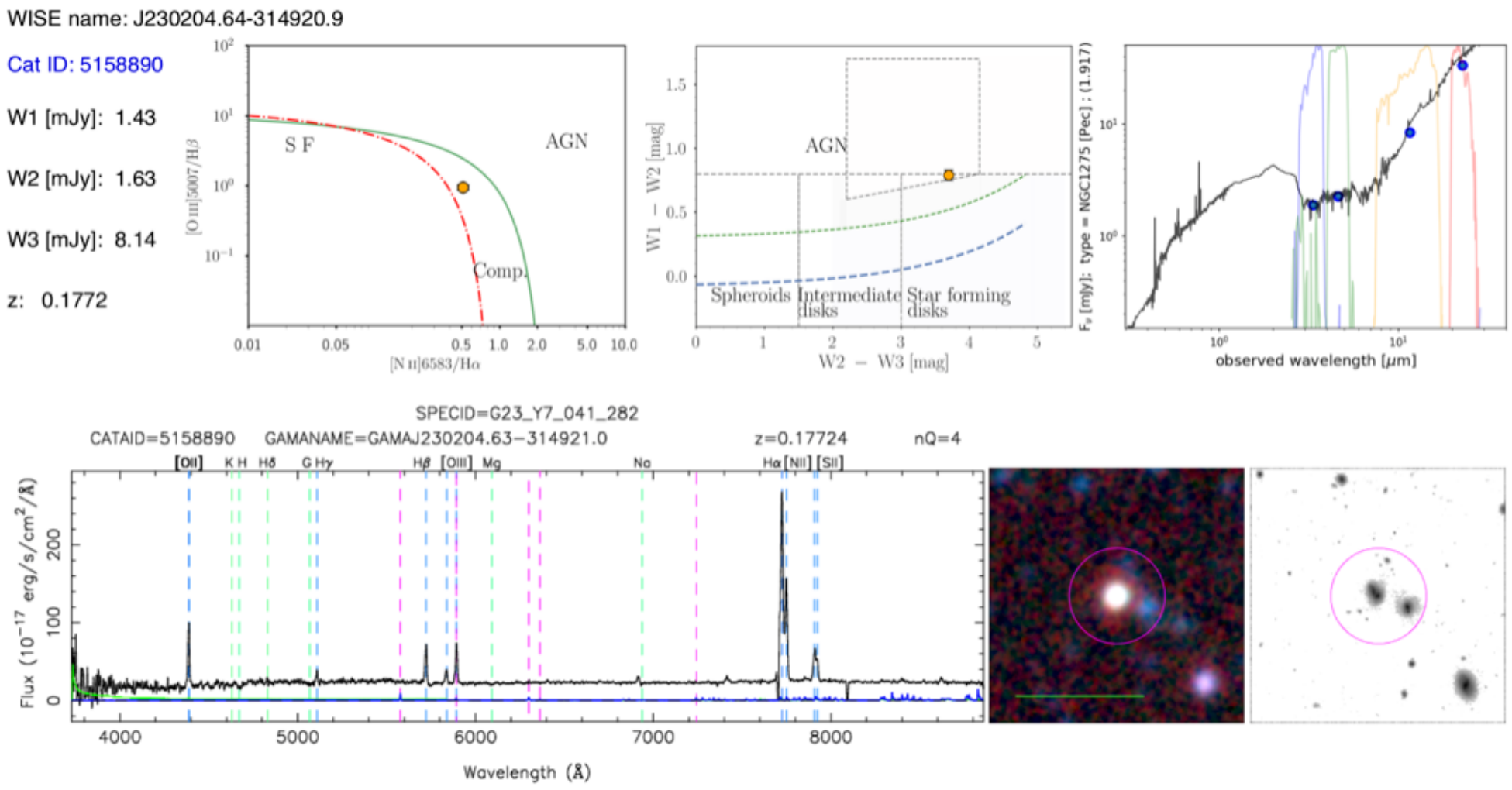}
    \caption{Galaxy 5158890. Classified as non-oAGN (mAGN)  (orange). }    \label{fig24}
\end{figure*}

\begin{figure*}
\centering
  \includegraphics[width= 17cm]{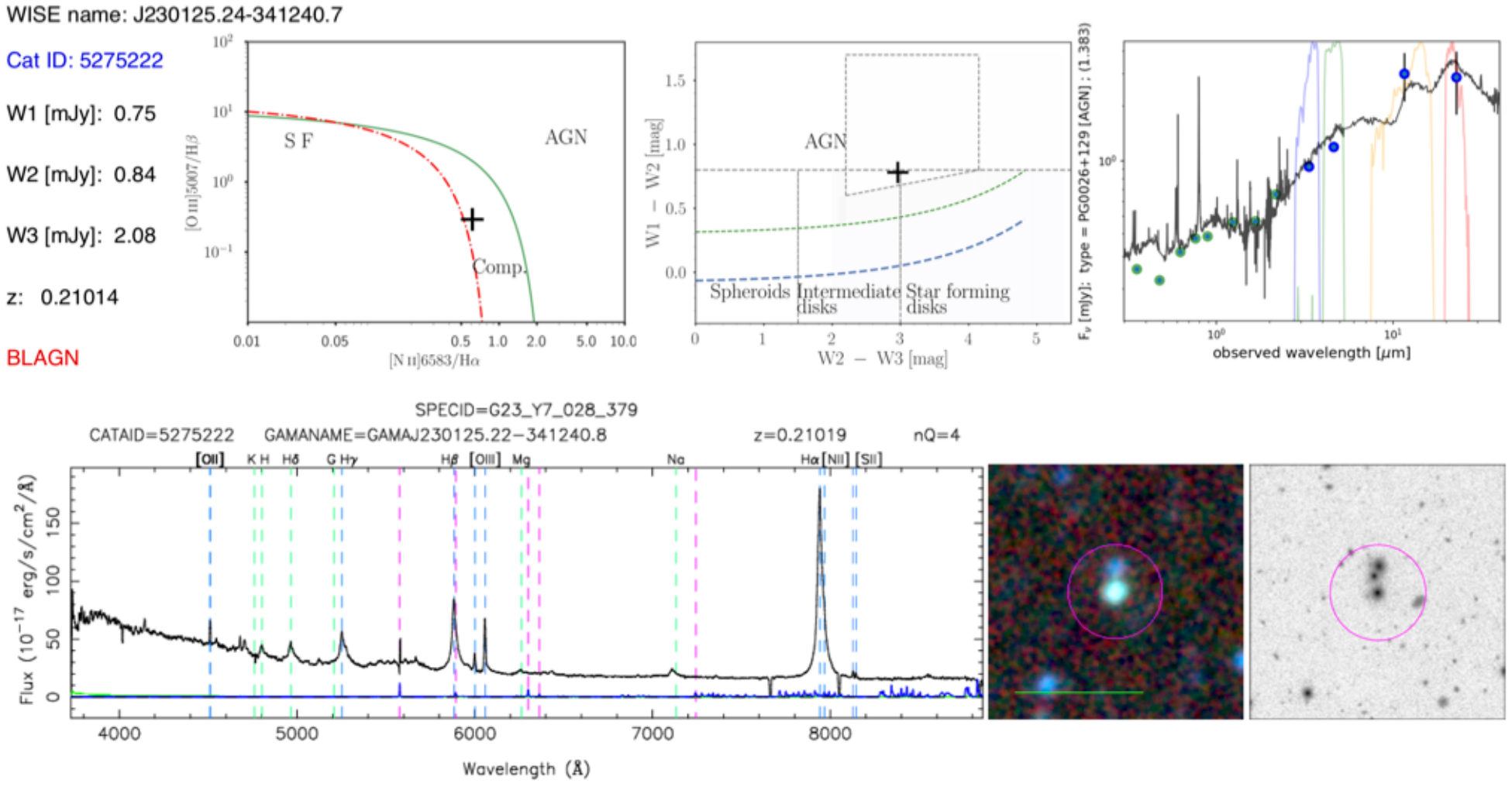}
    \caption{Galaxy 5275222. A broad-line AGN classified as a non-oAGN (mAGN). Note the [\ion{N}{ii}] line can't be distinguished from H$\alpha$.}    \label{fig25}
\end{figure*}

\begin{figure*}
\centering
  \includegraphics[width= 17cm]{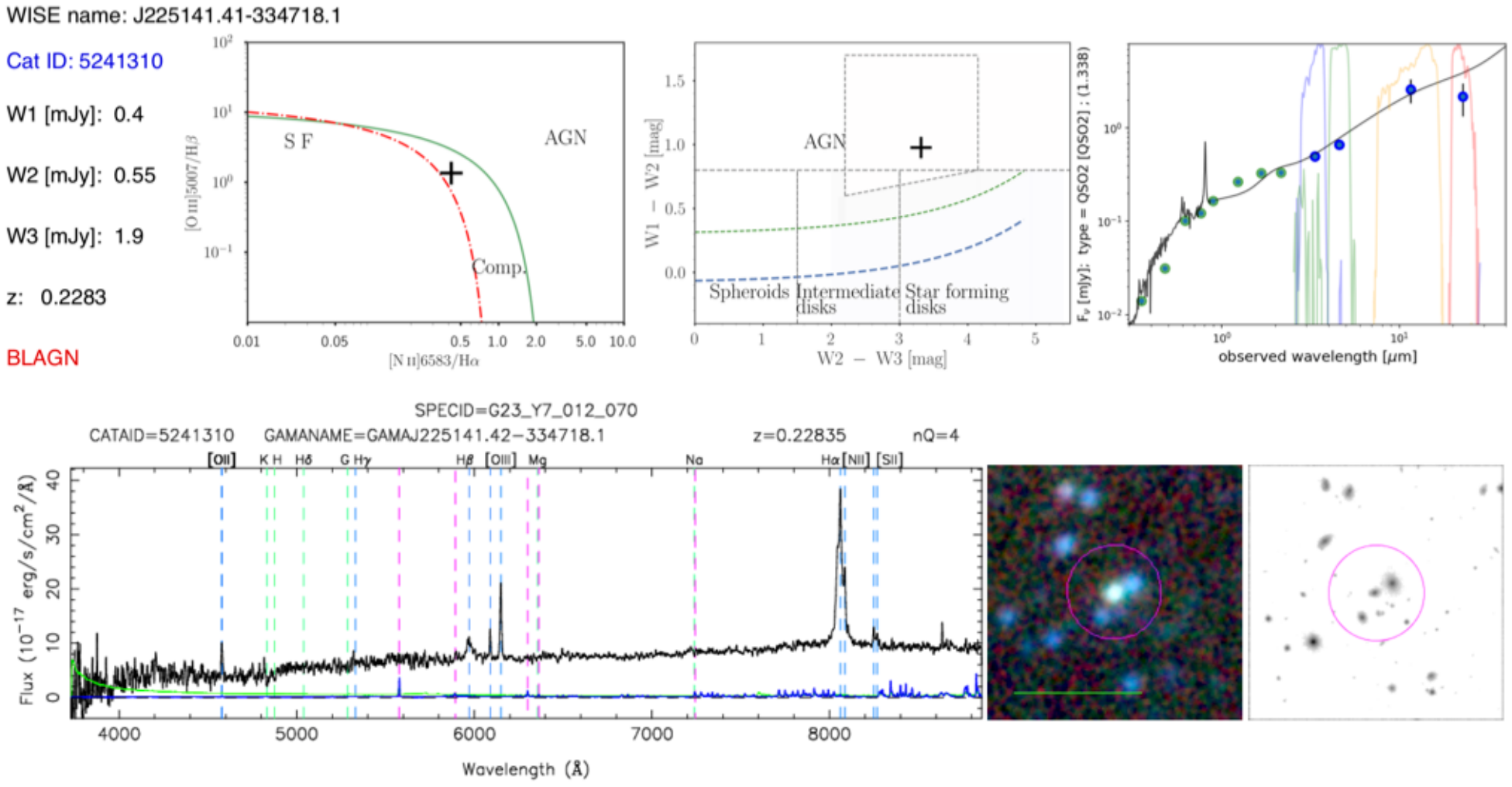}
    \caption{Galaxy 5241310. A broad-line AGN classified as a non-oAGN (mAGN).}    \label{fig26}
\end{figure*}

\begin{figure*}
\centering
  \includegraphics[width= 17cm]{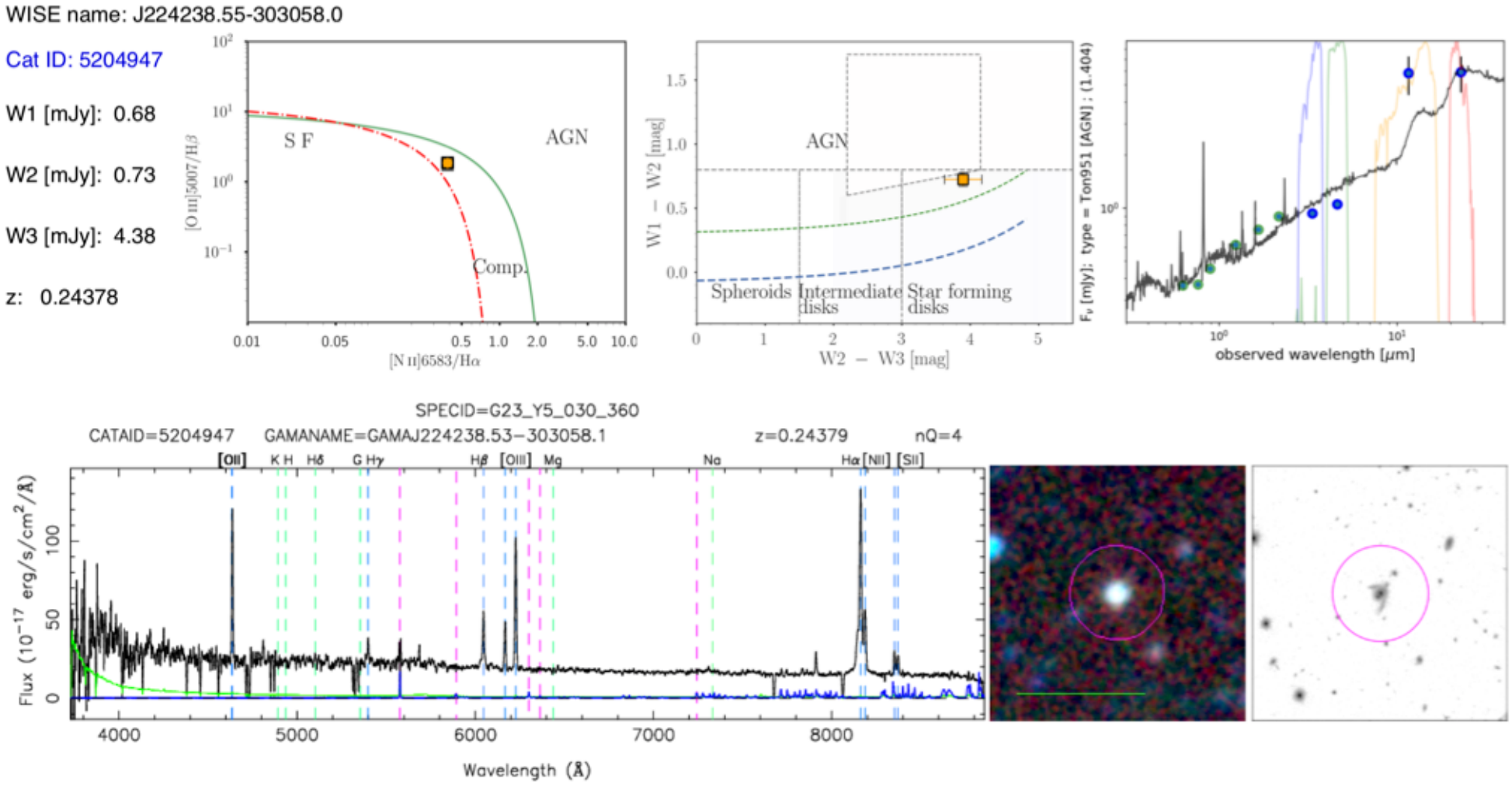}
    \caption{Galaxy 5204947. Classified as a non-oAGN (mWarm)  (orange).  Blended system seen as a single in $\textit{WISE}$ 3-band colour.}    \label{fig27}
\end{figure*}

\begin{figure*}
\centering
  \includegraphics[width= 17cm]{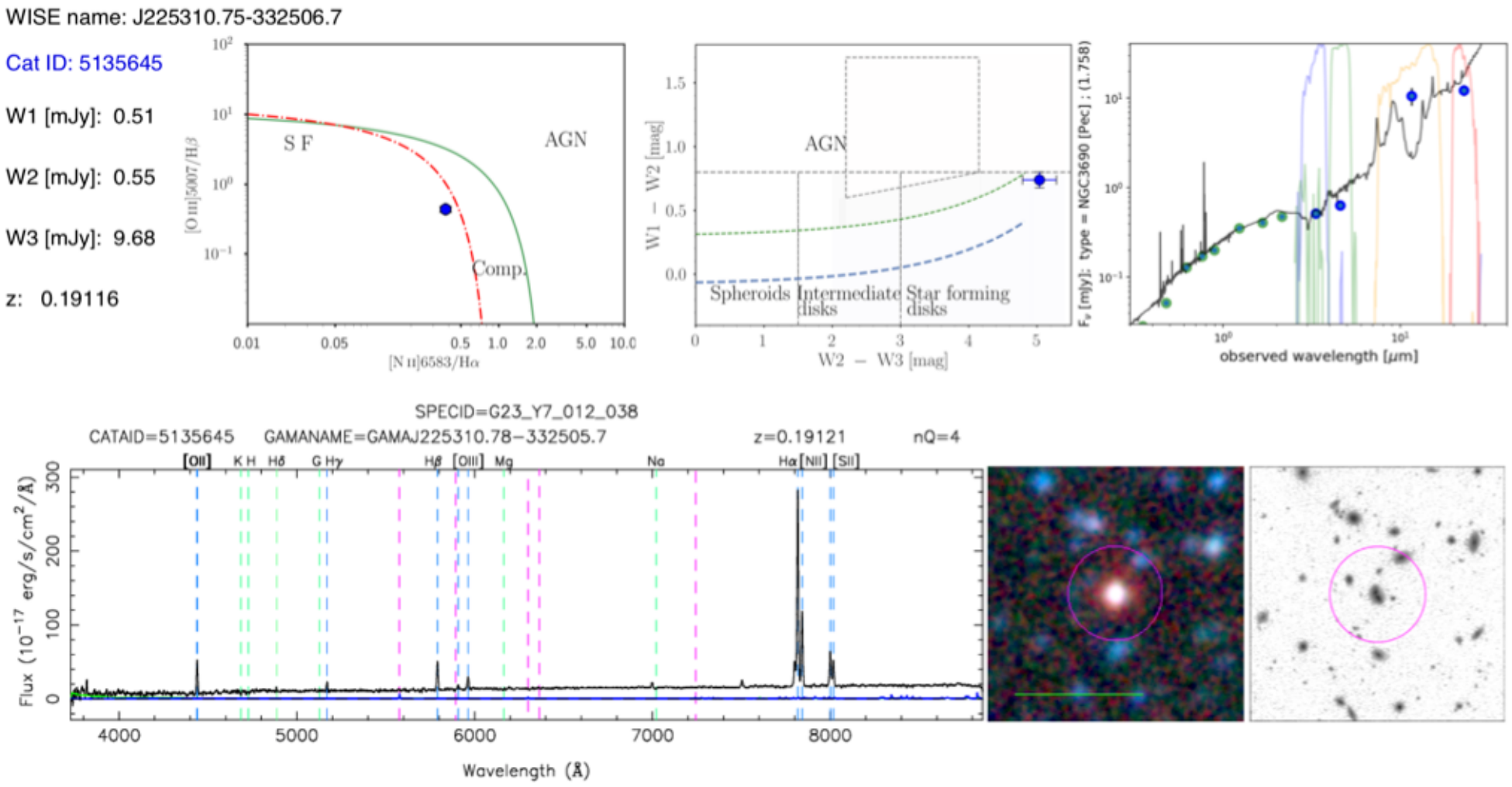}
    \caption{Galaxy 5135645.  Classified as a SF (blue). Extreme colours in $\textit{WISE}$. }    \label{fig28}
\end{figure*}

\begin{figure*}

\centering
\includegraphics[width= 4cm] {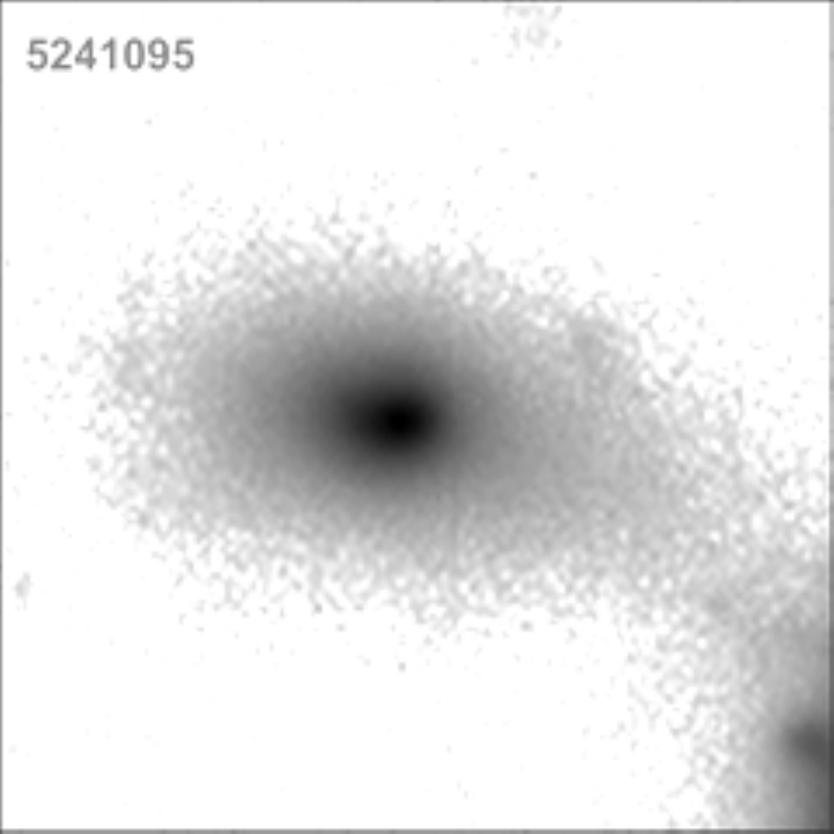} 
\includegraphics[width= 4cm] {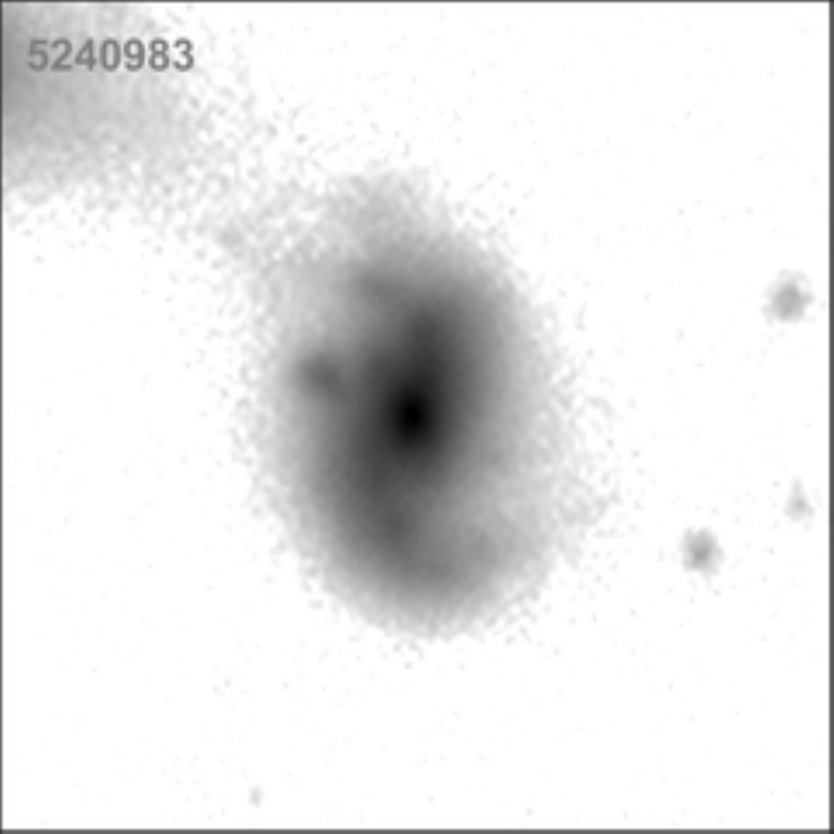} 
\includegraphics[width= 4cm] {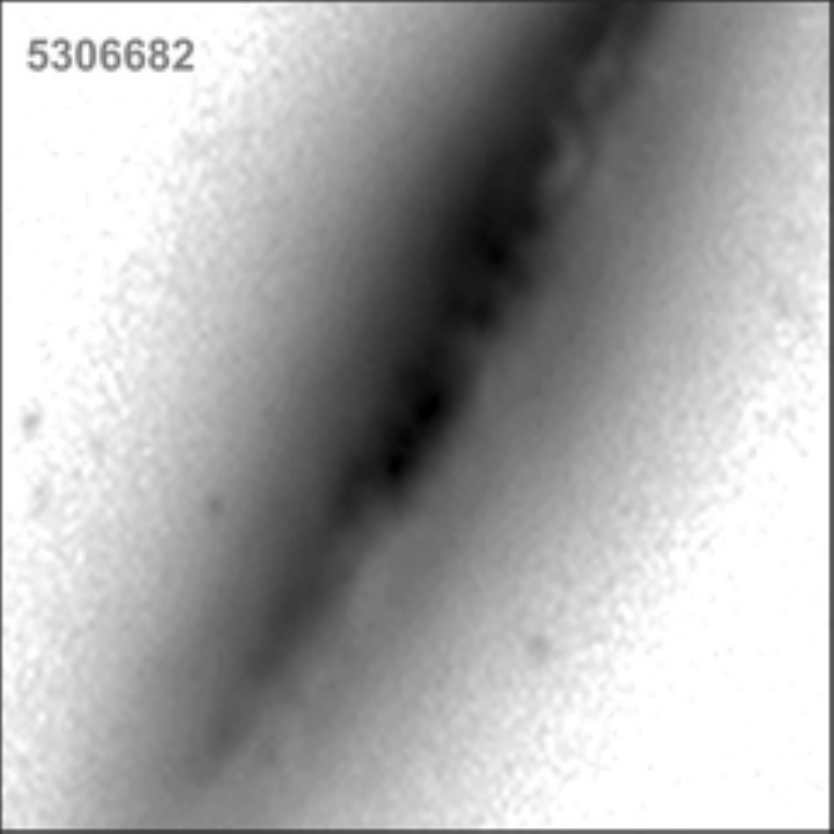} 
\includegraphics[width= 4cm] {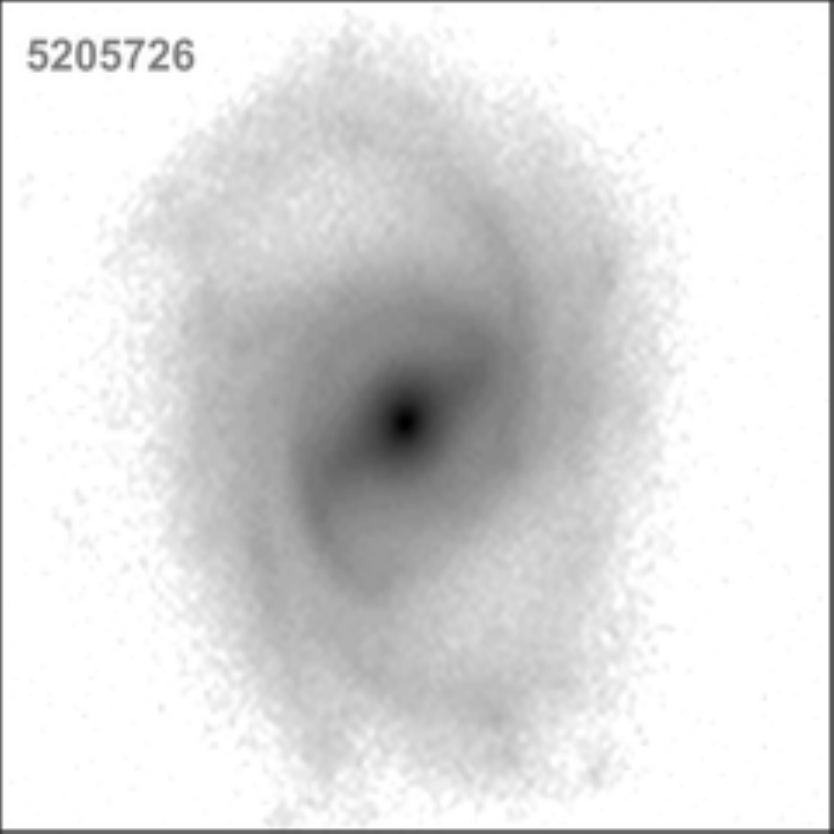}\\
\includegraphics[width= 4cm] {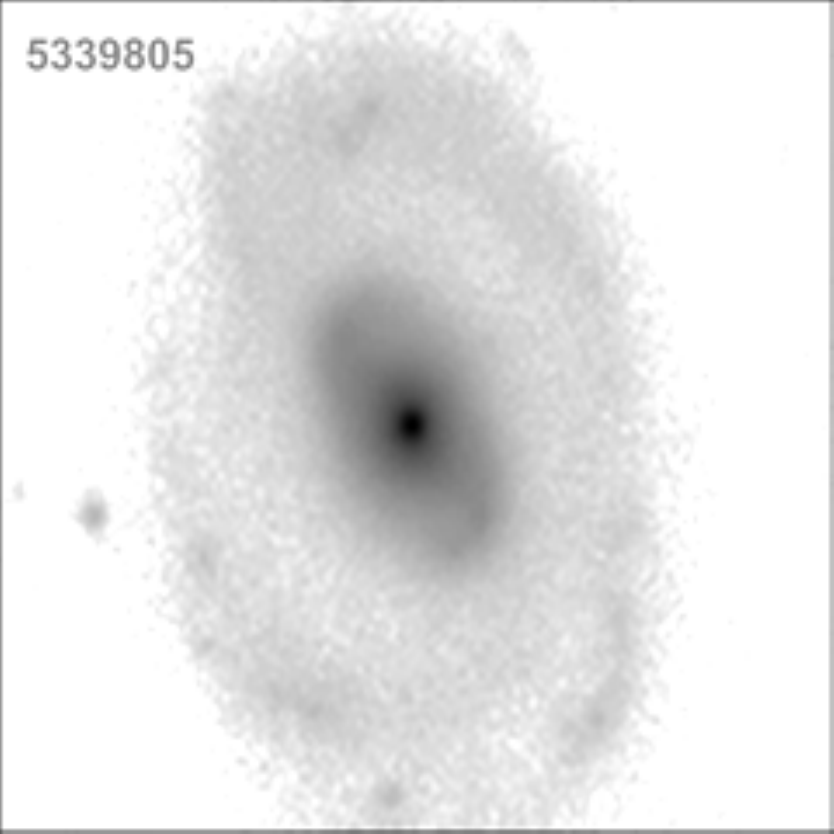} 
\includegraphics[width= 4cm] {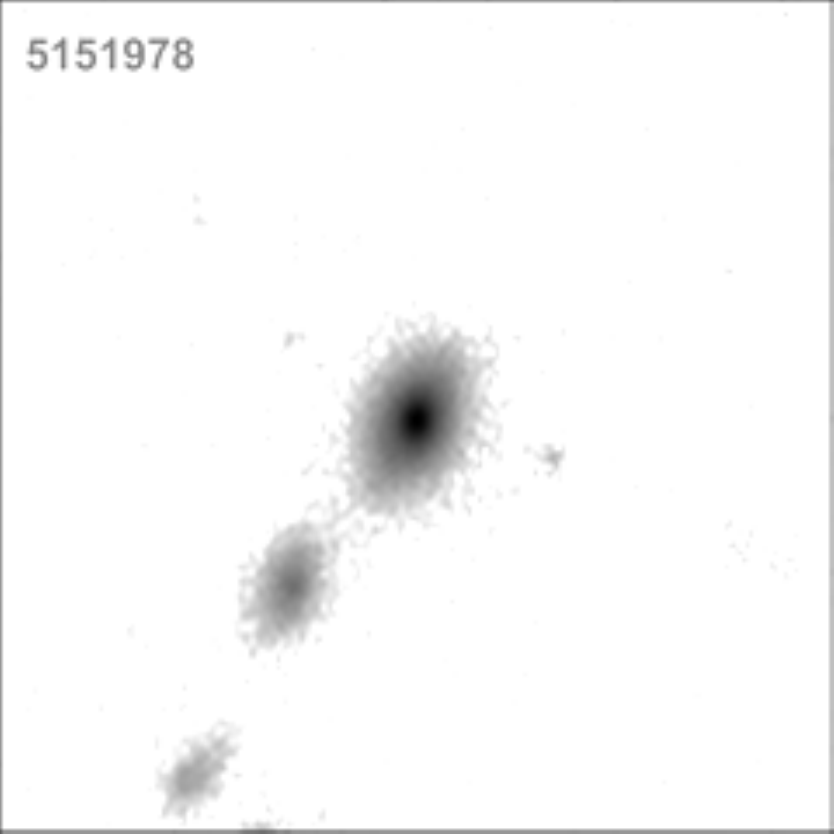}
\includegraphics[width= 4cm] {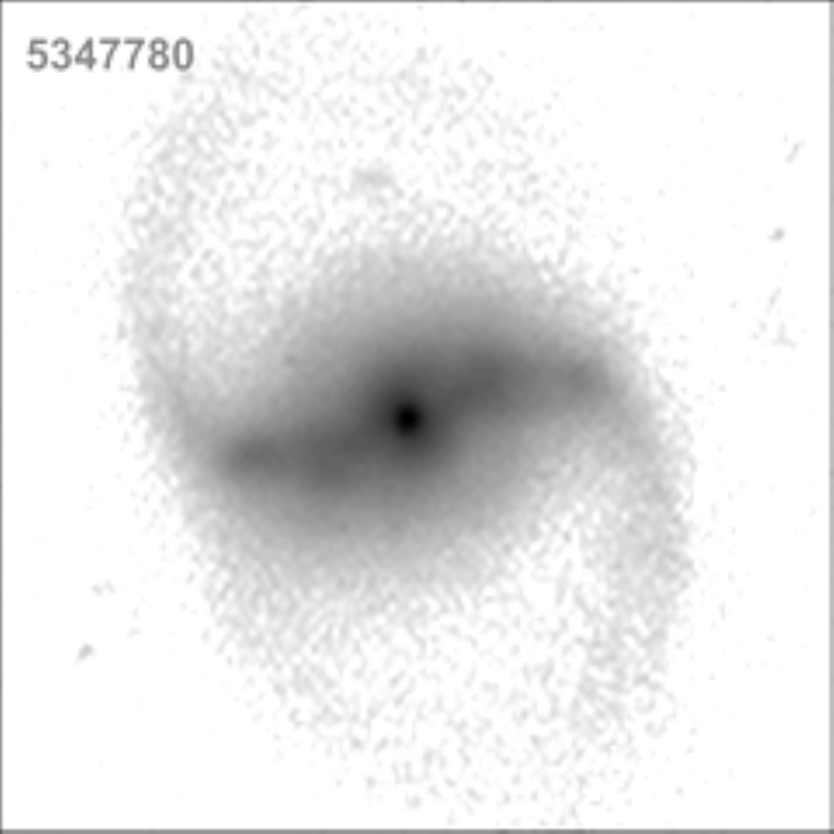}
\includegraphics[width= 4cm] {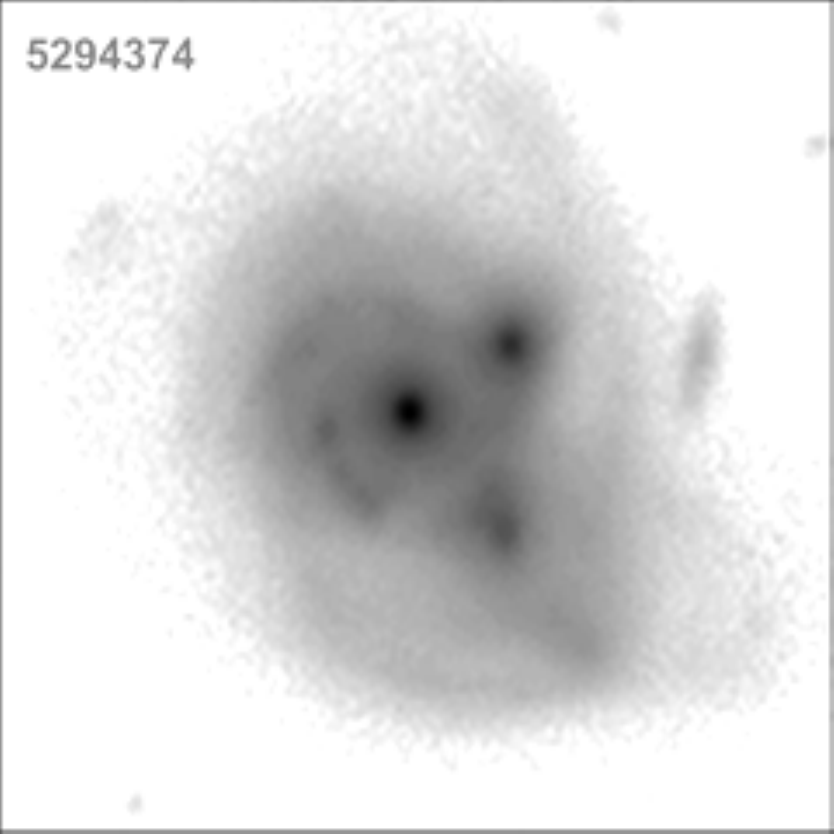}\\
\includegraphics[width= 4cm] {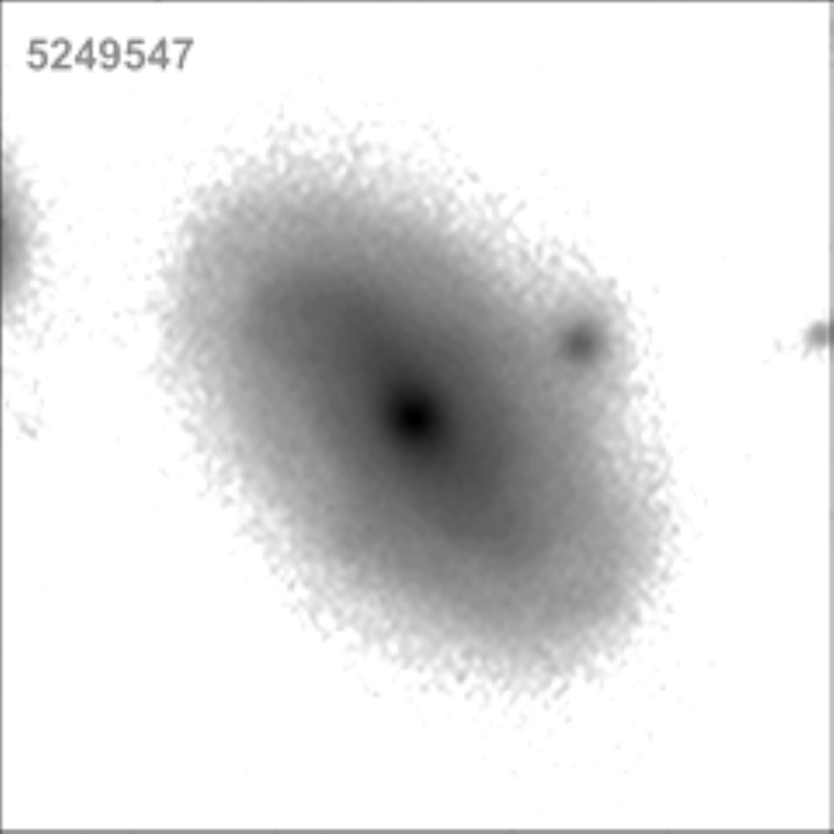} 
\includegraphics[width= 4cm] {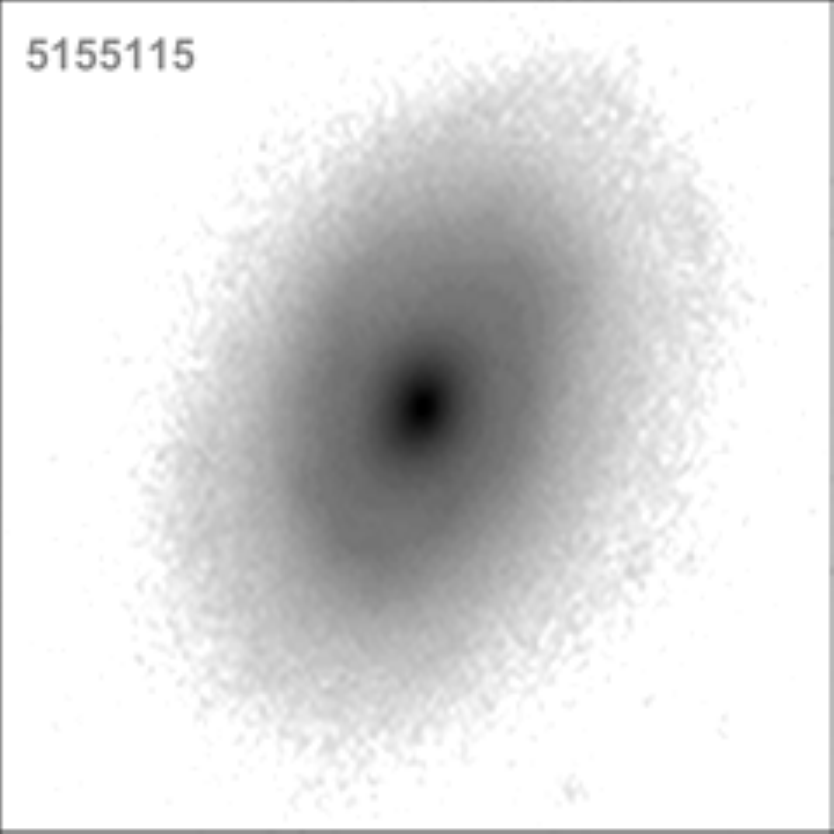} 
\includegraphics[width= 4cm] {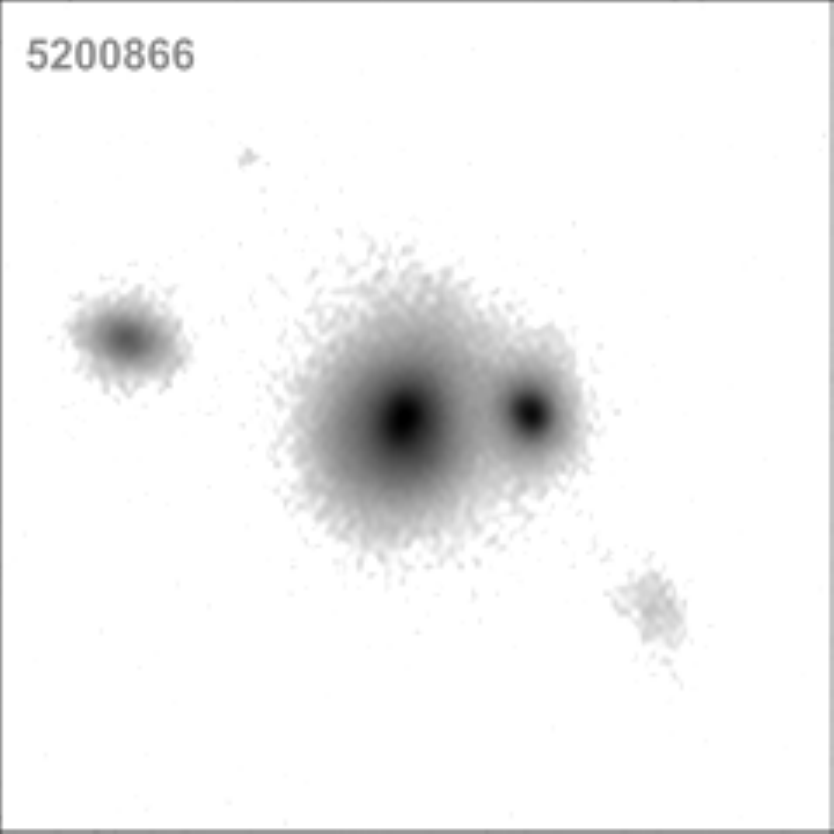}
\includegraphics[width= 4cm] {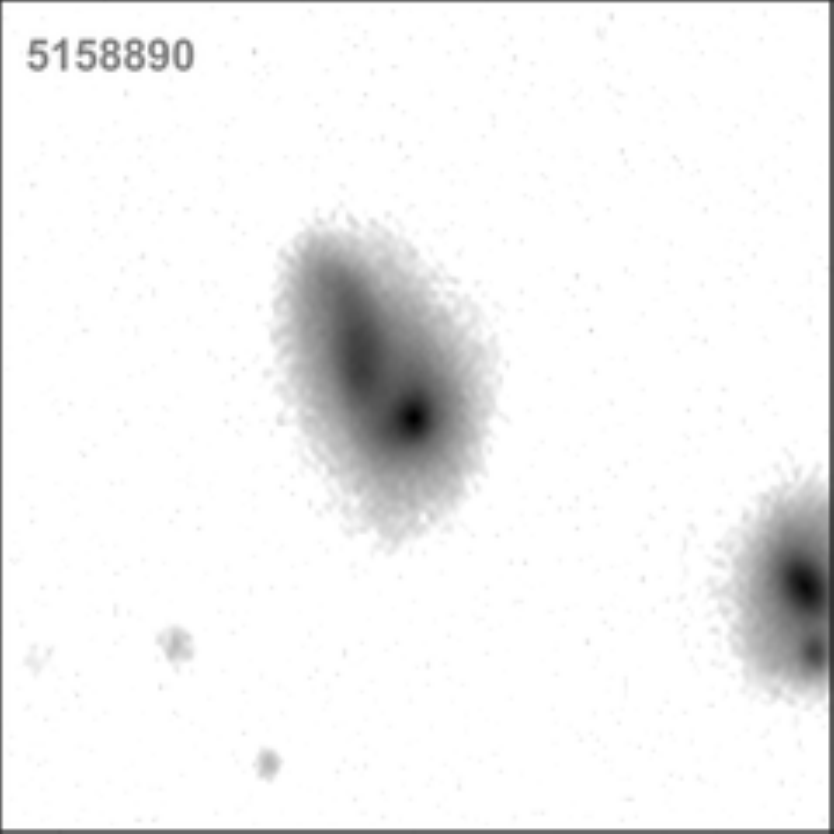}\\
\includegraphics[width= 4cm] {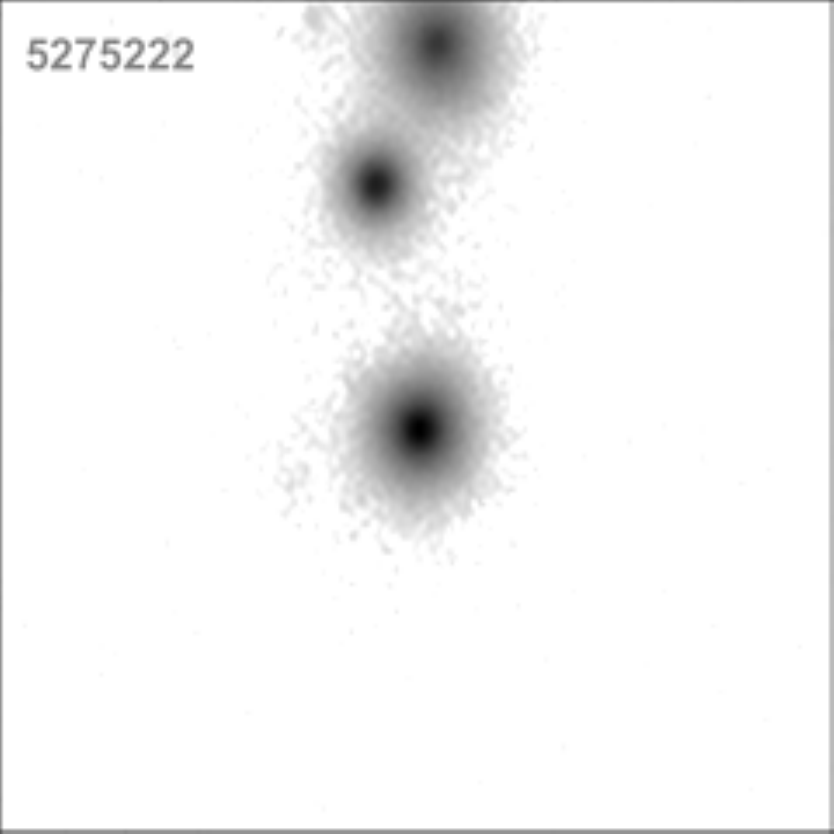}
\includegraphics[width= 4cm] {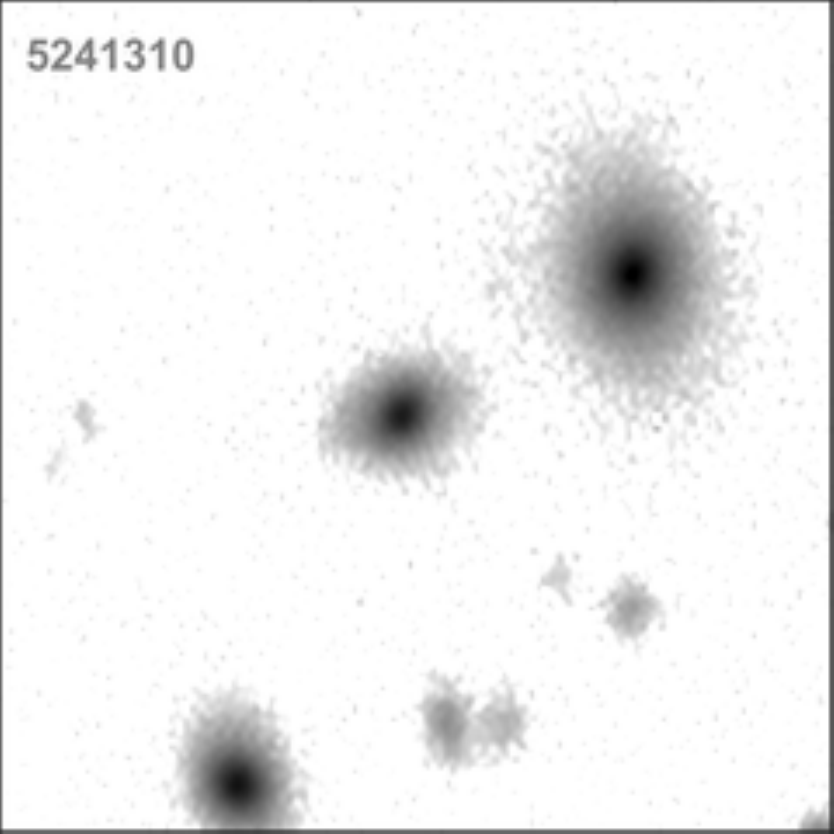}
\includegraphics[width= 4cm] {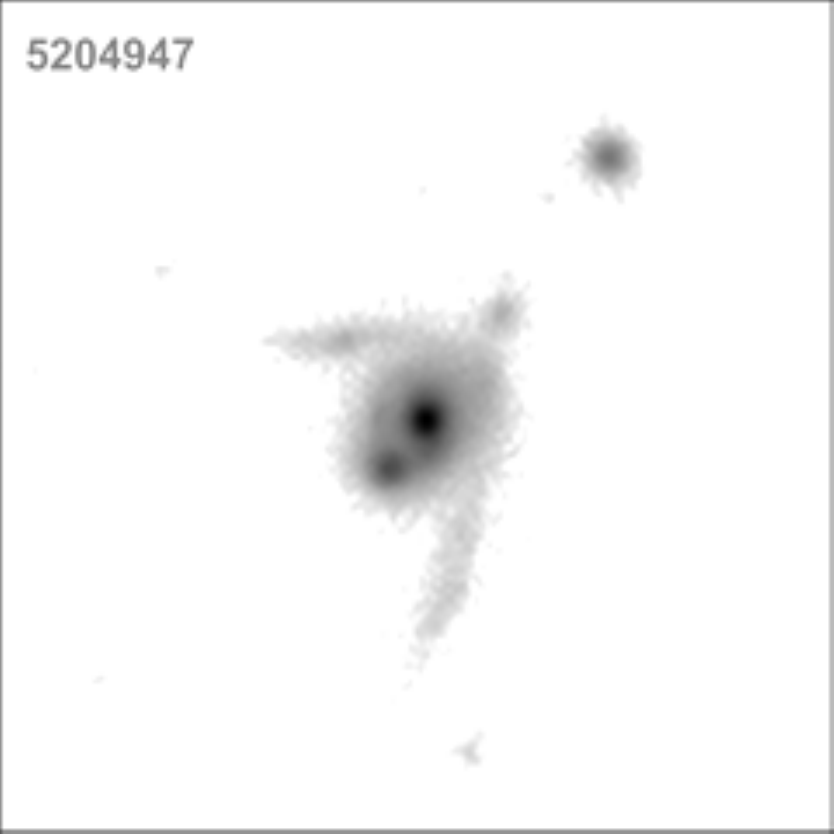}
\includegraphics[width= 4cm] {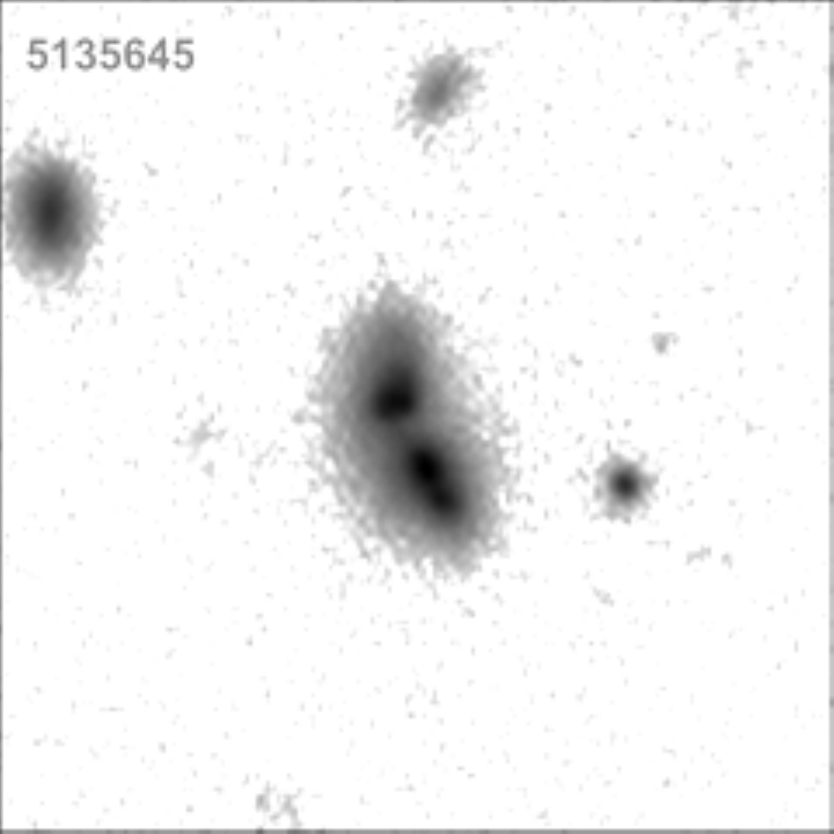}

\caption{Individual KiDS r-band images of the galaxies selected from Figure \ref{fig12} to \ref{fig28}. The GAMA names are on the upper left corner. All panels are 2 by 2 arcminutes.}  \label{fig29}
\end{figure*}

\begin{longrotatetable}
\begin{deluxetable}{l@{\hspace{-3cm}}cccccccc}

\tablecaption{Measured properties of a representative sample of the galaxies from each activity group.
  The fluxes are k-corrected and the rest frame fluxes in W1, W2, and W3 are given in mJy. W1-W2 and W2-W3 are rest-frame colours and the GAMA redshifts are also presented. The selected galaxies for case study in the next section are designated by a star next to their GAMA ID (ex 5241095$^{*}$: for the first selected galaxy in the SF  group etc.) and the broad-line AGN by a plus sign (ex 5340595$^{+}$: a broad-line AGN in the oAGN (mAGN) group etc.). The colours used for the different groups as presented in Figure \ref{fig8a} are added for clarity. \label{table1.3bb}}
\tablewidth{750pt}
\tabletypesize{\scriptsize}
\tablehead{
\colhead{CATAID } & \colhead{$\textit{z}$$_{\text{spec}}$ } &\colhead{W1} & \colhead{W2} &\colhead{ W3$_{\rm{PAH}}$} & \colhead{ W1-W2} &\colhead{ W2-W3} & \colhead{[\ion{O}{iii}]/H$\beta$} & 
\colhead{ [\ion{N}{ii}]/H$\alpha$} \\ 
\colhead{} & \colhead{} & \colhead{(mJy)} & \colhead{ (mJy)} & \colhead{mJy}&\colhead{(mag) } & \colhead{(mag)} & \colhead{} &\colhead{}  \\
} 
\startdata
SF~~(blue circles)&&&&&&&&\\ \hline                                                                                                                              
5241095$^{*}$ & 0.0274 & 0.42 $\pm$ 0.01 & 0.26 $\pm$ 0.02 & 0.7 $\pm$ 0.2 & 0.14 $\pm$ 0.08 & 3.08 $\pm$ 0.33 & 3.6 $\pm$ 0.09 & 0.07 $\pm$ 0.01 \\                 
5240983$^{*}$ & 0.0274 & 0.72 $\pm$ 0.02 & 0.49 $\pm$ 0.03 & 1.97 $\pm$ 0.59 & 0.24 $\pm$ 0.07 & 3.49 $\pm$ 0.33 & 2.37 $\pm$ 0.08 & 0.1 $\pm$ 0.0 \\    
5306682$^{*}$ & 0.0053 & 14.13 $\pm$ 0.16 & 7.99 $\pm$ 0.15 & 21.08 $\pm$ 3.24 & 0.03 $\pm$ 0.04 & 3.08 $\pm$ 0.17 & 0.78 $\pm$ 0.03 & 0.28 $\pm$ 0.01 \\            
5135645$^{*}$ & 0.1912 & 0.51 $\pm$ 0.01 & 0.55 $\pm$ 0.02 & 9.68 $\pm$ 2.16 & 0.74 $\pm$ 0.06 & 5.04 $\pm$ 0.25 & 0.44 $\pm$ 0.04 & 0.38 $\pm$ 0.01 \\              
5325430 ~ & 0.0648 & 0.59 $\pm$ 0.01 & 0.34 $\pm$ 0.02 & 1.93 $\pm$ 0.15 & 0.06 $\pm$ 0.07 & 3.86 $\pm$ 0.11 & 0.29 $\pm$ 0.06 & 0.33 $\pm$ 0.01 \\              
5327563 ~ & 0.0652 & 0.4 $\pm$ 0.01 & 0.27 $\pm$ 0.02 & 1.11 $\pm$ 0.43 & 0.22 $\pm$ 0.09 & 3.54 $\pm$ 0.43 & 1.01 $\pm$ 0.14 & 0.25 $\pm$ 0.03 \\               
5325339 ~ & 0.1247 & 0.66 $\pm$ 0.01 & 0.44 $\pm$ 0.02 & 1.71 $\pm$ 0.1 & 0.2 $\pm$ 0.07 & 3.46 $\pm$ 0.09 & 0.37 $\pm$ 0.09 & 0.38 $\pm$ 0.03 \\                
5123445 ~ & 0.0795 & 0.4 $\pm$ 0.01 & 0.29 $\pm$ 0.02 & 1.19 $\pm$ 0.18 & 0.28 $\pm$ 0.09 & 3.53 $\pm$ 0.19 & 0.41 $\pm$ 0.12 & 0.38 $\pm$ 0.03 \\               
5138489 ~ & 0.0852 & 0.24 $\pm$ 0.01 & 0.15 $\pm$ 0.02 & 0.7 $\pm$ 0.17 & 0.15 $\pm$ 0.13 & 3.63 $\pm$ 0.29 & 0.42 $\pm$ 0.15 & 0.35 $\pm$ 0.03 \\               
5153375 ~ & 0.0807 & 0.39 $\pm$ 0.01 & 0.28 $\pm$ 0.01 & 1.36 $\pm$ 0.16 & 0.28 $\pm$ 0.06 & 3.71 $\pm$ 0.14 & 0.67 $\pm$ 0.15 & 0.3 $\pm$ 0.02 \\               
5261290 ~ & 0.0834 & 1.16 $\pm$ 0.02 & 0.79 $\pm$ 0.03 & 5.57 $\pm$ 1.25 & 0.23 $\pm$ 0.05 & 4.08 $\pm$ 0.25 & 0.16 $\pm$ 0.04 & 0.4 $\pm$ 0.02 \\               
5187670 ~ & 0.1067 & 0.27 $\pm$ 0.01 & 0.16 $\pm$ 0.02 & 0.78 $\pm$ 0.12 & 0.08 $\pm$ 0.15 & 3.72 $\pm$ 0.23 & 0.18 $\pm$ 0.06 & 0.31 $\pm$ 0.04 \\              
5281879 ~ & 0.0286 & 2.89 $\pm$ 0.03 & 1.75 $\pm$ 0.04 & 6.64 $\pm$ 0.16 & 0.1 $\pm$ 0.04 & 3.44 $\pm$ 0.05 & 0.26 $\pm$ 0.03 & 0.41 $\pm$ 0.01 \\               
5281521 ~ & 0.0288 & 0.35 $\pm$ 0.01 & 0.18 $\pm$ 0.02 & 0.45 $\pm$ 0.17 & -0.06 $\pm$ 0.11 & 3.03 $\pm$ 0.43 & 1.6 $\pm$ 0.14 & 0.14 $\pm$ 0.01 \\              
5305483 ~ & 0.086 & 0.46 $\pm$ 0.01 & 0.33 $\pm$ 0.02 & 1.68 $\pm$ 0.43 & 0.28 $\pm$ 0.07 & 3.74 $\pm$ 0.29 & 0.36 $\pm$ 0.03 & 0.31 $\pm$ 0.01 \\               
5222747 ~ & 0.2084 & 0.22 $\pm$ 0.01 & 0.15 $\pm$ 0.01 & 0.56 $\pm$ 0.09 & 0.24 $\pm$ 0.07 & 3.4 $\pm$ 0.18 & 0.55 $\pm$ 0.15 & 0.37 $\pm$ 0.04 \\               
5266202 ~ & 0.0844 & 0.74 $\pm$ 0.02 & 0.49 $\pm$ 0.03 & 3.09 $\pm$ 0.15 & 0.19 $\pm$ 0.07 & 3.97 $\pm$ 0.09 & 0.18 $\pm$ 0.06 & 0.36 $\pm$ 0.01 \\              
5185570 ~ & 0.0594 & 0.53 $\pm$ 0.02 & 0.31 $\pm$ 0.02 & 0.3 $\pm$ 0.12 & 0.08 $\pm$ 0.09 & 2.15 $\pm$ 0.45 & 0.31 $\pm$ 0.04 & 0.31 $\pm$ 0.02 \\               
5256068 ~ & 0.223 & 0.67 $\pm$ 0.01 & 0.6 $\pm$ 0.01 & 6.5 $\pm$ 0.11 & 0.54 $\pm$ 0.04 & 4.52 $\pm$ 0.04 & 0.82 $\pm$ 0.1 & 0.36 $\pm$ 0.01 \\ \hline                   
Composites~~(green circles)&&&&&&&&\\ \hline                                                                                                                     
5205726$^{*}$& 0.0592 & 1.68 $\pm$ 0.02 & 0.99 $\pm$ 0.02 & 1.94 $\pm$ 0.14 & 0.08 $\pm$ 0.04 & 2.78 $\pm$ 0.09 & 0.38 $\pm$ 0.18 & 0.6 $\pm$ 0.02 \\               
5322507 ~ & 0.0608 & 1.38 $\pm$ 0.02 & 0.81 $\pm$ 0.03 & 1.05 $\pm$ 0.16 & 0.07 $\pm$ 0.05 & 2.41 $\pm$ 0.17 & 0.55 $\pm$ 0.11 & 0.76 $\pm$ 0.05 \\              
5246049 ~ & 0.1064 & 0.87 $\pm$ 0.01 & 0.64 $\pm$ 0.02 & 5.13 $\pm$ 1.16 & 0.32 $\pm$ 0.05 & 4.2 $\pm$ 0.25 & 0.29 $\pm$ 0.05 & 0.53 $\pm$ 0.01 \\               
5277048 ~ & 0.1088 & 0.21 $\pm$ 0.01 & 0.16 $\pm$ 0.02 & 1.31 $\pm$ 0.18 & 0.34 $\pm$ 0.12 & 4.25 $\pm$ 0.19 & 1.62 $\pm$ 0.28 & 0.36 $\pm$ 0.09 \\              
5251443 ~ & 0.2361 & 0.21 $\pm$ 0.01 & 0.11 $\pm$ 0.01 & 0.37 $\pm$ 0.13 & -0.08 $\pm$ 0.09 & 3.37 $\pm$ 0.39 & 0.41 $\pm$ 0.13 & 0.62 $\pm$ 0.04 \\             
5279625 ~ & 0.0842 & 0.63 $\pm$ 0.01 & 0.42 $\pm$ 0.02 & 2.61 $\pm$ 0.61 & 0.19 $\pm$ 0.06 & 3.96 $\pm$ 0.26 & 0.57 $\pm$ 0.1 & 0.46 $\pm$ 0.01 \\               
5285888 ~ & 0.2012 & 0.46 $\pm$ 0.01 & 0.32 $\pm$ 0.02 & 3.11 $\pm$ 0.77 & 0.25 $\pm$ 0.07 & 4.41 $\pm$ 0.28 & 1.36 $\pm$ 0.27 & 0.52 $\pm$ 0.06 \\              
5159722 ~ & 0.0656 & 1.84 $\pm$ 0.02 & 1.2 $\pm$ 0.04 & 7.04 $\pm$ 0.21 & 0.18 $\pm$ 0.05 & 3.88 $\pm$ 0.06 & 0.62 $\pm$ 0.11 & 0.46 $\pm$ 0.02 \\               
5154601 ~ & 0.0789 & 0.33 $\pm$ 0.01 & 0.18 $\pm$ 0.01 & 0.67 $\pm$ 0.13 & -0.04 $\pm$ 0.08 & 3.46 $\pm$ 0.22 & 0.8 $\pm$ 0.15 & 0.42 $\pm$ 0.03 \\    
5264991 ~ & 0.1897 & 0.3 $\pm$ 0.01 & 0.25 $\pm$ 0.01 & 2.09 $\pm$ 0.14 & 0.44 $\pm$ 0.07 & 4.24 $\pm$ 0.1 & 1.16 $\pm$ 0.34 & 0.55 $\pm$ 0.08 \\                
5364072 ~ & 0.0955 & 0.2 $\pm$ 0.01 & 0.1 $\pm$ 0.01 & 0.55 $\pm$ 0.12 & -0.13 $\pm$ 0.14 & 3.84 $\pm$ 0.28 & 1.09 $\pm$ 0.3 & 0.38 $\pm$ 0.04 \\                
5249313 ~ & 0.0799 & 0.9 $\pm$ 0.01 & 0.54 $\pm$ 0.02 & 0.86 $\pm$ 0.17 & 0.08 $\pm$ 0.06 & 2.61 $\pm$ 0.22 & 0.25 $\pm$ 0.04 & 0.61 $\pm$ 0.02 \\               
5280475 ~ & 0.0675 & 0.45 $\pm$ 0.01 & 0.27 $\pm$ 0.02 & 3.5 $\pm$ 0.95 & 0.1 $\pm$ 0.08 & 4.72 $\pm$ 0.3 & 1.35 $\pm$ 0.06 & 0.37 $\pm$ 0.01 \\                 
5154381 ~ & 0.0832 & 0.4 $\pm$ 0.02 & 0.25 $\pm$ 0.01 & 0.87 $\pm$ 0.13 & 0.15 $\pm$ 0.08 & 3.34 $\pm$ 0.18 & 1.16 $\pm$ 0.28 & 0.4 $\pm$ 0.04 \\ \hline                 
oAGN (mAGN)~~(red circles) &&&&&&&&\\ \hline                                                                                                                     
5339805$^{*}$$^{+}$ ~ & 0.0894 & 3.14 $\pm$ 0.04 & 3.43 $\pm$ 0.07 & 6.15 $\pm$ 0.14 & 0.74 $\pm$ 0.04 & 2.64 $\pm$ 0.05 & 0.39 $\pm$ 0.03 & 3.0 $\pm$ 0.09 \\         
5151978$^{*}$$^{+}$ ~ & 0.2349 & 0.36 $\pm$ 0.01 & 0.57 $\pm$ 0.02 & 4.41 $\pm$ 1.08 & 1.15 $\pm$ 0.06 & 4.16 $\pm$ 0.27 & 12.17 $\pm$ 0.73 & 0.92 $\pm$ 0.02 \\       
5145801 ~ & 0.1975 & 0.7 $\pm$ 0.01 & 0.8 $\pm$ 0.03 & 2.9 $\pm$ 0.76 & 0.79 $\pm$ 0.05 & 3.37 $\pm$ 0.29 & 2.96 $\pm$ 0.07 & 0.99 $\pm$ 0.06 \\                 
5252110 ~ & 0.1309 & 1.86 $\pm$ 0.03 & 2.53 $\pm$ 0.06 & 8.6 $\pm$ 1.89 & 0.98 $\pm$ 0.04 & 3.29 $\pm$ 0.24 & 3.89 $\pm$ 0.28 & 0.62 $\pm$ 0.02 \\               
5237160 ~ & 0.0604 & 0.59 $\pm$ 0.01 & 0.93 $\pm$ 0.03 & 10.49 $\pm$ 2.26 & 1.15 $\pm$ 0.05 & 4.56 $\pm$ 0.24 & 8.9 $\pm$ 0.46 & 0.56 $\pm$ 0.02 \\              
5340595$^{+}$ ~ & 0.1882 & 1.2 $\pm$ 0.02 & 1.44 $\pm$ 0.04 & 4.67 $\pm$ 1.05 & 0.84 $\pm$ 0.04 & 3.24 $\pm$ 0.25 & 0.69 $\pm$ 0.02 & 2.97 $\pm$ 0.14 \\         
5286102$^{+}$ ~ & 0.2425 & 1.24 $\pm$ 0.02 & 1.37 $\pm$ 0.03 & 3.01 $\pm$ 0.76 & 0.76 $\pm$ 0.04 & 2.84 $\pm$ 0.28 & 0.51 $\pm$ 0.05 & 2.43 $\pm$ 0.12 \\        
5108709 ~ & 0.1528 & 0.44 $\pm$ 0.01 & 0.63 $\pm$ 0.02 & 3.13 $\pm$ 0.16 & 1.04 $\pm$ 0.05 & 3.69 $\pm$ 0.07 & 3.41 $\pm$ 0.12 & 0.5 $\pm$ 0.01 \\               
5369226 ~ & 0.1367 & 1.28 $\pm$ 0.02 & 1.84 $\pm$ 0.05 & 12.08 $\pm$ 2.58 & 1.04 $\pm$ 0.04 & 3.98 $\pm$ 0.23 & 5.51 $\pm$ 0.83 & 0.8 $\pm$ 0.03 \\              
5162821$^{+}$ ~ & 0.2459 & 0.34 $\pm$ 0.01 & 0.42 $\pm$ 0.01 & 0.76 $\pm$ 0.11 & 0.86 $\pm$ 0.05 & 2.66 $\pm$ 0.17 & 4.71 $\pm$ 0.5 & 1.27 $\pm$ 0.08 \\         
5233898 ~ & 0.2039 & 0.82 $\pm$ 0.01 & 1.05 $\pm$ 0.03 & 4.38 $\pm$ 1.02 & 0.91 $\pm$ 0.05 & 3.51 $\pm$ 0.26 & 3.81 $\pm$ 0.63 & 0.63 $\pm$ 0.06 \\              
5197408$^{+}$ ~ & 0.2086 & 0.36 $\pm$ 0.01 & 0.41 $\pm$ 0.01 & 0.81 $\pm$ 0.13 & 0.79 $\pm$ 0.05 & 2.73 $\pm$ 0.17 & 2.81 $\pm$ 0.12 & 2.03 $\pm$ 0.08 \\        
5312967$^{+}$ ~ & 0.1963 & 0.27 $\pm$ 0.01 & 0.31 $\pm$ 0.01 & 0.53 $\pm$ 0.14 & 0.8 $\pm$ 0.06 & 2.58 $\pm$ 0.28 & 6.63 $\pm$ 0.63 & 0.9 $\pm$ 0.07 \\          
5305653$^{+}$ ~ & 0.2124 & 0.69 $\pm$ 0.01 & 0.71 $\pm$ 0.01 & 1.29 $\pm$ 0.11 & 0.67 $\pm$ 0.04 & 2.66 $\pm$ 0.1 & 2.74 $\pm$ 0.28 & 0.87 $\pm$ 0.04 \\         
5216684$^{+}$ ~ & 0.2096 & 1.13 $\pm$ 0.02 & 1.94 $\pm$ 0.05 & 8.59 $\pm$ 1.86 & 1.23 $\pm$ 0.04 & 3.56 $\pm$ 0.24 & 8.23 $\pm$ 0.36 & 0.83 $\pm$ 0.01 \\        
5342686$^{+}$ ~ & 0.1861 & 2.98 $\pm$ 0.03 & 4.1 $\pm$ 0.08 & 9.12 $\pm$ 1.93 & 0.99 $\pm$ 0.04 & 2.85 $\pm$ 0.23 & 4.95 $\pm$ 0.15 & 0.32 $\pm$ 0.02 \\  
5213139$^{+}$ ~ & 0.1228 & 2.43 $\pm$ 0.03 & 2.86 $\pm$ 0.06 & 9.16 $\pm$ 2.0 & 0.83 $\pm$ 0.04 & 3.23 $\pm$ 0.24 & 1.72 $\pm$ 0.06 & 1.37 $\pm$ 0.03 \\         
5155308$^{+}$ ~ & 0.209 & 4.63 $\pm$ 0.03 & 5.99 $\pm$ 0.03 & 21.22 $\pm$ 4.44 & 0.93 $\pm$ 0.03 & 3.33 $\pm$ 0.23 & 1.71 $\pm$ 0.08 & 2.33 $\pm$ 0.04 \\        
5240292$^{+}$ ~ & 0.2238 & 0.5 $\pm$ 0.01 & 0.66 $\pm$ 0.03 & 2.69 $\pm$ 0.7 & 0.93 $\pm$ 0.06 & 3.48 $\pm$ 0.29 & 0.43 $\pm$ 0.02 & 1.45 $\pm$ 0.03 \\ \hline           
oAGN (mWarm)~~(red squares) &&&&&&&&\\ \hline                                                                                                                    
5347780$^{*}$$^{+}$ ~ & 0.089 & 2.15 $\pm$ 0.03 & 1.89 $\pm$ 0.05 & 5.81 $\pm$ 0.14 & 0.51 $\pm$ 0.04 & 3.2 $\pm$ 0.05 & 1.17 $\pm$ 0.05 & 1.53 $\pm$ 0.07 \\          
5294374$^{*}$$^{+}$ ~ & 0.107 & 5.17 $\pm$ 0.06 & 4.63 $\pm$ 0.1 & 14.27 $\pm$ 2.81 & 0.53 $\pm$ 0.04 & 3.2 $\pm$ 0.22 & 1.09 $\pm$ 0.04 & 1.81 $\pm$ 0.03 \\          
5112784$^{+}$ ~ & 0.0578 & 1.74 $\pm$ 0.02 & 1.49 $\pm$ 0.04 & 2.31 $\pm$ 0.62 & 0.48 $\pm$ 0.04 & 2.52 $\pm$ 0.3 & 0.96 $\pm$ 0.18 & 2.52 $\pm$ 0.38 \\         
5153772 ~ & 0.189 & 0.44 $\pm$ 0.01 & 0.48 $\pm$ 0.02 & 4.79 $\pm$ 1.15 & 0.75 $\pm$ 0.06 & 4.43 $\pm$ 0.27 & 7.62 $\pm$ 0.4 & 0.8 $\pm$ 0.06 \\                 
5351862$^{+}$ ~ & 0.1103 & 0.75 $\pm$ 0.01 & 0.7 $\pm$ 0.02 & 3.62 $\pm$ 0.89 & 0.56 $\pm$ 0.05 & 3.74 $\pm$ 0.27 & 3.43 $\pm$ 0.17 & 1.19 $\pm$ 0.06 \\         
5369229 ~ & 0.1402 & 0.53 $\pm$ 0.02 & 0.48 $\pm$ 0.02 & 2.97 $\pm$ 0.17 & 0.55 $\pm$ 0.06 & 3.92 $\pm$ 0.08 & 7.45 $\pm$ 0.52 & 0.71 $\pm$ 0.06 \\              
5346487$^{+}$ ~ & 0.1898 & 0.45 $\pm$ 0.01 & 0.44 $\pm$ 0.02 & 3.46 $\pm$ 0.88 & 0.64 $\pm$ 0.06 & 4.17 $\pm$ 0.28 & 0.99 $\pm$ 0.04 & 1.57 $\pm$ 0.04 \\        
5110547$^{+}$ ~ & 0.0889 & 0.96 $\pm$ 0.01 & 0.78 $\pm$ 0.02 & 2.46 $\pm$ 0.65 & 0.41 $\pm$ 0.04 & 3.24 $\pm$ 0.29 & 6.17 $\pm$ 0.32 & 0.8 $\pm$ 0.02 \\         
5180136 ~ & 0.1056 & 0.61 $\pm$ 0.01 & 0.52 $\pm$ 0.02 & 1.76 $\pm$ 0.18 & 0.47 $\pm$ 0.05 & 3.32 $\pm$ 0.12 & 7.12 $\pm$ 0.67 & 0.7 $\pm$ 0.02 \\               
5137338$^{+}$ ~ & 0.2597 & 0.45 $\pm$ 0.01 & 0.43 $\pm$ 0.02 & 0.65 $\pm$ 0.14 & 0.6 $\pm$ 0.06 & 2.48 $\pm$ 0.24 & 1.77 $\pm$ 0.31 & 0.67 $\pm$ 0.04 \\         
5196019 ~ & 0.1258 & 0.69 $\pm$ 0.01 & 0.72 $\pm$ 0.03 & 5.65 $\pm$ 1.31 & 0.69 $\pm$ 0.05 & 4.18 $\pm$ 0.26 & 6.96 $\pm$ 0.68 & 0.78 $\pm$ 0.02 \\              
5211450 ~ & 0.2535 & 0.36 $\pm$ 0.01 & 0.34 $\pm$ 0.01 & 1.32 $\pm$ 0.41 & 0.6 $\pm$ 0.06 & 3.43 $\pm$ 0.34 & 8.64 $\pm$ 1.51 & 0.51 $\pm$ 0.05 \\               
5367800$^{+}$ ~ & 0.2627 & 0.43 $\pm$ 0.01 & 0.43 $\pm$ 0.02 & 0.41 $\pm$ 0.08 & 0.65 $\pm$ 0.06 & 2.04 $\pm$ 0.22 & 1.19 $\pm$ 0.1 & 1.34 $\pm$ 0.14 \\ \hline   
oAGN (mSF)~~(yellow circles) &&&&&&&&\\ \hline                                                                                                                   
5249547$^{*}$ ~ & 0.0834 & 1.36 $\pm$ 0.02 & 0.8 $\pm$ 0.02 & 2.13 $\pm$ 0.16 & 0.07 $\pm$ 0.04 & 3.09 $\pm$ 0.09 & 2.17 $\pm$ 0.21 & 0.79 $\pm$ 0.03 \\               
5155115$^{*}$ ~ & 0.0869 & 1.7 $\pm$ 0.02 & 0.99 $\pm$ 0.03 & 2.2 $\pm$ 0.18 & 0.06 $\pm$ 0.05 & 2.92 $\pm$ 0.1 & 5.56 $\pm$ 0.57 & 0.93 $\pm$ 0.05 \\                 
5163580 $^{+}$ ~ & 0.1228 & 0.51 $\pm$ 0.01 & 0.36 $\pm$ 0.02 & 0.39 $\pm$ 0.1 & 0.25 $\pm$ 0.06 & 2.22 $\pm$ 0.29 & 2.54 $\pm$ 0.2 & 2.71 $\pm$ 0.19 \\         
5368644 ~ & 0.1194 & 0.43 $\pm$ 0.01 & 0.34 $\pm$ 0.01 & 1.18 $\pm$ 0.13 & 0.37 $\pm$ 0.06 & 3.35 $\pm$ 0.13 & 2.55 $\pm$ 0.3 & 0.6 $\pm$ 0.08 \\                
5278828 ~ & 0.1388 & 0.22 $\pm$ 0.01 & 0.18 $\pm$ 0.01 & 1.15 $\pm$ 0.13 & 0.42 $\pm$ 0.1 & 3.97 $\pm$ 0.15 & 2.7 $\pm$ 0.46 & 0.59 $\pm$ 0.02 \\                
5139540 $^{+}$ ~ & 0.2323 & 0.36 $\pm$ 0.01 & 0.2 $\pm$ 0.01 & 0.59 $\pm$ 0.13 & 0.01 $\pm$ 0.07 & 3.18 $\pm$ 0.25 & 0.49 $\pm$ 0.09 & 1.89 $\pm$ 0.15 \\        
5305775 ~ & 0.133 & 0.63 $\pm$ 0.01 & 0.41 $\pm$ 0.02 & 2.32 $\pm$ 0.59 & 0.17 $\pm$ 0.06 & 3.86 $\pm$ 0.28 & 2.35 $\pm$ 0.34 & 0.54 $\pm$ 0.03 \\               
5355371 $^{+}$ ~ & 0.2006 & 0.25 $\pm$ 0.01 & 0.2 $\pm$ 0.01 & 0.92 $\pm$ 0.1 & 0.42 $\pm$ 0.06 & 3.62 $\pm$ 0.13 & 8.12 $\pm$ 0.28 & 0.58 $\pm$ 0.01 \\         
5188449 $^{+}$ ~ & 0.1169 & 0.2 $\pm$ 0.01 & 0.11 $\pm$ 0.01 & 0.24 $\pm$ 0.1 & -0.01 $\pm$ 0.14 & 2.95 $\pm$ 0.46 & 3.32 $\pm$ 0.2 & 0.68 $\pm$ 0.04 \\         
5427366 ~ & 0.0881 & 0.45 $\pm$ 0.01 & 0.32 $\pm$ 0.01 & 0.87 $\pm$ 0.11 & 0.27 $\pm$ 0.05 & 3.11 $\pm$ 0.15 & 1.84 $\pm$ 0.17 & 0.68 $\pm$ 0.04 \\              
5163246 ~ & 0.2064 & 0.37 $\pm$ 0.01 & 0.25 $\pm$ 0.01 & 0.3 $\pm$ 0.07 & 0.2 $\pm$ 0.06 & 2.34 $\pm$ 0.26 & 4.53 $\pm$ 0.32 & 0.63 $\pm$ 0.02 \\                
5258350 ~ & 0.0768 & 0.89 $\pm$ 0.01 & 0.6 $\pm$ 0.01 & 1.94 $\pm$ 0.17 & 0.21 $\pm$ 0.04 & 3.28 $\pm$ 0.1 & 3.07 $\pm$ 0.16 & 0.59 $\pm$ 0.01 \\                
5154519 ~ & 0.2123 & 0.21 $\pm$ 0.01 & 0.17 $\pm$ 0.01 & 0.95 $\pm$ 0.12 & 0.43 $\pm$ 0.08 & 3.83 $\pm$ 0.15 & 3.61 $\pm$ 0.73 & 0.79 $\pm$ 0.05 \\        
5271798 ~ & 0.152 & 0.36 $\pm$ 0.01 & 0.29 $\pm$ 0.01 & 3.34 $\pm$ 0.81 & 0.4 $\pm$ 0.06 & 4.61 $\pm$ 0.27 & 4.59 $\pm$ 0.16 & 0.54 $\pm$ 0.01 \\ \hline                 
non-oAGN (mAGN)~~(orange circles)&&&&&&&&\\ \hline                                                                                                              
5200866$^{*}$ ~ & 0.1218 & 0.97 $\pm$ 0.01 & 1.38 $\pm$ 0.04 & 17.47 $\pm$ 3.64 & 1.04 $\pm$ 0.04 & 4.68 $\pm$ 0.23 & 0.97 $\pm$ 0.02 & 0.35 $\pm$ 0.0 \\              
5158890$^{*}$ ~ & 0.1772 & 1.43 $\pm$ 0.02 & 1.63 $\pm$ 0.04 & 8.14 $\pm$ 0.15 & 0.79 $\pm$ 0.04 & 3.7 $\pm$ 0.04 & 0.95 $\pm$ 0.07 & 0.51 $\pm$ 0.01 \\               
5275222$^{*}$$^{+}$ ~ & 0.2101 & 0.75 $\pm$ 0.01 & 0.84 $\pm$ 0.03 & 2.08 $\pm$ 0.62 & 0.78 $\pm$ 0.06 & 2.96 $\pm$ 0.33 & 0.29 $\pm$ 0.01 & 0.61 $\pm$ 0.01 \\        
5241310$^{*}$$^{+}$ ~ & 0.2283 & 0.4 $\pm$ 0.01 & 0.55 $\pm$ 0.04 & 1.9 $\pm$ 0.54 & 0.98 $\pm$ 0.09 & 3.31 $\pm$ 0.32 & 1.35 $\pm$ 0.1 & 0.42 $\pm$ 0.02 \\           
5362108 ~ & 0.1669 & 0.36 $\pm$ 0.01 & 0.46 $\pm$ 0.02 & 3.47 $\pm$ 0.84 & 0.92 $\pm$ 0.06 & 4.13 $\pm$ 0.27 & 0.93 $\pm$ 0.23 & 0.28 $\pm$ 0.05 \\              
5154472 ~ & 0.1107 & 0.25 $\pm$ 0.01 & 0.31 $\pm$ 0.01 & 1.59 $\pm$ 0.12 & 0.91 $\pm$ 0.07 & 3.71 $\pm$ 0.1 & 0.45 $\pm$ 0.15 & 0.39 $\pm$ 0.03 \\               
5129662 ~ & 0.2045 & 0.44 $\pm$ 0.01 & 0.52 $\pm$ 0.02 & 2.18 $\pm$ 0.56 & 0.83 $\pm$ 0.06 & 3.51 $\pm$ 0.28 & 1.73 $\pm$ 0.1 & 0.5 $\pm$ 0.01 \\                
5246095$^{+}$ ~ & 0.2184 & 0.7 $\pm$ 0.01 & 0.84 $\pm$ 0.03 & 3.08 $\pm$ 0.73 & 0.83 $\pm$ 0.05 & 3.38 $\pm$ 0.26 & 0.78 $\pm$ 0.04 & 0.71 $\pm$ 0.01 \\         
5119859 ~ & 0.1865 & 0.2 $\pm$ 0.01 & 0.25 $\pm$ 0.01 & 1.41 $\pm$ 0.14 & 0.92 $\pm$ 0.08 & 3.81 $\pm$ 0.13 & 0.2 $\pm$ 0.06 & 0.34 $\pm$ 0.03 \\          
5317117$^{+}$ ~ & 0.2038 & 0.79 $\pm$ 0.01 & 0.92 $\pm$ 0.01 & 2.35 $\pm$ 0.12 & 0.81 $\pm$ 0.04 & 3.0 $\pm$ 0.07 & 0.57 $\pm$ 0.02 & 0.26 $\pm$ 0.01 \\         
5247018$^{+}$ ~ & 0.209 & 0.22 $\pm$ 0.01 & 0.49 $\pm$ 0.03 & 0.78 $\pm$ 0.21 & 1.52 $\pm$ 0.07 & 2.48 $\pm$ 0.3 & 0.3 $\pm$ 0.02 & 1.01 $\pm$ 0.05 \\ \hline            
non-oAGN (mWarm)~~(orange squares)&&&&&&&&\\ \hline                                                                                                              
5204947$^{*}$ ~ & 0.2438 & 0.68 $\pm$ 0.01 & 0.73 $\pm$ 0.03 & 4.38 $\pm$ 1.08 & 0.72 $\pm$ 0.05 & 3.9 $\pm$ 0.27 & 1.86 $\pm$ 0.4 & 0.39 $\pm$ 0.01 \\                
5219936 ~ & 0.0935 & 0.63 $\pm$ 0.01 & 0.66 $\pm$ 0.03 & 6.72 $\pm$ 1.51 & 0.7 $\pm$ 0.06 & 4.47 $\pm$ 0.25 & 1.02 $\pm$ 0.06 & 0.3 $\pm$ 0.01 \\                
5282867 ~ & 0.0846 & 0.25 $\pm$ 0.01 & 0.21 $\pm$ 0.01 & 0.85 $\pm$ 0.15 & 0.46 $\pm$ 0.09 & 3.49 $\pm$ 0.2 & 0.71 $\pm$ 0.14 & 0.26 $\pm$ 0.02 \\               
5310140 ~ & 0.2038 & 0.26 $\pm$ 0.01 & 0.23 $\pm$ 0.01 & 1.41 $\pm$ 0.12 & 0.54 $\pm$ 0.07 & 3.9 $\pm$ 0.11 & 0.84 $\pm$ 0.16 & 0.34 $\pm$ 0.02 \\               
5100400 ~ & 0.2733 & 0.52 $\pm$ 0.01 & 0.41 $\pm$ 0.02 & 0.65 $\pm$ 0.11 & 0.37 $\pm$ 0.06 & 2.57 $\pm$ 0.19 & 0.11 $\pm$ 0.04 & 0.42 $\pm$ 0.02 \\              
5115102 ~ & 0.2989 & 0.22 $\pm$ 0.01 & 0.2 $\pm$ 0.01 & 0.85 $\pm$ 0.1 & 0.55 $\pm$ 0.08 & 3.55 $\pm$ 0.15 & 0.61 $\pm$ 0.13 & 0.48 $\pm$ 0.03 \\                
5154420 ~ & 0.1993 & 0.28 $\pm$ 0.01 & 0.25 $\pm$ 0.01 & 1.56 $\pm$ 0.13 & 0.52 $\pm$ 0.07 & 3.96 $\pm$ 0.11 & 0.3 $\pm$ 0.09 & 0.4 $\pm$ 0.02 \\ 
5228724$^{+}$ ~ & 0.2367 & 0.44 $\pm$ 0.01 & 0.43 $\pm$ 0.01 & 1.48 $\pm$ 0.09 & 0.63 $\pm$ 0.04 & 3.3 $\pm$ 0.08 & 1.36 $\pm$ 0.14 & 0.54 $\pm$ 0.02 \\         
5350683 ~ & 0.2508 & 0.27 $\pm$ 0.01 & 0.26 $\pm$ 0.01 & 1.84 $\pm$ 0.14 & 0.61 $\pm$ 0.06 & 4.06 $\pm$ 0.1 & 0.68 $\pm$ 0.15 & 0.54 $\pm$ 0.03 \\
5133213 ~ & 0.2059 & 0.29 $\pm$ 0.01 & 0.28 $\pm$ 0.01 & 1.7 $\pm$ 0.12 & 0.61 $\pm$ 0.07 & 3.92 $\pm$ 0.09 & 0.24 $\pm$ 0.07 & 0.45 $\pm$ 0.02 \\
\enddata
\end{deluxetable}
\end{longrotatetable}

\newpage

\startlongtable
\begin{deluxetable*}{lcccc}

\tablecaption{Derived properties of a representative sample of the galaxies from each activity group.   
 The stellar masses are derived based on the M/L ratio from Cluver et al.  (\citeyear{Cluver2014}). The star formation rate (SFR$_{12 \textit{$\mu$} \text{m}}$) are derived using the $\nu$L$\nu$(12\,$\mu$m) and the calibration from Cluver et al. (\citeyear{Cluver2017}).  \label{table1.3k}}
\tabletypesize{\scriptsize}
\tablehead{\colhead{CATAID } &  \colhead{ $\mathrm{LogM_{stellar}\;[ M_{\odot}]}$} & \colhead{SFR$_{12 \textit{$\mu$} \text{m}}$}  &\colhead{Log (L[\ion{O}{iii}]) }& \colhead{Log (sSFR)} \\ 
\colhead{} & \colhead{ ($\mathrm{ M_{\odot}}$)} & \colhead{ (M$_{\odot}$yr$^{-1}$)} &  \colhead{(erg s$^{-1}$) } & \colhead{ (yr$^{-1}$)}\\
} 
\startdata
SF~~(blue circles)&&&&\\ \hline
5241095$^{*}$ ~ & 9.05 $\pm$ 0.20 & 0.2 $\pm$ 0.09 & 41.14 $\pm$ 0.0 & -9.75 $\pm$ 0.27 \\
5240983$^{*}$ ~ & 9.14 $\pm$ 0.19 & 0.49 $\pm$ 0.21 & 40.94 $\pm$ 0.01 & -9.45 $\pm$ 0.27 \\ 
5306682$^{*}$ ~ & 9.43 $\pm$ 0.10 & 0.21 $\pm$ 0.08 & 39.44 $\pm$ 0.01 & -10.11 $\pm$ 0.19 \\
5135645$^{*}$ ~ & 10.78 $\pm$ 0.16 & 73.44 $\pm$ 29.20 & 41.25 $\pm$ 0.03 & -8.91 $\pm$ 0.23 \\
5325430 ~ & 10.19 $\pm$ 0.18 & 2.3 $\pm$ 0.81 & 40.23 $\pm$ 0.08 & -9.83 $\pm$ 0.24 \\
5327563 ~ & 9.66 $\pm$ 0.24 & 1.43 $\pm$ 0.69 & -- & -9.51 $\pm$ 0.32 \\
5325339 ~ & 10.48 $\pm$ 0.17 & 7.0 $\pm$ 2.45 & -- & -9.64 $\pm$ 0.23 \\
5123445 ~ & 9.85 $\pm$ 0.23 & 2.2 $\pm$ 0.82 & -- & -9.51 $\pm$ 0.28 \\
5138489 ~ & 9.82 $\pm$ 0.33 & 1.57 $\pm$ 0.64 & 40.23 $\pm$ 0.14 & -9.62 $\pm$ 0.38 \\
5153375 ~ & 9.85 $\pm$ 0.16 & 2.53 $\pm$ 0.91 & 40.29 $\pm$ 0.08 & -9.44 $\pm$ 0.23 \\
5261290 ~ & 10.35 $\pm$ 0.13 & 9.29 $\pm$ 3.70 & 40.35 $\pm$ 0.11 & -9.39 $\pm$ 0.22 \\
5187670 ~ & 10.25 $\pm$ 0.38 & 2.63 $\pm$ 0.98 & 39.94 $\pm$ 0.13 & -9.83 $\pm$ 0.42 \\
5281879 ~ & 10.04 $\pm$ 0.11 & 1.54 $\pm$ 0.54 & 39.87 $\pm$ 0.05 & -9.85 $\pm$ 0.19 \\
5281521 ~ & 9.51 $\pm$ 0.28 & 0.15 $\pm$ 0.07 & 40.01 $\pm$ 0.02 & -10.35 $\pm$ 0.35 \\
5305483 ~ & 9.98 $\pm$ 0.19 & 3.42 $\pm$ 1.42 & 40.54 $\pm$ 0.04 & -9.45 $\pm$ 0.26 \\
5222747 ~ & 10.50 $\pm$ 0.18 & 7.13 $\pm$ 2.65 & 40.76 $\pm$ 0.11 & -9.65 $\pm$ 0.24 \\
5266202 ~ & 10.20 $\pm$ 0.18 & 5.67 $\pm$ 1.98 & 40.09 $\pm$ 0.14 & -9.44 $\pm$ 0.24 \\
5185570 ~ & 10.01 $\pm$ 0.23 & 0.38 $\pm$ 0.19 & 39.86 $\pm$ 0.05 & -10.43 $\pm$ 0.31 \\
5256068 ~ & 11.05 $\pm$ 0.10 & 70.22 $\pm$ 24.32 & 41.47 $\pm$ 0.04 & -9.2 $\pm$ 0.18 \\ \hline
Composites~~(green circles)&&&&\\ \hline
5205726$^{*}$ ~ & 10.51 $\pm$ 0.11 & 1.94 $\pm$ 0.68 & 40.29 $\pm$ 0.21 & -10.22 $\pm$ 0.19 \\
5322507 ~ & 10.47 $\pm$ 0.13 & 1.19 $\pm$ 0.44 & -- & -10.39 $\pm$ 0.2 \\
5246049 ~ & 10.45 $\pm$ 0.14 & 13.63 $\pm$ 5.44 & 40.57 $\pm$ 0.07 & -9.32 $\pm$ 0.22 \\
5277048 ~ & 9.85 $\pm$ 0.31 & 4.27 $\pm$ 1.56 & -- & -9.22 $\pm$ 0.35 \\
5251443 ~ & 11.23 $\pm$ 0.24 & 6.3 $\pm$ 2.90 & 40.62 $\pm$ 0.13 & -10.43 $\pm$ 0.31 \\
5279625 ~ & 10.11 $\pm$ 0.15 & 4.86 $\pm$ 1.95 & 40.32 $\pm$ 0.06 & -9.43 $\pm$ 0.23 \\
5285888 ~ & 10.78 $\pm$ 0.17 & 29.91 $\pm$ 12.21 & 41.5 $\pm$ 0.07 & -9.31 $\pm$ 0.25 \\
5159722 ~ & 10.37 $\pm$ 0.13 & 7.32 $\pm$ 2.54 & 40.64 $\pm$ 0.07 & -9.51 $\pm$ 0.20 \\
5154601 ~ & 10.37 $\pm$ 0.21 & 1.31 $\pm$ 0.50 & 40.18 $\pm$ 0.07 & -10.25 $\pm$ 0.27 \\ 
5264991 ~ & 10.55 $\pm$ 0.19 & 18.81 $\pm$ 6.60 & 41.2 $\pm$ 0.07 & -9.27 $\pm$ 0.24 \\
5364072 ~ & 10.35 $\pm$ 0.35 & 1.55 $\pm$ 0.62 & 40.56 $\pm$ 0.08 & -10.16 $\pm$ 0.39 \\
5249313 ~ & 10.50 $\pm$ 0.14 & 1.66 $\pm$ 0.64 & 40.01 $\pm$ 0.07 & -10.28 $\pm$ 0.22 \\
5280475 ~ & 9.99 $\pm$ 0.21 & 4.17 $\pm$ 1.75 & 40.54 $\pm$ 0.02 & -9.38 $\pm$ 0.28 \\
5154381 ~ & 10.01 $\pm$ 0.20 & 1.8 $\pm$ 0.67 & 40.64 $\pm$ 0.1 & -9.75 $\pm$ 0.26 \\ \hline
oAGN (mAGN)~~(red circles) &&&&\\ \hline
5339805$^{*}$ + & 10.85 $\pm$ 0.10 & 11.54 $\pm$ 4.00 & 41.4 $\pm$ 0.03 & -9.79 $\pm$ 0.18 \\
5151978$^{*}$ + & 10.83 $\pm$ 0.14 & 55.33 $\pm$ 22.54 & 42.37 $\pm$ 0.0 & -9.08 $\pm$ 0.23 \\
5145801 ~ & 10.95 $\pm$ 0.13 & 27.1 $\pm$ 11.27 & -- & -9.51 $\pm$ 0.22 \\
5252110 ~ & 10.97 $\pm$ 0.11 & 31.75 $\pm$ 12.57 & -- & -9.48 $\pm$ 0.20 \\
5237160 ~ & 9.76 $\pm$ 0.13 & 8.9 $\pm$ 3.51 & -- & -8.81 $\pm$ 0.21 \\
5340595 + ~ & 11.13 $\pm$ 0.11 & 37.46 $\pm$ 14.92 & -- & -9.56 $\pm$ 0.21 \\
5286102 + & 11.40 $\pm$ 0.11 & 42.13 $\pm$ 17.32 & 41.63 $\pm$ 0.04 & -9.77 $\pm$ 0.21 \\
5108709 ~ & 10.50 $\pm$ 0.13 & 17.61 $\pm$ 6.15 & 42.11 $\pm$ 0.01 & -9.25 $\pm$ 0.20 \\
5369226 ~ & 10.85 $\pm$ 0.11 & 46.57 $\pm$ 18.33 & 41.73 $\pm$ 0.02 & -9.19 $\pm$ 0.20 \\
5162821 + & 10.85 $\pm$ 0.12 & 12.99 $\pm$ 4.80 & 41.89 $\pm$ 0.03 & -9.74 $\pm$ 0.20 \\
5233898 ~ & 11.05 $\pm$ 0.12 & 41.49 $\pm$ 16.68 & 42.15 $\pm$ 0.02 & -9.43 $\pm$ 0.21 \\
5197408 + & 10.71 $\pm$ 0.14 & 9.86 $\pm$ 3.67 & 41.97 $\pm$ 0.01 & -9.72 $\pm$ 0.21 \\
5312967 + & 10.53 $\pm$ 0.16 & 6.06 $\pm$ 2.50 & 41.68 $\pm$ 0.01 & -9.75 $\pm$ 0.24 \\
5305653 + & 11.01 $\pm$ 0.11 & 15.34 $\pm$ 5.42 & 41.92 $\pm$ 0.01 & -9.83 $\pm$ 0.19 \\
5216684 + & 11.21 $\pm$ 0.11 & 79.31 $\pm$ 31.31 & 42.99 $\pm$ 0.0 & -9.31 $\pm$ 0.20 \\
5342686 + & 11.52 $\pm$ 0.10 & 66.05 $\pm$ 25.95 & 42.78 $\pm$ 0.0 & -9.70 $\pm$ 0.20 \\ 
5213139 + & 11.03 $\pm$ 0.10 & 29.78 $\pm$ 11.78 & 42.17 $\pm$ 0.01 & -9.56 $\pm$ 0.20 \\
5155308 + & 11.82 $\pm$ 0.08 & 174.43 $\pm$ 68.32 & 42.13 $\pm$ 0.01 & -9.58 $\pm$ 0.19 \\
5240292 + & 10.93 $\pm$ 0.15 & 32.53 $\pm$ 13.51 & 41.52 $\pm$ 0.01 & -9.41 $\pm$ 0.24 \\ \hline
oAGN (mWarm)~~(red squares) &&&&\\ \hline
5347780$^{*}$ + ~ & 10.68 $\pm$ 0.11 & 10.86 $\pm$ 3.77 & -- & -9.64 $\pm$ 0.19 \\
5294374$^{*}$ + & 11.23 $\pm$ 0.10 & 33.8 $\pm$ 13.08 & 42.0 $\pm$ 0.01 & -9.7 $\pm$ 0.2 \\
5112784 + ~ & 10.19 $\pm$ 0.11 & 2.17 $\pm$ 0.91 & -- & -9.86 $\pm$ 0.21 \\
5153772 ~ & 10.70 $\pm$ 0.15 & 38.63 $\pm$ 15.66 & -- & -9.12 $\pm$ 0.23 \\
5351862 + ~ & 10.42 $\pm$ 0.13 & 10.73 $\pm$ 4.37 & -- & -9.39 $\pm$ 0.22 \\
5369229 ~ & 10.50 $\pm$ 0.16 & 14.24 $\pm$ 4.98 & -- & -9.34 $\pm$ 0.22 \\
5346487 + & 10.71 $\pm$ 0.15 & 29.29 $\pm$ 12.05 & 41.71 $\pm$ 0.01 & -9.25 $\pm$ 0.23 \\
5110547 + & 10.33 $\pm$ 0.11 & 5.11 $\pm$ 2.13 & 41.7 $\pm$ 0.01 & -9.62 $\pm$ 0.21 \\
5180136 ~ & 10.29 $\pm$ 0.12 & 5.25 $\pm$ 1.88 & 41.75 $\pm$ 0.01 & -9.57 $\pm$ 0.2 \\
5137338 + & 11.03 $\pm$ 0.15 & 12.56 $\pm$ 4.96 & 41.51 $\pm$ 0.04 & -9.93 $\pm$ 0.23 \\
5196019 ~ & 10.51 $\pm$ 0.13 & 20.38 $\pm$ 8.18 & 41.68 $\pm$ 0.01 & -9.2 $\pm$ 0.22 \\
5211450 ~ & 10.90 $\pm$ 0.15 & 22.25 $\pm$ 9.79 & 42.56 $\pm$ 0.01 & -9.55 $\pm$ 0.24 \\
5367800 + & 11.01 $\pm$ 0.14 & 8.59 $\pm$ 3.32 & 41.8 $\pm$ 0.03 & -10.08 $\pm$ 0.22 \\ \hline
oAGN (mSF)~~(yellow circles) &&&&\\ \hline
5249547$^{*}$ ~ & 10.74 $\pm$ 0.11 & 3.99 $\pm$ 1.40 & 41.19 $\pm$ 0.02 & -10.14 $\pm$ 0.19 \\
5155115$^{*}$ ~ & 10.92 $\pm$ 0.12 & 4.44 $\pm$ 1.57 & -- & -10.28 $\pm$ 0.19 \\
5163580 + ~ & 10.36 $\pm$ 0.16 & 1.85 $\pm$ 0.77 & -- & -10.09 $\pm$ 0.24 \\
5368644 ~ & 10.26 $\pm$ 0.15 & 4.66 $\pm$ 1.67 & -- & -9.59 $\pm$ 0.22 \\
5278828 ~ & 10.11 $\pm$ 0.25 & 6.08 $\pm$ 2.18 & 41.0 $\pm$ 0.07 & -9.33 $\pm$ 0.29 \\
5139540 + & 11.31 $\pm$ 0.18 & 9.17 $\pm$ 3.65 & 41.15 $\pm$ 0.07 & -10.35 $\pm$ 0.25 \\
5305775 ~ & 10.60 $\pm$ 0.15 & 10.37 $\pm$ 4.28 & 41.35 $\pm$ 0.06 & -9.59 $\pm$ 0.23 \\
5355371 + & 10.51 $\pm$ 0.16 & 10.19 $\pm$ 3.67 & 42.5 $\pm$ 0.0 & -9.5 $\pm$ 0.22 \\
5188449 + & 10.44 $\pm$ 0.35 & 1.12 $\pm$ 0.55 & 40.99 $\pm$ 0.02 & -10.39 $\pm$ 0.41 \\
5427366 ~ & 9.99 $\pm$ 0.13 & 2.01 $\pm$ 0.73 & -- & -9.69 $\pm$ 0.21 \\
5163246 ~ & 10.71 $\pm$ 0.15 & 4.06 $\pm$ 1.63 & 42.42 $\pm$ 0.01 & -10.11 $\pm$ 0.23 \\
5258350 ~ & 10.16 $\pm$ 0.10 & 3.15 $\pm$ 1.12 & 41.24 $\pm$ 0.01 & -9.66 $\pm$ 0.19 \\
5154519 ~ & 10.49 $\pm$ 0.20 & 11.78 $\pm$ 4.28 & 41.87 $\pm$ 0.02 & -9.42 $\pm$ 0.26 \\ 
5271798 ~ & 10.41 $\pm$ 0.16 & 18.45 $\pm$ 7.49 & 42.1 $\pm$ 0.0 & -9.14 $\pm$ 0.24 \\ \hline
non-oAGN (mAGN)~~(orange circles)&&&&\\ \hline
5200866$^{*}$ ~ & 10.63 $\pm$ 0.11 & 51.72 $\pm$ 20.24 & 41.69 $\pm$ 0.01 & -8.91 $\pm$ 0.2 \\
5158890$^{*}$ ~ & 11.15 $\pm$ 0.11 & 54.4 $\pm$ 18.85 & 41.75 $\pm$ 0.02 & -9.42 $\pm$ 0.19 \\
5275222$^{*}$ + & 11.04 $\pm$ 0.14 & 22.92 $\pm$ 9.91 & 41.78 $\pm$ 0.01 & -9.68 $\pm$ 0.24 \\
5241310$^{*}$ + & 10.85 $\pm$ 0.22 & 24.95 $\pm$ 10.65 & 41.39 $\pm$ 0.02 & -9.45 $\pm$ 0.29 \\
5362108 ~ & 10.49 $\pm$ 0.16 & 22.85 $\pm$ 9.29 & -- & -9.14 $\pm$ 0.24 \\
5154472 ~ & 9.94 $\pm$ 0.17 & 5.25 $\pm$ 1.85 & 40.42 $\pm$ 0.13 & -9.22 $\pm$ 0.23 \\
5129662 ~ & 10.78 $\pm$ 0.15 & 22.59 $\pm$ 9.33 & 41.87 $\pm$ 0.01 & -9.43 $\pm$ 0.24 \\
5246095 + & 11.05 $\pm$ 0.13 & 34.86 $\pm$ 14.10 & 41.68 $\pm$ 0.02 & -9.51 $\pm$ 0.22 \\
5119859 ~ & 10.34 $\pm$ 0.19 & 12.84 $\pm$ 4.59 & 40.56 $\pm$ 0.12 & -9.23 $\pm$ 0.25 \\ 
5317117 + & 11.03 $\pm$ 0.09 & 23.98 $\pm$ 8.37 & 41.89 $\pm$ 0.01 & -9.65 $\pm$ 0.18 \\
5247018 + & 10.49 $\pm$ 0.18 & 9.58 $\pm$ 4.03 & 41.4 $\pm$ 0.03 & -9.51 $\pm$ 0.26 \\ \hline
non-oAGN (mWarm)~~(orange squares)&&&&\\ \hline
5204947$^{*}$ ~ & 11.14 $\pm$ 0.14 & 59.25 $\pm$ 24.19 & 42.19 $\pm$ 0.07 & -9.37 $\pm$ 0.22 \\
5219936 ~ & 10.19 $\pm$ 0.14 & 13.57 $\pm$ 5.41 & 41.36 $\pm$ 0.02 & -9.06 $\pm$ 0.22 \\
5282867 ~ & 9.70 $\pm$ 0.22 & 1.84 $\pm$ 0.70 & 40.28 $\pm$ 0.07 & -9.44 $\pm$ 0.27 \\
5310140 ~ & 10.54 $\pm$ 0.19 & 15.28 $\pm$ 5.42 & 41.19 $\pm$ 0.06 & -9.36 $\pm$ 0.24 \\
5100400 ~ & 11.14 $\pm$ 0.15 & 13.99 $\pm$ 5.27 & 40.77 $\pm$ 0.14 & -9.99 $\pm$ 0.22 \\
5115102 ~ & 10.84 $\pm$ 0.22 & 21.03 $\pm$ 7.63 & 41.0 $\pm$ 0.07 & -9.52 $\pm$ 0.27 \\
5154420 ~ & 10.55 $\pm$ 0.17 & 16.05 $\pm$ 5.67 & 40.57 $\pm$ 0.13 & -9.35 $\pm$ 0.23 \\
5228724 + & 10.92 $\pm$ 0.12 & 21.46 $\pm$ 7.52 & 41.47 $\pm$ 0.03 & -9.59 $\pm$ 0.19 \\
5350683 ~ & 10.77 $\pm$ 0.15 & 29.26 $\pm$ 10.31 & 41.02 $\pm$ 0.09 & -9.3 $\pm$ 0.22 \\
5133213 ~ & 10.60 $\pm$ 0.16 & 18.44 $\pm$ 6.48 & 40.78 $\pm$ 0.11 & -9.34 $\pm$ 0.22 \\
\enddata
\end{deluxetable*}

\end{document}